# Electronic Properties of Nano and Molecular Quantum Devices

Oday Arkan Abbas Al-Owaedi

Ph.D. Thesis in Nanoelectronics

Department of Physics, Lancaster University, UK

Lancaster University

This Thesis is submitted in partial fulfilment of the requirements for degree of Doctor of Philosophy

November 2016

# Declaration

I hereby declare that the thesis is my own work and effort and has not been submitted in substantially the same form for the award of a higher degree elsewhere. Other sources of information have been used, they have been acknowledged. This thesis documents work carried out between May 2013 and October 2016 at Lancaster University, UK, under the supervision of Prof. Colin J. Lambert and funded by Ministry of Higher Education and Scientific Research of Iraq in partnership with University of Babylon, Iraq.

Oday A. Al-Owaedi

November 15 2016

*I would like to dedicate my thesis to the memory of my late beloved brother Qusai*

# Abstract

The exploring and understanding the electronic properties of molecules connected to metallic leads is a vital part of nanoscience if molecule is to have a future. This thesis documents a study for various families of organic and organometallic molecules, which offer unique concepts and new insights into the electronic properties of molecular junctions. Different families of molecules were studied using a combination of density functional theory (DFT) and non-equilibrium Green's function formalism of transport theory. The main results of this thesis are as follows:

A quantum circuit rule for combining quantum interference effects in the conductive properties of oligo(phenyleneethynylene) (OPE)-type molecules possessing three aromatic rings was investigated both theoretically and experimentally. Molecules were of the type X-Y-X, where X represents pyridyl anchors with para (p), meta (m) or ortho (o) connectivities and Y represents a phenyl ring with p and m connectivities. The conductances $G_{xmx}$ ($G_{xpx}$) of molecules of the form X-m-X (X-p-X), with meta (para) connections in the central ring, were predominantly lower (higher), irrespective of the meta, para or ortho nature of the anchor groups X, demonstrating that conductance is controlled by the nature of quantum interference in the central ring Y. The single-molecule conductances were found to satisfy the quantum circuit rule Gppp/Gpmp=Gmpm/Gmmm. This demonstrates that the contribution to the conductance from the central ring is independent of the para versus meta nature of the anchor groups.

The conductance and the decay of conductance as a function of molecular length within a homologous series of oligoynes, Me3Si— (C≡C)n—SiMe3 (n = 2, 3, 4, or 5), is shown to depend strongly on the solvent medium. Single molecule junction conductance measurements have been made with the I(s) method for each member of the series Me3Si—(C≡C)n—SiMe3 (n = 2, 3, 4, and 5) in mesitylene (MES), 1,2,4-trichlorobenzene (TCB), and propylene carbonate (PC). In mesitylene, a lower conductance is obtained across the whole series with a higher length decay ($\beta \approx 1$ nm−1). In contrast, measurements in 1,2,4-trichlorobenzene and propylene carbonate give higher conductance values with lower length decay ($\beta \approx 0.1$ and 0.5 nm−1 respectively). This behaviour is rationalized through theoretical investigations, where β values are found to be higher when the contact Fermi energies are close to the middle of the HOMO−LUMO gap but decrease as the Fermi energies approach resonance with either the occupied or unoccupied frontier orbitals. The different conductance and β values between MES, PC, and TCB have been further explored using DFT-based models of the molecular junction, which include solvent molecules interacting with the oligoyne backbone. Good agreement between the experimental results and these “solvated” junction models is achieved, giving new insights into how solvent can influence charge transport in oligoyne-based single molecule junctions.

The single molecule conductances of a series of bis-2,2′:6′,2″-terpyridine complexes featuring Ru(II), Fe(II), and Co(II) metal ions and trimethylsilylethynyl (Me3SiC≡C−) or thiomethyl (MeS-) surface contact groups have been determined theoretically and experimentally.

The single molecule conductance of metal complexes of general form trans-Ru(C≡CArC≡CY)2(dppe)2 and trans-Pt(C≡CArC≡CY)2(PPh3)2 (Ar = 1,4-C6H2-2,5-(OC6H13)2; Y = 4-C5H4N, 4-C6H4SMe) have been determined theoretically and experimentally. The complexes display high conductance (Y = 4-C5H4N, M = Ru (0.4±0.18 nS), Pt (0.8±0.5 nS); Y = 4-C6H5SMe, M = Ru (1.4±0.4 nS), Pt (1.8±0.6 nS)) for molecular structures of ca. 3 nm in length, which has been attributed to transport processes arising from tunneling through the tails of LUMO states.

# Acknowledgment

This thesis would never completed without the excellent supervision and guidance I received from Prof. Colin J. Lambert. Colin, There are no words in English books can express my gratitude to you. Really you were excellent supervisor.

I would like also to express my grateful admiration to Dr. Steven Bailey, Dr. Iain Grace, Dr. Tom Pope and Dr. Laith Al-Gharagholy. Also I would like to thank my sponsors, the Iraqi Ministry of Higher Education and Scientific Research and the University of Babylon for funding my PhD study.

The majority of the work was carried out in collaboration with experimentalists and I thank the participants: Prof. Paul J. Low, Prof. Richard J. Nichols, Prof. Martin R. Bryce, Prof. Simon J. Higgins, Prof. Jaime Ferrier and Prof. Thomas Wandlowski.

My great thanks to the wonderful participant Dr. David C. Milan. Also I would like to thank Dr. Víctor M. García-Suárez, Dr. Marie-Christine Oerthel, Dr. Santiago Marqués-González, Dr. Pilar Cea, Dr. Cancan Huang, Dr. Masoud Baghernejad, Dr. Veerabhadrarao Kaliginedi and Sören Bock.

My regardful respect and thanks to the group members Zain, Mohsin, Ali, Mohammed, Alaa and Nasser. Also I would like to pay my regards to all relatives and friends for the help they showed

I would like to express my very deep respect and sincere appreciation to my family, whose patience and moral encouragement gave me so much hope and support.

Above all, my great thanks to **ALLAH** for his mercy and blessing.

# List of Publications

During my PhD studies I published the following journal articles:

**1-** Oday A. Al-Owaedi, David C. Milan, Marie-Christine Oerthel, Sören Bock, Dmitry S. Yufit, Judith A. K. Howard, Simon J. Higgins, Richard J. Nichols, Colin J. Lambert, Martin R. Bryce, Paul J. Low. Experimental and computational studies of the single molecule conductance of Ru(II) and Pt(II) trans-bis(acetylide) complexes. Organometallics, *2016, 35(17): p. 2955 – 2954.*

**2-** Milan David C.; Oday A. Al-Owaedi, Oerthel Marie-Christine, Marques-Gonzalez Santiago, Brooke Richard J.; Bryce Martin R.; Cea, Pilar, Ferrer Jaime, Higgins Simon J.; Lambert Colin J.; Low Paul J.; Manrique David Zsolt, Martin Santiago, Nichols Richard J.; Schwarzacher Walther, Garcia-Suarez Victor M. Solvent Dependence of the Single Molecule Conductance of Oligoyne-Based Molecular Wires. J. Phys. Chem. C, 2016. 120(29): p. 15666–15674.

**3-** Ross D.; Oday A. Al-Owaedi, David C. M.; Qiang Z.; Joanne T.; František, H.; Simon J. H.; Richard J. N.; Colin J. L.; Paul J. L. Effects of Electrode−Molecule Binding and Junction Geometry on the Single-Molecule Conductance of bis-2,2′:6′,2″-Terpyridine-based Complexes. Inorg. Chem. 2016. 55(6): p. 2691–2700.

**4-** Hiroaki Ozawa, Masoud Baghernejad, Oday A. Al-Owaedi, Veerabhadrarao Kaliginedi, Takumi Nagashima, Jaime Ferrer, Thomas Wandlowski, Víctor M. García-Suárez, Peter Broekmann, Colin J. Lambert, Masa-aki Haga. Synthesis and Single-Molecule Conductance Study of Redox Active Ruthenium Complexes with Pyridyl and Dihydrobenzo[b]thiophene Anchoring Groups. Chem. Eur. J. 2016. 22(36): p. 12732–12740.

**5-** Oday A. Al-Owaedi, Sören Bock, David C. Milan, Marie-Christine Oerthel, Michael Inkpen, Nick Long, Tim Albrecht, Simon J. Higgins, Martin R. Bryce, Richard J. Nichols, Colin J. Lambert, Paul J. Low. Insulated molecular wires: Inhibiting orthogonal contacts in metal complex based molecular junctions. Journal of the American Chemical Society, Submitted.

**6-** Sören Bock, Oday A. Al-Owaedi, Samantha G. Eaves, David C. Milan, Mario Lemmer, Brian W. Skelton, Henrry M. Osorio, Richard J. Nichols, Simon J. Higgins, Pilar Cea, Nicholas J. Long, Tim Albrecht, Santiago Martín Colin J. Lambert, Paul J. Low. Single Molecule Conductance Studies of Organometallic Complexes Bearing 3-Thienyl Contacting Groups. Chem. Eur. J. Accepted, 10.1002/chem.201604565.

**7-** Manrique D. Z.; Huang C.; Baghernejad M.; Zhao X.; Oday A. Al-Owaedi, Sadeghi H.; Kaliginedi V.; Hong W.; Gulcur M.; Wandlowski T.; Bryce M. R.; Lambert C. J. A Quantum Circuit Rule for Interference Effects in Single-Molecule Electrical Junctions. Nat. Commun. 2015. 6(6389): p. 1–8.

# Contents

# Chapter 1

# Introduction

## 1.1. General

Nanoscale science and engineering offer the possibility for revolutionary advances in both fundamental science and technology and may have an impact on our life. It is comparable in scale and scope to transistor based electronics. At its most basic level, nanoscale science is the study of novel phenomena and properties of materials that occur at extremely small length scales, at the nanoscale this is the size of atoms and molecules [1]. The nanoscience revolution has created an urgent need for a more robust quantitative understanding of matter at the nanoscale by modeling and simulation, since the absence of quantitative models that describe newly observed phenomena increasingly limits the progress in the field.

Recently, new insights in the field of nanoscience have been obtained from the application of fundamental modeling techniques such as density functional theory (DFT), and molecular dynamics [2]. Advances in computer technology have led to an increase in computational capability which has made possible the modeling and simulation of complex systems with millions of degrees of freedom. However, the full potential of novel theoretical and modeling tools has not been reached yet.

## 1.2. Molecular Electronics

Molecular electronics is presented as the field of science that investigates the electronic transport properties of systems in which individual molecules are used as a basic building block. The dimensions of some molecular systems are a few nanometers, and therefore molecular electronic should be viewed as a subfield of nanotechnology [3]. In terms of a potential technology, molecular electronics is based on the bottom-up approach where the idea is to assemble specific and designed molecules to form more complex structures, active components and connecting wires.

The remarkable predictions of Gordon Moore in 1965, that the number of transistors per square centimetre on a silicon chip doubles every 18 months [4], has encouraged the constant quest for new technologies that could complement the silicon-based electronics, and molecular electronics is one such technology.

Many decades ago, there were many fundamental questions such as how does the electrical current flow through a single molecule? [5, 6]. The concept of electrons passing through a single molecule comes via two different approaches [7]. The first is electron transfer, which involves charge moving from one end of the molecule to the other. The second one is the charge transport, which involves current passing through a single molecule that is connected between electrodes [8]. The two are closely related, because both of them attempt to answer the previous question. From the point of view of fundamental science, molecular electronics could provide an answer for previous inquiry, and perhaps be an ideal way to explore the electronic and thermal conductance through the smallest molecular circuits, where quantum mechanical effects completely dominate [9].

The variety of molecules (both organic and organometallic molecules), together with their various properties could lead to the discovery of new physical phenomena. In addition, molecular junctions could also be promising systems to investigate the basic principles of electron transfer mechanisms. However, there are also other motivations in terms of a technological viewpoint which is the use of molecules as electronically active elements for many applications. One of these reasons is the size, since the typical size of molecules (between 1 and 10 nm) could lead to a higher packing density on a device with subsequent advantages in cost, efficiency and power dissipation [10]. These concepts and many others make molecular electronics an attractive field of science.

This thesis involves theoretical and experimental studies on a range of organic and organometallic molecules. Theoretically, two main techniques have been used to study the systems in this thesis; the first is density functional theory (DFT), which is implemented in the SIESTA code [11], and the nonequilibrium Green's function formalism of transport theory, which is implemented in the GOLLUM code [12]. Experimentally, scanning tunnelling microscopy break junction (STM-BJ), mechanically controllable break-junction (MCBJ) [13], and the current-distance (*I(s)*) technique [14], have been used to study the transport characteristics of the molecules that are the subject of investigation.

## 1.3. Thesis Outline

The outline of this thesis can be summarized as follows; this chapter is followed by chapter 2 which presents a brief overview of density functional theory (DFT), which is one of the main theoretical techniques that has been used in this thesis to study and understand the electronic properties of single-molecule junctions. Chapter 3 describes the single particle transport theory. This chapter involves a Green's function scattering formalism and all related topics such as the Landauer formula, Green's function of infinite leads, some of examples of scattering and a general approach to solving the surface lead Green's function.

The main goal of this work is to explore and understand novel fundamentals of different molecular junctions, which is the link between the chapters of this thesis, since these studies aim to provide a model of the electronic properties of $\mathbf{Au}\left|\begin{matrix}\textbf{organic molecules}\\ \textbf{OR}\\ \textbf{organometallic molecules}\end{matrix}\right|\mathbf{Au}$ molecular junctions. In other words, chapters 4 and 5 focus on organic molecules to explore some novel principles such as the quantum transport rule (chapter 4), and the influence of different environments surrounding Oligoyne-based molecular wires on their electronic properties (chapter 5).

Chapters 6 and 7 are focused on organometallic molecules, to probe and obtain a deeper understanding of the electronic properties of this kind of molecule. In what follows, chapter 6 seeks to extend the former studies and arrive at a more detailed understanding of the role of the anchor unit and metal complex fragment on the electrical behaviour of bis-2,2′:6′,2″-terpyridine based complexes.

The results in chapter 7 provide more insight into the transport mechanisms that have been reported in chapter 6, and open new avenues for the design of metal-complex based molecular wires. Finally, chapter 8 presents conclusions and future works.

# Chapter 2

# Density Functional Theory

## 2.1. Introduction

In an attempt to explore and understand the electronic properties of molecules, many theories have emerged; one of the most important of these theories and most common is density functional theory (DFT). Nowadays, DFT can be presented as a powerful tool for computations of the quantum state of atoms, molecules and solids, and of ab-initio molecular dynamics. In 1927, immediately after the foundation of quantum mechanics, an initial and approximate version of density functional theory was conceived by Thomas and Fermi [1]. Later, using the basics of quantum mechanics, Hohenberg, Kohn, and Sham, developed density functional theory of the quantum ground state to be superior to both Thomas-Fermi and Hartree-Fock theories, which opened a wide door to applications for realistic physical systems [2, 3]. From that time on, density functional theory (DFT) has grown vastly, and it has become one of the main tools in theoretical physics and molecular chemistry.

This chapter presents a brief summary of DFT and SIESTA (Spanish Initiative for Electronic Simulations with Thousands of Atoms) code [4], which has been used to study the electronic properties of the molecules that are the subject of research in this thesis. SIESTA is an implementation of DFT that can be used to perform calculations to investigate the characteristics of systems that involve a huge number of atoms ~ 1000.

## 2.2. The Many-Body Problem

This is an approach which aims to solve any system consisting of a large number of interacting particles. In a microscopic system consisting of charged nuclei surrounded by electron clouds these interactions such as electron-nuclei, electron-electron, nuclei-nuclei and electron correlations are described via Schrödinger equation.

The full Hamiltonian operator of a general system describing these interactions is

$$H = \sum_{n} -\frac{\hbar^2}{2m_e}\nabla^2_{r_n} + \frac{1}{8\pi\varepsilon_0}\sum_{n\neq m}\frac{e^2}{|r_n - r_m|}$$

$$-\sum_{n}\frac{\hbar^2}{2m_n}\nabla^2_{R_n} + \frac{1}{8\pi\varepsilon_0}\sum_{n\neq m}\frac{Z_n Z_m e^2}{|R_n - R_m|} - \frac{1}{4\pi\varepsilon_0}\sum_{nm}\frac{Z_m e^2}{|r_n - R_m|} \tag{2.1}$$

Here, $m_n$, $Z_n$ and $R_n$ represent the mass, atomic number and position of the *n-th* atom in the solid respectively. The position of *n-th* electron is denoted by the symbols $r_{n,}r_m$ and $m_e$ is the mass of a single electron. This Hamiltonian consists of five parts; the electron kinetic energy, electron-electron interactions, the nucleons kinetic energy, nucleon-nucleon interactions and electron-nucleon interactions respectively.

Approximately, the mass of nucleons is a few orders of magnitude higher than that of electron, and in terms of their velocities, the nuclei could be considered as a classical particle which creates an external potential, and the electrons as quantum particles are subjected to this potential.

This concept is known as the Born-Oppenheimer approximation [5], together with an assumption that the nucleon wavefunction is independent of the electron position, equation (2.1) can then be written as follows:

$$H = T_e + U_{e-e} + V_{e-nuc} \quad (2.2)$$

The first part of equation (2.2) presents the kinetic energy of all electrons, which is described by;

$$T_e = \sum_n \frac{\hbar^2}{2m_e} \nabla_n^2 \quad (2.3)$$

The electron-electron interaction is represented in the second part of equation (2.2), which is given by;

$$U_{e-e} = \sum_{n,m,n\neq m} \frac{e^2}{4\pi\varepsilon_0} \frac{1}{|r_n - r_m|} \quad (2.4)$$

$U_{e-e}$ describes the sum of all potentials acting on a given electron at position $r_n$ by all other electrons at position $r_m$ .

The third part of equation (2.2) describes the interactions between electrons and nuclei, which is expressed by;

$$V_{e-nuc} = \sum_N \sum_n v_{nuc}(r_n - R_N) \quad (2.5)$$

$V_{e-nuc}$ is the interaction between electrons and nuclei; it depends on the positions of electrons $r_n$ and nuclei $R_N$ .

The employment of a Born-Oppenheimer approximation [5], allows the electron and nucleon degrees of freedom to be decoupled.

# 2.3. The Hohenberg-Kohen theorems

Essentially, density functional theory (DFT) evolved significantly depending on two ingeniously simple theorems put forward and proved by Hohenberg and Kohn in 1964 [2]. These theorems are two powerful statements:

**Theorem I:** For any system of interacting particles in an external potential $V_{ext}$, the density is uniquely determined. In other words, the external potential is a unique functional of the density.

To prove this theorem, assume that there are two external potentials $V_{ext}^{(1)}$ and $V_{ext}^{(2)}$ differing by more than a constant, and giving rise to the same ground state density, $\rho_{0\,(r)}$. It is clear that these potentials belong to different Hamiltonians, which are denoted $H^{(1)}$ and $H^{(2)}$, and they give rise to distinct ground-state wavefunctions $\Psi^{(1)}$ and $\Psi^{(2)}$. Since $\Psi^{(2)}$ is not a ground state of $H^{(1)}$, so:

$$E^{(1)} = \langle \Psi^{(1)} \middle| H^{(1)} \middle| \Psi^{(1)} \rangle < \langle \Psi^{(2)} \middle| H^{(1)} \middle| \Psi^{(2)} \rangle \tag{2.6}$$

and, similarly:

$$E^{(2)} = \langle \Psi^{(2)} \middle| H^{(2)} \middle| \Psi^{(2)} \rangle < \langle \Psi^{(1)} \middle| H^{(2)} \middle| \Psi^{(1)} \rangle \tag{2.7}$$

Assuming that the ground states are non-degenerate [6, 7], one could rewrite equation (2.6) as follows:

$$\begin{aligned} \langle \Psi^{(2)} \middle| H^{(1)} \middle| \Psi^{(2)} \rangle &= \langle \Psi^{(2)} \middle| H^{(2)} \middle| \Psi^{(2)} \rangle + \langle \Psi^{(2)} \middle| H^{(1)} - H^{(2)} \middle| \Psi^{(2)} \rangle \\ &= E^{(2)} + \int dr \left( V_{ext}^{(1)}(r) - V_{ext}^{(2)}(r) \right) \rho_0(r) \end{aligned} \tag{2.8}$$

and assuming that $\left|\Psi^{(1)}\right\rangle$ has the same density $\rho_0(r)$ as $\left|\Psi^{(2)}\right\rangle$:

$$\langle\Psi^{(1)}|H^{(2)}|\Psi^{(1)}\rangle = E^{(1)} + \int dr\left(V_{ext}^{(2)}(r) - V_{ext}^{(1)}(r)\right)\rho_0(r) \quad (2.9)$$

Combining of equations (2.8) and (2.9) leads to,

$$E^{(1)} + E^{(2)} < E^{(1)} + E^{(2)} \quad (2.10)$$

This equation proves the two different external potentials cannot produce the same ground-state density.

**Theorem II:** A universal functional $F[\rho]$ for the energy $E[\rho]$ could be defined in terms of the density. The exact ground state is the global minimum value of this functional. In other words, the ground state energy of the system is given by the functional $F[\rho]$. If the input density and ground-state density are the same, the functional $F[\rho]$ would deliver the lowest energy. Hence, the functional could be minimised by varying the density to obtain the ground-state energy for the external potential.

The second theorem could be proven by considering the expression for the total energy, $E$, of the system with density $\rho$.

$$E[\rho] = T[\rho] + E_{int}[\rho] + \int drV_{ext}(r)\rho(r) \quad (2.11)$$

The kinetic term, $T$, and internal interaction of the electrons, $E_{int}$, depend only on the charge density, and so are universal.

The first theorem reported that the ground-state density $\rho_0$ for a system with external potential $V_{ext}$ and wavefunction $\Psi_0$, determines the Hamiltonian of that system, so for any density, $\rho$, and wavefunction, $\Psi$, other than the ground state, it could be found:

$$E_0 = \langle \Psi_0 | H | \Psi_0 \rangle < \langle \Psi | H | \Psi \rangle = E \tag{2.12}$$

Hence, the ground-state density, $\rho_0$, minimizes the functional (equation 2.11). If the functional $T[\rho] + E_{int}[\rho]$ is known, then by minimizing equation 2.11, the ground-state of the system could be obtained, and then all ground-state characteristics could be calculated, which are the subject of the interest.

## 2.4. The Kohn-Sham Method

The Kohn-Sham method has been used in solid state physics for about fifty years. By now, largely due to the development of increasingly accurate density functionals, the method has also gained a large popularity among physicist and chemists, especially as it allows in many cases, accurate treatments of molecular systems unattainable by the more traditional quantum mechanical methods [8, 9].

It has been reported in the previous section that obtaining the ground-state density leads to calculating the ground-state energy. However, the precise form of the functional shown in equation (2.11) is not known. Generally, the kinetic term and internal energies of the interacting particles cannot be expressed as functionals of the density.

Kohn and Sham in 1965 [3], came up with the idea, that it possible to replace the original Hamiltonian of the system by an effective Hamiltonian of non-interacting particles in an effective external potential, which gives rise to the same ground state density as the original system [10, 11].

The form of energy functional of the Kohn-Sham ansatz is:

$$E_{KS}[\rho] = T_{KS}[\rho] + \int drV_{ext}(r)\rho(r) + E_H[\rho] + E_{xc}[\rho] \qquad (2.13)$$

Here, $T_{KS}$ is the kinetic energy of the non-interacting system. The kinetic energy, $T$, in equation (2.11) has been used for the interacting system. This discrimination is due to the exchange correlation, $E_{xc}$, functional, which is described in equation (2.15). $E_H$ is the Hartree functional, which describes the electron-electron interaction using the Hartree-Fock method [12 – 16], and it is given by:

$$E_H[\rho] = \frac{1}{2}\iint \frac{\rho(r)\rho(r')}{|r - r'|}drdr' \qquad (2.14)$$

This is an approximated version of the internal interaction of the electrons, $E_{int}$, as defined previously. Again, the difference referred to the exchange correlation, $E_{xc}$. Therefore, the differences between the exact and approximate solutions for the kinetic energy, and electron-electron interaction terms were represented via $E_{xc}$, which is expressed by:

$$E_{xc}[\rho] = (E_{int}[\rho] - E_H[\rho]) + (T[\rho] - T_{KS}[\rho]) \qquad (2.15)$$

Consequently, the Kohn-Sham method could be a powerful approach to obtain an accurate ground-state density if the exchange correlation, $E_{xc}$, is known precisely.

## 2.5. The Exchange Correlation Functionals

The biggest challenge of Kohn-Sham DFT is the finding of accurate approximations to the exchange correlation energy, $E_{xc}$. The best understanding of exact functional could be obtained by the best approximation could be designed it [17]. Many efforts have been spent to find the best approximation for the exchange-correlation functional, and numerous forms have been proposed. This section presents a brief summary of two of the most popular approximation forms. The first one is the local density approximation (LDA) [18]. Secondly, the generalized gradient approximation (GGA) [19]. The comparison in terms of the accuracy between LDA and GGA, reported that the GGA is more accurate approximation, because it is designed based on density and the density gradients, while LDA is the simplest, because it is based on the local density.

### 2.5.1. Local Density Approximation

The simplest approximation is to assume that the density can be treated as a uniform electron gas. Based on this approximation, which was initially proposed by Kohn and Sham [3], the exchange-correlation energy for a density ρ is given by:

$$E_{xc}^{LDA}[\rho] = \int dr\rho(r)\left(\in_{x}^{hom}\left(\rho(r)\right) + \in_{c}^{hom}\left(\rho(r)\right)\right) \qquad (2.16)$$

Here, the terms $\in_{x}^{hom}$ and $\in_{c}^{hom}$ are the exchange and correlation energy densities for the homogeneous electron gas. The analytical exchange energy, $\in_{x}^{hom}$ , can be found in the literature [20]:

$$\in_{x}^{hom} = -\frac{3}{4\pi}\sqrt[3]{3\pi^2\rho} \qquad (2.17)$$

Ceperley and Alder [21] calculated numerically the correlation energy $\in_c^{hom}$ using the quantum Monte-Carlo method, then Perdew and Zunger [22] fitted the numerical data to analytical expressions, and found:

$$\in_c^{hom} = \begin{cases} -0.048 + 0.031 ln r_{(s)} - 0.0116 r_s + 0.002 r_s \ln(r_s) & r_s < 1 \\ -\dfrac{0.1423}{\left(1 + 1.9529\sqrt{r_s} + 0.3334 r_s\right)} & r_s > 1 \end{cases} \tag{2.18}$$

Here, $r_s = \left(\frac{3}{4\pi\rho}\right)^{1/3}$ is the radius of a sphere in a homogeneous electron gas of density, ρ that contains one electron.

The LDA is often surprisingly accurate and for systems such as graphene and carbon nanotubes or where the electron density slowly varies, generally gives very good results. Despite the remarkable success [23, 24], of the LDA, care should be taken in its application. For example, LDA predicts a wrong ground state for the titanium atom and it gave a very poor description for hydrogen bonding [25, 26], as well as it gives an incorrect value of the band gap in semiconductors and insulators [27, 28].

## 2.5.2. Generalized Gradient Approximation

The pitfalls of the local density approximation (LDA) and a fact that real systems are inhomogeneous, means that there is a need to find an alternative approximation, which is the generalized gradient approximation (GGA).

Basically, there is no analytical form for the exchange energy in GGA, and therefore it has been calculated along with the correlation term, numerically. Nowadays, there are various parameterizations which are used in this approximation; one of the most popular and most reliable is the PBE functional form, which was proposed in 1996 by Perdew, Burke and Ernzherhof [29]:

$$E_{xc}^{GGA} = E_x^{GGA}[\rho] + E_c^{GGA}[\rho] \tag{2.19}$$

The exchange part is given by:

$$E_x^{GGA}[\rho] = \int \in_x \big(\rho(r)\big)\, V_x\big(\rho(r)\, \nabla\rho(r)\big)\rho(r)dr \tag{2.20}$$

$$V_x(\rho, \nabla\rho) = 1 + k - \frac{k}{1 + \frac{\mu s^2}{k}}$$

Here, $k$ = 0.804, $\mu$ = 0.21951 and $s = \frac{|\nabla\rho|}{2k_F\rho}$ is the dimensionless density gradient and $k_F$ is the Fermi wavelength and $V_x(\rho, \nabla\rho)$ represents the enhancement factor.

The correlation energy form is given by:

$$E_c^{GGA}[\rho] = \int \rho(r)[\in_c (\rho(r)) + F(\rho(r), \nabla\rho(r))]\, dr \tag{2.21}$$

$$F(\rho, \nabla\rho) = \frac{\gamma e^2}{a_0} ln\left[1 + \frac{\beta t^2}{\gamma}\left(\frac{1 + At^2}{1 + At^2 + A^2t^4}\right)\right], \qquad A = \frac{\beta}{\gamma}\frac{1}{(e^{-\in_c(\rho)/\gamma} - 1)}$$

β = 0.066725, $\gamma = \frac{(1 - ln2)}{\pi^2}\gamma$, $a_0 = \frac{\hbar}{m^2}$, and the dimensionless gradient is $= \frac{|\nabla\rho|}{2k_{TF}\rho}$, where $K_{TF} = \sqrt[3]{12/\pi}/\sqrt{r_s}$ is known as the Thomas-Fermi screening wavelength and $r_s$ is defined as the local Seitz radius. The PBE-GGA functional has been extremely influential, both for performing actual calculations and as the basis for functionals involving higher derivatives and exact exchange [30]. It has been used in all studies of this thesis, and it gives a good agreement with experiment [31 – 33].

# 2.6. SIESTA

SIESTA is an acronym derived from the Spanish Initiative for Electronic Simulations with Thousands of Atoms [4] is both a method and computer program implementation, to perform electronic structure calculations and *ab initio* molecular dynamics simulations of molecules and solids. One of the main characteristics of SIESTA, that it uses the standard Kohn-Sham self-consistent density functional method in the local density (LDA) and, or generalized gradient (GGA) approximations. In addition, it utilizes norm-conserving pseudopotentials in their fully nonlocal form, and a linear combination of atomic orbital basis set to achieve efficient calculations [4].

In this thesis, the SIESTA code has been used to perform all DFT calculations. It is used to obtain the optimized geometries of the molecules which are the subject of this research, and a Hamiltonian describing their electronic properties.

## 2.6.1. The Pseudopotential Approximation

In terms of time and computer memory, the investigation of the electronic properties of typical systems of molecules consist of a large number of atoms containing complex potentials could be expensive. Although, the splitting of a large interacting problem into a large effective non-interacting problem as shown previously simplifies the problem; still there is a need for more simplification, which could be obtained by using the proposed pseudopotential approximation by Fermi in 1934 [34, 35]. The fundamentals of this concept are the removing of the core electrons, which lie within filled atomic shells, and replace them by an external potential known as a pseudopotential, while valance electrons are arranged in the partially filled outer shells, and they are only contribute in the formation of molecular orbitals. The advantages of this kind of approximations could be summarized in two points; first it decreases the electrons number of the system significantly, and that lowers the cost (time and memory) to perform the calculations of the system. The second benefit is the numerical stability, because these pseudopotentials are smooth.

In this section, a special type of *ab-initio* pseudopotential is presented, which is used in the SIESTA code, a norm-conserving pseudopotential [36]. The calculation of the pseudopotential based on the Kohn-Sham formalism to solve the many-electron problem for a single atom, could replace the effect of the core electrons. The definition of the valance electron wavefunctions is the product of radial and spherical harmonic wavefunctions:

$$\Psi_{nlm}^{ae}(r) = \frac{1}{r} R_{nl}^{ae}(r) Y_{lm}(\varphi, \vartheta) \tag{2.22}$$

Here, $n$ = 1, 2, …, $l$ = 0, …$n$ – 1 and $m$ = – $l$, …$l$ are quantum numbers, $Y_{lm}(\varphi,\vartheta)$ indicates to normalized spherical harmonics and $R_{nl}^{ae}$ is the solution to the radial Schrödinger equation (equation 2.23) that contains the all-electron potential , $V_{nl}^{ae}$, which includes all core-and valence-electron interactions.

$$\epsilon_{nl}^{ae}\ R_{nl}^{ae}(r) = \left[-\frac{1}{2}\frac{d^2}{dr^2} + \frac{l(l+1)}{2r^2} + V_{nl}^{ae}(r)\right] R_{nl}^{ae}(r) \tag{2.23}$$

To reduce the size of the system, the core electrons were removed and replaced the all-electron potential with an effective potential: $V_{nl}^{eff}[\rho](r)$, where $\rho(r)$ is the electron density given by the filled Kohn-Sham orbitals:

$$\rho(r) = \rho(r, \varphi, \vartheta) = \sum_{nlm} f_{nlm} |\Psi_{nlm}(r, \varphi, \vartheta)|^2 \tag{2.24}$$

$f_{nlm} = 0, 1, 2$ is the occupancy factor indicating whether an orbital is empty, half-filled or full. In general, $\rho(r)$ is not spherically symmetric, but for an isolated atom, it is possible to integrate out the angular dependence:

$$\rho(r) = \int \rho(r, \varphi, \vartheta)\, r^2 sin\vartheta d\vartheta d\varphi$$

$$= \sum_{nlm} f_{nlm} |R_{nl}(r)|^2 \int |Y_{lm}(\varphi, \vartheta|^2\, sin\vartheta d\vartheta d\varphi = \sum_{nlm} f_{nlm} |R_{nl}(r)|^2 \tag{2.25}$$

This purely radially-dependent density generates an effective potential that is also only radially dependent. Hence, the Kohn-Sham equation takes the form:

$$\in_{nl} R_{nl}(r) = \left[ -\frac{1}{2}\frac{d^2}{dr^2} + \frac{l(l+1)}{2r^2} + V_{nl}^{eff}[\rho](r) \right] R_{nl}(r) \tag{2.26}$$

The attractive nuclear potential is also included into the effective potential. This is the Kohn-Sham orbitals and the self-consistent effective potential that involves all of the interaction between electrons. The pseudowavefunction, $R_{nl}^{ps}(r)$, and eigen-energies, $\in_{nl}^{ps}$ are obtained from the solution to equation (2.26), after replacing the central potential $V_{nl}^{eff}$ by a pseudopotential $V_{nl}^{ps}$, which is given by:

$$V_{nl}^{ps} = \in_{nl}^{ps} - \frac{l(l+1)}{2r^2} + \frac{1}{2R_{nl}^{ps}(r)} \frac{d^2}{dr^2} R_{nl}^{ps}(r) \tag{2.27}$$

The formula (2.27) is obtained by inverting the Schrödinger equation for known valence wavefunctions (or pseudowavefunction), $R_{nl}^{ps}(r)$ , and eigenvalues, $\in_{nl}^{ps}$. Therefore, the pseudopotential depends on the $n$ and $l$ quantum numbers and are parameterized to be smooth and continuous outside the given cut-off radius.

In this thesis, all DFT calculations were achieved using SIESTA with pseudopotentials generated using the Troullier-Martins method [37, 38]. In this method, the radial part of the pseudowavefunction described by two formulas and depends on a given cut-off radius $r_c$ as follows:

$$R_{nl}^{ps} = \begin{cases} R_{nl}\,(r) & r > r_c \\ r^1 e^{p(r)} & r > r_c \end{cases} \tag{2.28}$$

Here, $p(r)$ is given by:

$$p(r) = a_0 + a_2 r^2 + a_4 r^4 + a_6 r^6 + a_8 r^8 + a_{10} r^{10} + a_{12} r^{12} \tag{2.29}$$

The $a_i$ coefficients are determined by the following conditions:

1. Norm-conservation refers to the charge in as sphere less than $r_c$ has to be same for the pseudopotential and all valence electron wavefunctions.

$$\int_0^{r_c} |R_{nl}(r)|^2 = \int_0^{r_c} \left|R_{nl}^{ps}(r)\right|^2$$

2. The corresponding valence eigenvalues are the same.

$$\epsilon_{nl}^{ps} = \epsilon_{nl}$$

3. Smoothness of the pseudowavefunction leads to a smooth pseudopotential

$R_{nl}(r_c) = R_{nl}^{ps}(r_c)$, and for $i$ = 1, 2, 3, 4: $\left[\frac{d^i R_{nl}(r)}{dr^i}\right]_{r=r_c} = \left[\frac{d^i R_{nl}^{ps}(r)}{dr^i}\right]_{r=r_c}$

From these conditions, it is possible to parameterize $R_{nl}^{ps}(r)$ (equation (2.28)), and then substitute this into equation (2.27), to obtain the explicit form of the pseudopotential:

$$V_{nl}^{ps}(r) = \begin{cases} V_{nl}^{eff}[n](r) & r > r_c \\ \in_{nl}^{ps} + \dfrac{(l+1)p'(r)}{r} + \dfrac{1}{2}\left(p'(r) + p''(r)\right) & r > r_c \end{cases} \qquad (2.30)$$

The resulting pseudopotential would be smooth and nodeless, if these conditions are satisfied [37]. $V_{nl}^{ps}(r)$ is known as a screened pseudopotential because it involves the effects of both core and valence electrons. The employing of this pseudopotential in other environment such as molecules, any screening from the valence electrons should be removed, which could be performed by subtracting the exchange-correlation and Hartree potentials, and that yielded the bare ion potential, $V_{nl}^{ion}$, which would be transferable to different environments.

$$V_{nl}^{ion}(r) = V_{nl}^{ps}(r) - V_H[n^{val}(r)] - V_{xc}[n^{val}(r)] \qquad (2.31)$$

Here, $n^{val}(r)$ denotes the valence components of the self-consistent charge density.
A huge number of real-space points is required to calculate the potential matrix directly. Secondly, the number of matrix elements per $l$ value is scaled by $n(n+1)$, where $n$ represents the orbital number. To optimize this procedure, the pseudopotential could be consider in two parts; the first one is called a local potential, $V^{loc}$, which is the same for all $l$ components. The second one is called a semi-local potential, $V_n^{sl} = V^{ion} - V^{loc}$, which differs for each $l$, and it is constructed inside region less than the given cut-off radius, $r_c$, otherwise it is zero outside this region.

The expression of this potential is given by:

$$V_n^{ion} = V_n^{loc}(r) + \sum_{l=0}^{n-1} \sum_{m=-1}^{l} \delta V_{nl} |Y_{lm}\rangle\langle Y_{lm}| \tag{2.32}$$

$V_n^{sl}$ is the second part of equation (2.32), where $\delta V_{nl}$ is constructed in a way that it is zero beyond the cut-off radius. This is reasonable since, beyond the cut-off radius, the pseudopotential is the original effective potential, which is local.

The semi-local part of equation (2.32) has been represented as a fully non-local potential in terms of Kleinmann-Bylander projectors [39]:

$$V_n^{sl} = \sum \frac{|\delta V_{lm} \Psi_{lm}^{ps}\rangle\langle \Psi_{lm}^{ps} \delta V_{lm}|}{\langle \Psi_{lm}^{ps} | \delta V_{lm} | \Psi_{lm}^{ps}\rangle} = V^{KB} \tag{2.33}$$

Here, $|\Psi_{lm}^{ps}\rangle$, $\delta V_{lm}$ and $|\delta V_{lm} \Psi_{lm}^{ps}\rangle$ are the spherical harmonic, the semi-local part of the pseudopotential, and the pseudowavefunction respectively.

Hence, the non-local parts could be calculated by implementing Kleinmann-Bylander projectors, which decreases the number of matrix elements per *l* from *n*(*n*+1) to *n* [39]. In addition, the non-local pseudopotential matrix elements could be calculated in *k*-space rather than a real space grid, which dramatically lowers the computational expenses for the large systems.

## 2.6.2. SIESTA Basis Sets

A basis set is a mathematical description of the orbital within a system used to perform the theoretical calculations. One elegant and popular choice of basis sets in periodic system calculations is the plane-wave basis set. However, one of the main reasons for applying the SIESTA code for my calculations is that it used localised basis sets (which are not implicitly periodic) and therefore can be used to construct a tight-binding Hamiltonian, this is not easy to achieve using a plane wave based code.

The type of basis set is one of the most important aspects for calculations using SIESTA. For example, to perform efficient calculations, the Hamiltonian should be sparse, and therefor SIESTA utilizes a linear combination of atomic orbital basis sets (LCAOs), which are constrained to be zero outside of a certain radius (cut-off radius), and are constructed from the orbitals of the atoms. This generates the desired sparse form for the Hamiltonian, and it reduces the overlap between basis functions, and therefore a minimal size basis set creates characteristics similar to that of the system under investigation.

The simplest basis set for an atom is called a single-$\zeta$ basis, which corresponds to a single basis function, $\Psi_{nlm}(r)$, per electron orbital (i.e. 1 for an *s*-orbital, 3 for a *p*-orbital, etc...). In this case each basis function consists of a product of one radial wavefunction, $\phi^1_{nl}$, and one spherical harmonic, $Y_{lm}$:

$$\Psi_{nlm}(r) = \phi^1_{nl}(r) Y_{lm}(\varphi, \vartheta) \tag{2.34}$$

The small number of the expected basis function is one, and therefore a single-$\zeta$ basis uses single atomic orbitals as a basis function. Therefore, the radial part in equation (2.34) is found by using the Sankey method [40], because the component of the real orbital is described by an infinitely long tail, which is not suitable for a localized basis function.

This method uses a modified version of the Schrödinger equation, which is solved for an atom placed inside a spherical box, and the radial wavefunction equals zero at the cut-off radius, $r_c$ . This restriction yields an energy shift $\delta E$ within the Schrödinger equation such that eigenfunction has a node at the cut-off radius, $r_c$ :

$$\left[-\frac{d^2}{dr^2}+\frac{l(l+1)}{2r^2}+V_{nl}^{ion}(r)\right]\phi_{nl}^1(r)=(\epsilon_{nl}+\delta E)\phi_{nl}^1(r) \qquad (2.35)$$

The energy shift $\delta E$ satisfies the previous constraint and the corresponding pseudopotential is $V_{nl}^{ion}(r)$. To obtain high accuracy basis sets (multiple-$\zeta$), additional radial wavefunctions could be included for each electron orbital. The split-valence method has been used to calculate the additional radial wavefunctions, $\phi_{nl}^i$, for $i$ > 1. This involves defining a split-valence cut off for each additional wavefunction $r_s^i$. Thus, it is split into two piecewise functions: a polynomial below the cut-off and the former wavefunction above it:

$$\phi_{nl}^i(r)=\begin{cases} r^l(a_{nl}-b_{nl}r^2) & r>r_s^i \\ \phi_{nl}^{i-1} & r_s^i<r<r_s^{i-1} \end{cases} \qquad (2.36)$$

The additional parameters are found at the point $r_s^i$ where the wavefunction and its derivative are assumed continuous.

The polarization of a real orbital due to the electrical field of the neighbour atoms, is taken into account to calculate the basis set function. This kind is called double-$\zeta$ polarized, which is used to achieve all calculations in this thesis. Table (2.1) shows the number of basis of orbitals for a selected number of atoms for single-$\zeta$ polarized, double-$\zeta$ and double-$\zeta$ polarized.

*Table 2.1: Examples of the radial basis functions per atom as used within the SIESTA for different degrees of precisions.*

| Atom/Basis Function | Single-ζ (SZ) | Double-ζ (DZ) | Single-ζ Polarised (SZP) | Double-ζ Polarised (DZP) |
|---|---|---|---|---|
| $H^1/1s$ | 1 | 2 | 4 | 5 |
| $C^6/1s\ 2s\ 2p_x\ 2p_y\ 2p_z$ | 4 | 8 | 9 | 13 |
| $N^7/1s\ 2s\ 2p_x\ 2p_y\ 2p_z$ | 4 | 8 | 9 | 13 |

## 2.6.3. The Basis Set Superposition Error Correction (BSSE) and Counterpoise Correction (CP)

The employment of DFT to compute the ground state energy of various molecular junctions, permits to calculate binding energies and optimal geometries. However, these calculations are subject to errors, due to the employing of localized basis sets, which are concentrated on the nuclei. At the point when atoms are sufficiently close to each other so that their basis functions will overlap. This might cause an artificial strengthening of the atomic interaction and an artificial shortening of the atomic distances and hence this could influence the aggregate energy of the system.

The solution of this kind or errors has been demonstrated by the basis set superposition error correction (BSSE) [41] or the counterpoise correction [42]. Assuming two molecular systems, denoted *a* and *b*, the energy of the interaction may be expressed as:

$$\Delta E(ab) = E_{ab} - \left(E_a^{ab} + E_b^{ab}\right) \tag{2.37}$$

The total energy of the combined $a$ and $b$ system is $E_{ab}$, while the total energy of isolated systems $a$ and $b$ is $E_a$ and $E_b$ respectively with keeping identical basis sets for the three energies. To execute these amendments inside SIESTA, ghost states have been used to assess the aggregate energy of segregated systems $a$ or $b$ in dimer basis. In terms of ghost states, there are two principles; first, the basis should be preserved of one part of a dimer on atomic centres and disregard its electrons and nuclear charge. Second, the other part of dimer should be kept without neglecting of anything. Precise results [43, 44] have been obtained by using this method, which will be shown in chapters 5, 6 and 8 of this thesis.

### 2.6.4. The Electron Hamiltonian

The electron Hamiltonian created by SIESTA follows the Kohn-Sham formalism and incorporates the local and non-local parts of pseudopotential:

$$H = \hat{T} + \sum_i V_i^{KB}(r) + \sum_i V_i^{loc}(r) + V_H(r) + V_{xc}(r) \tag{2.38}$$

Here, $T$ is the kinetic energy operator, $V_i^{loc}$ and $V_i^{KB}$ are the local and non-local parts of the pseudopotential for atom $i$, and $V_H$ and $V_{xc}$ are Hartree and exchange-correlation potentials. By using two centre integrals in $k$-space, the first two terms of equation (2.38) have been calculated, which are expressed as follows:

$$\langle \Psi_1 | \hat{O} | \Psi_2 \rangle = \int \Psi_1^*(k)\, \hat{O} \Psi_1(k) e^{-ik.R} dk \tag{2.39}$$

These are found by taking the Fourier transforms in $k$-space with $\Psi_\alpha$ corresponding to either the basis orbitals (for $\hat{O} = \hat{T}$) or the Kleinmann-Bylander pseudopotential projects (for $\hat{O} = V_i^{KB}$). The last three terms of equation (2.38) are calculated on a three-dimensional real space grid with a fineness $\Delta x$ controlled a grid cut off energy, $\in_c$, which is equivalent to a plane-wave cut off $\epsilon_c = \frac{\pi^2}{2\Delta x}$. A cut-off energy of 250 Ry has been used in all calculations of this thesis.

## 2.7. Calculation in Practice

In view of previous sections, everything is obtained to start the computation. After building the atomic structure of the system, suitable pseudopotentials are required for every component, which can be distinctive for each exchange-correlation functional. Additionally, a reasonable choice of the basis set must be made for every element present in calculation. In terms of time and memory, it is well known that more precise calculations are more computationally expensive. For example, the larger basis set gives an accurate calculation, but it takes a longer time and uses a larger memory.

The creation of the initial charge density based on an assuming that there is no interaction between atoms could be the next step. This step is simple and the sum of charge densities is the total charge density, because the pseudopotentials are known.

Figure 2.1 shows the self-consistent calculation, which begins by calculating the Hartree potential and exchange-correlation potential. Since the density is represented in real space, the Hartree potential is obtained by solving the Poisson equation with the multi-grid [45, 46] or fast Fourier-transform [45, 47] method, and the exchange-correlation potential is obtained by performing the integrals shown in section 2.5.

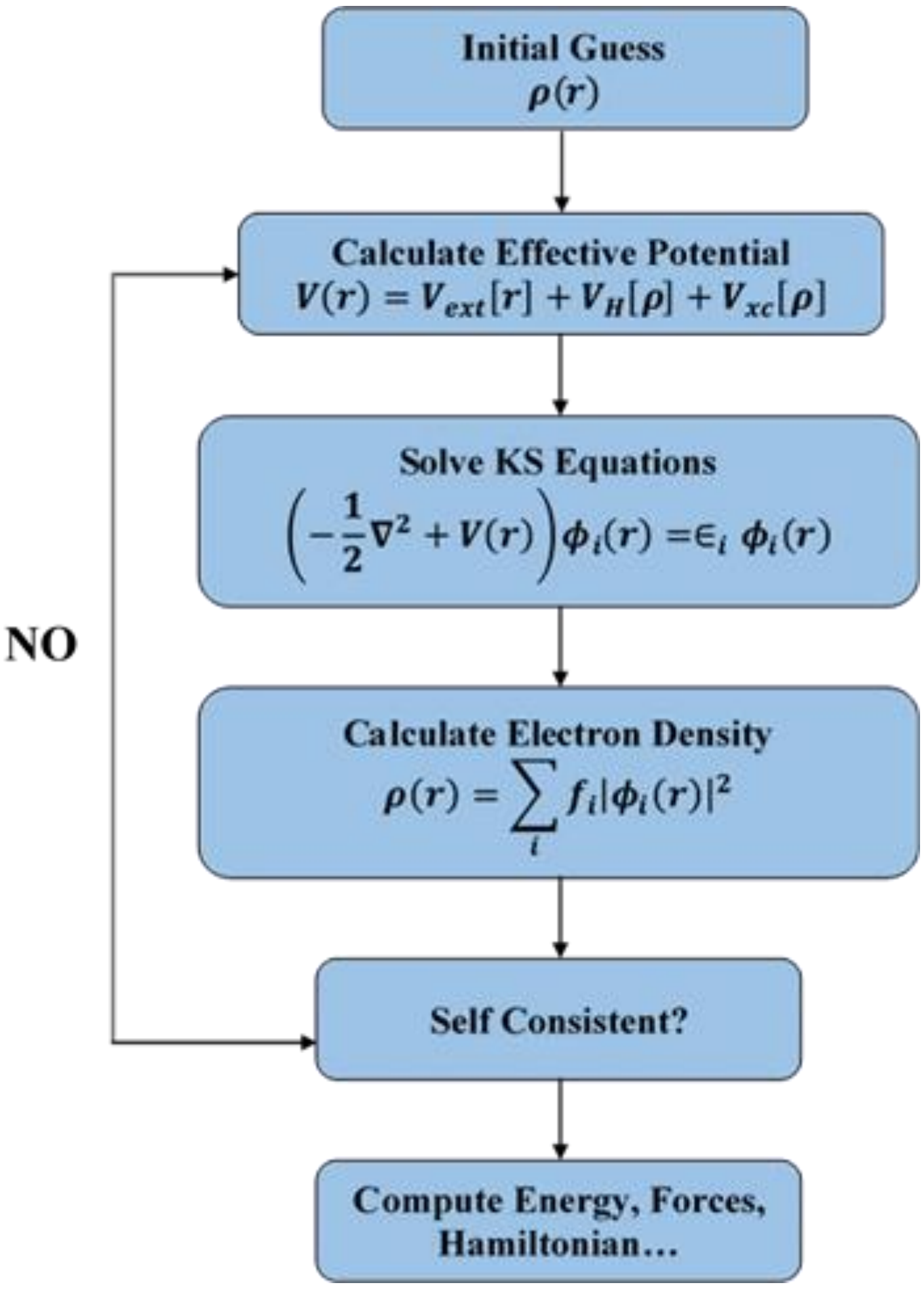

*Figure 2.1: Schematic of the self-consistency process within SIESTA.*

The next iteration as shown in figure 2.1 starts after solving the Kohn-Sham equations and obtaining a new density $\rho(r)$. The reaching of the necessary convergence criteria is the end of this iteration. Consequently, the ground state Kohn-Sham orbitals and the ground state energy for a given atomic configuration are obtained. For geometrical optimization, the conjugate gradient method [45, 48] is used to obtain the minimal ground state and the corresponding atomic configuration. Finally, if the self-consistency is performed, the Hamiltonian and overlap matrices could be extracted.

# Chapter 3

## Single Particle Transport

## 3.1. Introduction

Essentially, the aim of molecular electronics is to explore and understand the electrical behaviour and properties of molecular junctions; where a molecule of a few nanometers is attached to bulk electrodes so that the ballistic transport can happen through its energy levels. Indeed, there are some of fundamentals, which should be considered as a first step to understand the electronic properties of the molecular junctions, such as the scattering process of electrode|molecule|electrode structure. One of the most powerful approaches to understand the scattering process in the electrode junction and the molecular bridge is the Green's function formalism.

The first section of this chapter presents a brief overview of the Landauer formula. The retarded Green's function for a one-dimensional tight binding chain will be introduced in the second section. Following this, a break in the periodicity of this lattice at a single connection would be shown, and the Green's function is related directly to the transmission coefficient across the scattering region. Then, the methods that have been used to calculate the transmission coefficient of mesoscopic conductors with complex geometries would be presented. In this chapter, the method assumes the negligible interaction between carriers, the absence of inelastic processes and zero temperature.

## 3.2. The Landauer Formula

The Landauer formula [1, 2] has become the standard theoretical model to describe the transport phenomena in phase coherent, ballistic mesoscopic systems. The main principle of this approach is the assumption that the system under investigation is coupled to large reservoirs as shown in figure 3.1, where all inelastic processes take place [3]. Consequently, the transport through the systems can be formulated as a quantum mechanical scattering problem. Another important assumption is that the system is connected to reservoirs by ideal quantum wires, which behave as waveguides for the electron waves.

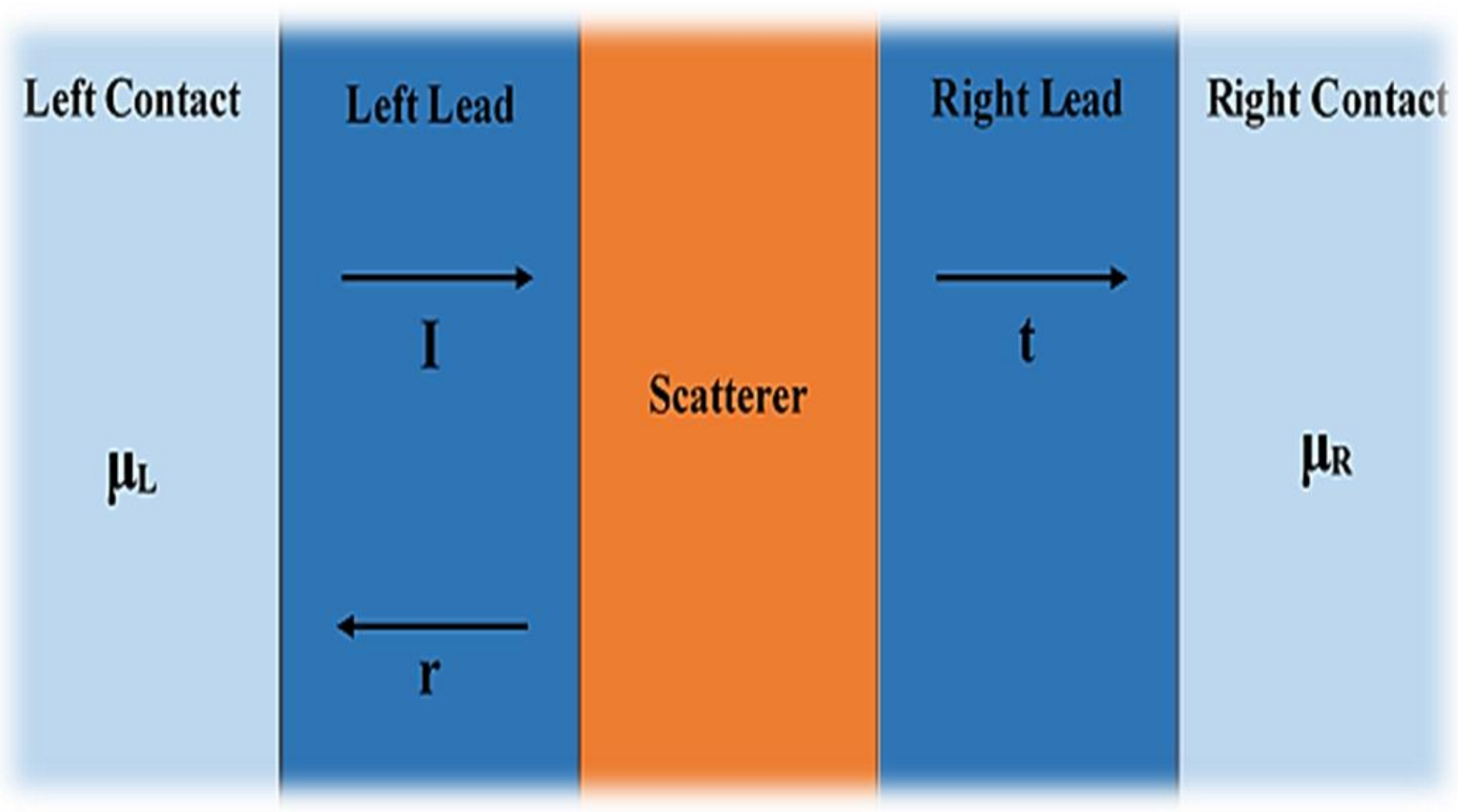

*Figure 3.1: A mesoscopic scatterer connected to contacts by ballistic leads. The chemical potential in the contacts is $\mu_L$ and $\mu_R$ respectively. If an incident wave packet hits the scatterer from the left, it will be transmitted with probability $\mathcal{T} = tt^*$ and reflected with probability $R = rr^*$. Charge conservation requires $\mathcal{T} + R = 1$.*

Figure 3.1, shows a mesoscopic scatterer connected between two large electron reservoirs, by means of two ideal ballistic leads; all inelastic relaxation processes are limited to the reservoirs [3]. The assumption that the reservoirs have a small chemical potential difference $\mu_L - \mu_R = \delta E > 0$, allows the movement of electrons from the left to the right reservoir. For one open channel, the incident electrical current, $\delta I$, is given by:

$$\delta I = e\upsilon_g \frac{\partial n}{\partial E} \delta E = e\upsilon_g \frac{\partial n}{\partial E} (\mu_L - \mu_R) \tag{3.1}$$

Here, $e$ is the electronic charge, $\upsilon_g$ is the group velocity – i.e. the velocity of electron, and $\frac{\partial n}{\partial E}$ is the density of states per unit length in the lead in the energy window given by the chemical potential of the contacts:

$$\frac{\partial n}{\partial E} = \frac{\partial n}{\partial k}\frac{\partial k}{\partial E} = \frac{\partial n}{\partial k}\frac{1}{\upsilon_g \hbar} \tag{3.2}$$

In one dimension, after including a factor of 2 for spin dependency, $\frac{\partial n}{\partial k} = \frac{1}{\pi}$. Substituting this into equation (3.2), finds that, $\frac{\partial n}{\partial E} = \frac{1}{\upsilon_g h}$. This simplifies equation (3.1) to:

$$\partial I = \frac{2e}{h}(\mu_L - \mu_R) = \frac{2e^2}{h}\partial V \tag{3.3}$$

Here, $\partial V$ is the voltage generated by the potential mismatch. From equation (3.3), it is obvious that in the absence of a scattering region, the conductance of a quantum wire with one open channel is $\frac{e^2}{h}$, which is approximately *77.5μS* or a resistance of *12.9kΩ*. If a scattering region is considered, the current collected in the right contact would be:

$$\delta I = \frac{2e^2}{h}\mathcal{T}\delta V \Rightarrow \frac{\delta I}{\delta V} = G = \frac{2e^2}{h}\mathcal{T} \tag{3.4}$$

This is the well-known Landauer formula, relating the conductance, *G*, of a mesoscopic scatterer to the transmission probability, $\mathcal{T}$, of the electrons traveling through it. It describes the linear response conductance, hence it only holds for small bias voltages, $\delta V \approx 0$.

Büttiker [2] has generalised the Landauer formula for more than one open channel. In view of that, the total of all transmission amplitudes, instead of transmission coefficient has been used to describe the transport of electrons from left contact to the right. Hence, the Landauer formula for more than one open channel is given by:

$$\frac{\delta I}{\delta V} = G = \frac{2e^2}{h}\sum_{i,j}\left|t_{i,j}\right|^2 = \frac{2e^2}{h}Tr(tt^\dagger) \tag{3.5}$$

Here, $t_{i,j}$ is the transmission amplitude describing scattering from the $j^{th}$ channel of the left lead to $i^{th}$ channel of the right lead. Based on the definition of transmission amplitudes, the reflection amplitudes, $r_{i,j}$, can be introduced, which describe scattering processes, where the particle is scattered from the $j^{th}$ channel of the left lead to the $i^{th}$ channel of the same lead. The definition of the *S* matrix can be obtained via combination of the transmission and reflection amplitudes. The *S* matrix, which connects the states coming from the left lead to the right and vice versa, is given by:

$$S = \begin{pmatrix} r & t' \\ t & r' \end{pmatrix} \tag{3.6}$$

Here, *r* and *t* describe the electrons transferring from the left, while *r'* and *t'* describe the electrons transferring from the right. Equation (3.5) suggests that *r*, *t*, *r'* and *t'* are matrices for more than one open channel, and in presence of magnetic field they could be complex. The *S* matrix is useful not only in describing linear transport, but also in other problems, such as adiabatic pumping [4].

# 3.3. One Dimension

The calculation of the scattering matrix for a simple one-dimensional system can be a helpful step to clarify the outline of the generalised methodology. A Green's function based approach is used in the derivation. Therefore, the form of the Green's function for a simple one dimensional lattice will be discussed in section (3.3.1), followed by the calculation of the scattering matrix for a one-dimensional scatterer in section (3.3.2).

## 3.3.1. Perfect One-Dimensional Lattice

This section presents the form of the Green's function for a simple one-dimensional lattice with on-site energies $\varepsilon_0$ and real hopping parameters $-\gamma$ as shown in figure (3.2). The tight-binding approximation has been used, which assumes that the wavefunction of the system can be described as a superposition of wavefunctions for isolated atoms. Another assumption that is taken into account is that only nearest-neighbour coupling is non-zero.

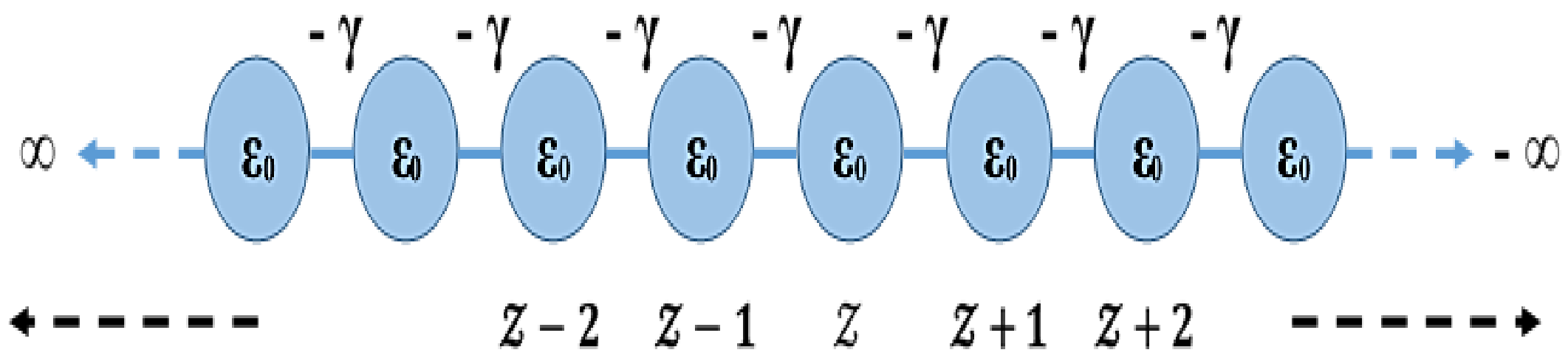

*Figure 3.2: Tight-binding model of a one-dimensional periodic lattice with on-site energies $\varepsilon_0$ and couplings $\gamma$.*

The matrix form of the Hamiltonian can be simply written:

$$H = \begin{pmatrix} \ddots & -\gamma & 0 & 0 \\ -\gamma & \varepsilon_0 & -\gamma & 0 \\ 0 & -\gamma & \varepsilon_0 & -\gamma \\ 0 & 0 & -\gamma & \ddots \end{pmatrix} \qquad (3.7)$$

The Schrödinger equation (3.8), can be expanded at a lattice site $\mathcal{Z}$ in terms of the energy and wavefunction $\Psi_{\mathcal{Z}}$ (equation (3.9)):

$$(E - H)\Psi = 0 \qquad (3.8)$$

$$\varepsilon_0 \Psi_{\mathcal{Z}} - \gamma \Psi_{\mathcal{Z}+1} - \gamma \Psi_{\mathcal{Z}-1} = E\Psi_{\mathcal{Z}} \qquad (3.9)$$

The wavefunction for this perfect lattice takes the form of propagating Bloch state (equation (3.10)), normalised by its group velocity, $\upsilon_g$, in order for it to carry unit current flux. The substitution of this into equation (3.9) leads to the well-known one dispersion relation (equation (3.11)).

$$\Psi_{\mathcal{Z}} = \frac{1}{\sqrt{\upsilon_g}} e^{ik\mathcal{Z}} \qquad (3.10)$$

$$E = \varepsilon_0 - 2\gamma cosk \qquad (3.11)$$

Here, $k$ refers to the wavenumber. It is true that the retarded Green's function, $g(\mathcal{Z},\mathcal{Z}')$, is closely related to the wavefunction and is the solution to an equation very similar to that of the Schrödinger equation:

$$(E-H)g(\mathcal{Z},\mathcal{Z}') = \delta_{\mathcal{Z},\mathcal{Z}'} \quad (3.12)$$

The retarded Green's function, $g(\mathcal{Z},\mathcal{Z}')$, describes the response of a system at a point $\mathcal{Z}$ due to a source at a point $\mathcal{Z}'$.

Naturally, it can be anticipated that such an excitation to give rise to two waves, traveling outwards from a point of the excitation, with amplitudes A and B as shown in figure (3.3).

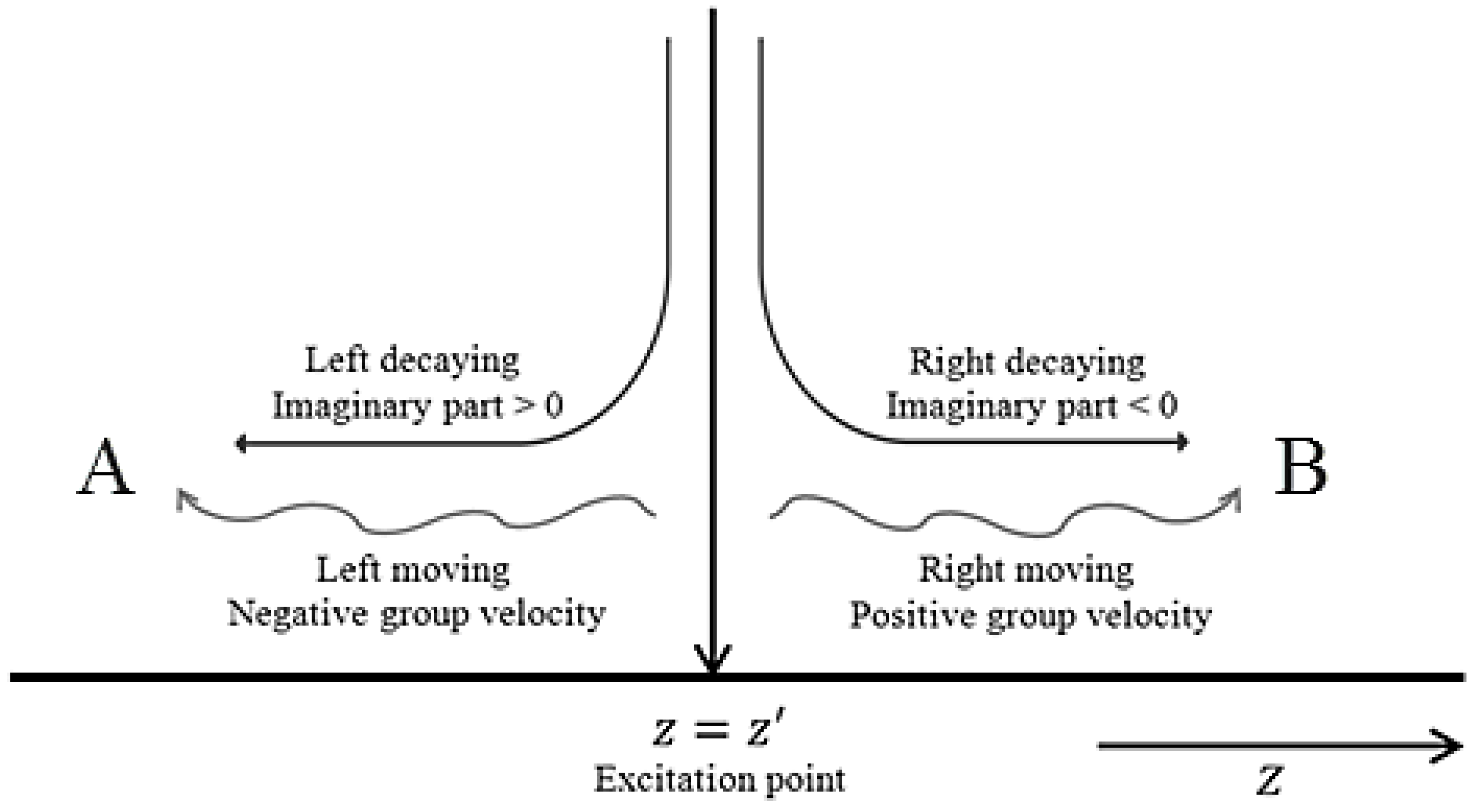

*Figure 3.3: Retarded Green's function of an infinite one-dimensional lattice. The excitation at $\mathcal{Z} = \mathcal{Z}'$ causes waves to propagate left and right with amplitudes A and B respectively.*

These waves can be expressed as:

$$g(\mathcal{Z}',\mathcal{Z}) = Be^{ik\mathcal{Z}} \qquad \mathcal{Z} > \mathcal{Z}'$$
$$g(\mathcal{Z}',\mathcal{Z}) = Ae^{-ik\mathcal{Z}} \qquad \mathcal{Z} < \mathcal{Z}' \qquad (3.13)$$

This solution satisfies equation (3.12) at every point, but $\mathcal{Z} = \mathcal{Z}'$. To overcome this, the Green's function must be continuous (equation (3.14)), and therefore the two are equated at $\mathcal{Z} = \mathcal{Z}'$:

$$[g(\mathcal{Z},\mathcal{Z}')]_{\mathcal{Z}=\mathcal{Z}'\,left} = [g(\mathcal{Z},\mathcal{Z}')]_{\mathcal{Z}=\mathcal{Z}'\,right} \qquad (3.14)$$

$$Be^{ik\mathcal{Z}'} = Ae^{-ik\mathcal{Z}'} \implies A = Be^{2ik\mathcal{Z}'} \qquad (3.15)$$

Substituting equation (3.15) into Green's function (equation (3.13)) yields:

$$g(\mathcal{Z}',\mathcal{Z}) = Be^{ik\mathcal{Z}} \qquad = Be^{ik\mathcal{Z}'}e^{ik(\mathcal{Z}-\mathcal{Z}')} \qquad \mathcal{Z} \geq \mathcal{Z}'$$
$$g(\mathcal{Z}',\mathcal{Z}) = Be^{2ik\mathcal{Z}'}e^{-ik\mathcal{Z}} \qquad = Be^{ik\mathcal{Z}'}e^{ik(\mathcal{Z}'-\mathcal{Z})} \qquad \mathcal{Z} \leq \mathcal{Z}' \qquad (3.16)$$

Equation (3.16) can be rewritten simply as:

$$g(\mathcal{Z},\mathcal{Z}') = Be^{ik\mathcal{Z}'}e^{ik|\mathcal{Z}-\mathcal{Z}'|} \qquad (3.17)$$

To define the constant *B*, the Green's function (equation (3.12)) must be considered, then *H* can be written as $-\frac{\hbar^2}{2m}\nabla^2 \; or \; -\frac{\hbar v_g}{2k}\nabla^2$ (where $v_g = \frac{\hbar k}{m}$ is the group velocity), and then substituted into the Green's function (equation (3.17)), so that the equation becomes:

$$\left(E + \frac{\hbar v_g}{2k}\frac{\partial^2}{\partial \mathcal{Z}^2}\right)\left(Be^{ik\mathcal{Z}'}e^{ik|\mathcal{Z}-\mathcal{Z}'|}\right) = \delta_{\mathcal{Z},\mathcal{Z}'} \qquad (3.18)$$

The integral of this function over a small distance, centred on $\mathcal{Z}'$, of width $2\omega^+$ gives:

$$\int_{\mathcal{Z}'-\omega^+}^{\mathcal{Z}'+\omega^+}\left(E+\frac{\hbar v_g}{2k}\frac{\partial^2}{\partial \mathcal{Z}^2}\right)\left(Be^{ik\mathcal{Z}'}e^{ik|\mathcal{Z}-\mathcal{Z}'|}\right)d\mathcal{Z}=\int_{\mathcal{Z}'-\omega^+}^{\mathcal{Z}'+\omega^+}\delta_{\mathcal{Z},\mathcal{Z}'}\,d\mathcal{Z}$$

$$Be^{ik\mathcal{Z}'}\left(E\int_{\mathcal{Z}'-\omega^+}^{\mathcal{Z}'+\omega^+}e^{ik|\mathcal{Z}-\mathcal{Z}'|}\,d\mathcal{Z}+\int_{\mathcal{Z}'-\omega^+}^{\mathcal{Z}'+\omega^+}\frac{\hbar v_g}{2k}\frac{\partial^2}{\partial \mathcal{Z}^2}e^{ik|\mathcal{Z}-\mathcal{Z}'|}d\mathcal{Z}\right)=1$$

$$Be^{ik\mathcal{Z}'}\left(\frac{\hbar v_g}{2k}\frac{\partial}{\partial \mathcal{Z}}e^{ik|\mathcal{Z}-\mathcal{Z}'|}\right)_{\mathcal{Z}'-\omega^+}^{\mathcal{Z}'+\omega^+}=Be^{ik\mathcal{Z}'}\left(\frac{\hbar v_g}{2k}ike^{ik|\mathcal{Z}-\mathcal{Z}'|}\right)_{\mathcal{Z}'-\omega^+}^{\mathcal{Z}'+\omega^+}=1$$

$$Be^{ik\mathcal{Z}'}\frac{\hbar v_g}{2k}2ik=1\quad\Longrightarrow\quad Be^{ik\mathcal{Z}'}=\frac{1}{i\hbar v_g} \tag{3.19}$$

Hence, the retarded Green's function can be written as:

$$g^R(\mathcal{Z}-\mathcal{Z}')=\frac{1}{i\hbar v_g}e^{ik|\mathcal{Z}-\mathcal{Z}'|} \tag{3.20}$$

The group velocity, which is found from the dispersion relation is:

$$v_g=\frac{1}{\hbar}\frac{\partial E(k)}{\partial k}=2\gamma sink \tag{3.21}$$

The retarded Green's function will be used in this thesis, and for more simplicity, *R* has been excluded from its representation. Therefore, $g^R(\mathcal{Z}-\mathcal{Z}')=g(\mathcal{Z},\mathcal{Z}')$; more details of this derivation can be found in literature [3, 5, 6].

## 3.3.2. One-Dimensional Scattering

This section presents an attempt to obtain the Green's function of a system, which involves two segments of one-dimensional tight binding semi-infinite leads connected by a coupling element α, as shown in figure 3.4.

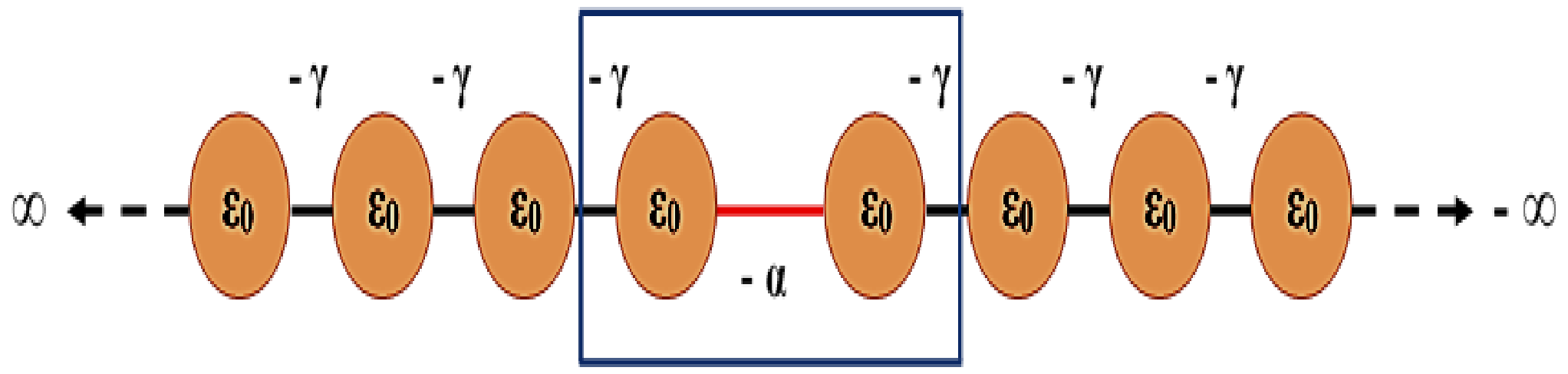

*Figure 3.4: Simple tight-binding model of a one dimensional scatterer attached to one dimensional leads.*

The system in figure 3.4 is tricky because, though it seems simple, all one-dimensional setups can be reduced back to this topology. Taking this into account, the analytical solutions for the transmission and reflection coefficient would be very valuable. The definition of the Hamiltonian, which takes the form of an infinite matrix is:

$$H = \begin{pmatrix} \ddots & -\gamma & 0 & 0 & 0 & 0 \\ -\gamma & \varepsilon_0 & -\gamma & 0 & 0 & 0 \\ 0 & -\gamma & \varepsilon_0 & \alpha & 0 & 0 \\ 0 & 0 & \alpha & \varepsilon_0 & -\gamma & 0 \\ 0 & 0 & 0 & -\gamma & \varepsilon_0 & -\gamma \\ 0 & 0 & 0 & 0 & -\gamma & \ddots \end{pmatrix} = \begin{pmatrix} H_L & V_c \\ & \\ V_c^{\dagger} & H_R \end{pmatrix} \quad (3.22)$$

Here, $H_L$ and $H_R$ indicate to Hamiltonians of the leads, which are the semi-infinite equivalent of the Hamiltonian shown in equation (3.7). $V_c$ denotes the coupling parameter. For real γ, the dispersion relation corresponding to the leads introduced above, are given in equation (3.11), and the group velocity was given in equation (3.21). The scattering amplitudes can be obtained by calculating the Green's function of the system. The formal solution to equation (3.12), can be written as:

$$G = (E - H)^{-1} \tag{3.23}$$

It is singular if the energy $E$ is equal to an eigenvalues of the Hamiltonian $H$. To circumvent this problem, it is practical to consider the limit:

$$G_{\mp} = \lim_{\eta \to 0} (E - H \pm i\eta)^{-1} \tag{3.24}$$

Here, $\eta$ is a positive number and $G_{\mp}$ is the retarded (advanced) Green's function. The retarded Green's function has only been used in this thesis, and therefore the positive sign has been chosen. The retarded Green's function for an infinite one-dimensional chain is defined in equation (3.20).

$$g_{jl}^{\infty} = \frac{1}{i\hbar v_g} e^{ik|j-l|} \tag{3.25}$$

Here, $j$ and $l$ are the labels of the sites in the chain and adequate boundary conditions are needed to obtain the Green's function of a semi-infinite lead. The lattice is semi-infinite, and therefore the chain must be terminated at a given point, $i_0$. In order to satisfy the boundary conditions, the source is anticipated at $i_0$ – i.e. when $l = i_0$, there is no effect on the chain. In other words: $g = (j, i_0) = 0$ for $j \leq l$. In addition, it is expected that boundary to give rise to a reflected wave, $De^{-ik|j-l|}$:

$$g_{j,i_0} = \frac{1}{i\hbar v_g} e^{ik(i_0 - j)} + De^{-ik(i_0 - j)} = 0 \quad \Rightarrow \quad D = -\frac{1}{i\hbar v_g} e^{2ik(i_0 - j)} \tag{3.26}$$

Substituting this back into the Green's function leads to:

$$g_{j,l} = \frac{1}{i\hbar v_g} e^{ik(l-j)} - \frac{1}{i\hbar v_g} e^{2ik(i_0-j)} e^{-ik(l-j)}$$

$$g_{j,l} = \frac{1}{i\hbar v_g}\left(e^{ik(l-j)} - e^{ik(2i_0-j-l)}\right) \tag{3.27}$$

It is known that there is no propagation of a wave beyond the boundary, and that means at any point beyond $i_0 - 1$ there is no impact of the source on the chain. Therefore, if $j \geq l$ and $j \geq i_0$, it is expected $g(i_0, l) = 0$. According to this condition:

$$g_{i_0,l} = \frac{1}{i\hbar v_g} e^{ik(i_0-l)} + De^{-ik(i_0-l)} = 0 \;\; \Rightarrow \;\; D = -\frac{1}{i\hbar v_g} e^{2ik(i_0-l)} \tag{3.28}$$

Substituting this back into the Green's function leads to:

$$g_{j,l} = \frac{1}{i\hbar v_g} e^{ik(j-l)} - \frac{1}{i\hbar v_g} e^{2ik(i_0-l)} e^{-ik(j-l)}$$

$$g_{j,l} = \frac{1}{i\hbar v_g}\left(e^{ik(j-l)} - e^{ik(2i_0-j-l)}\right) \tag{3.29}$$

To summarize:

$$g_{j,l} = \begin{cases} \dfrac{1}{i\hbar v_g}\left(e^{ik(j-l)} - e^{ik(2i_0-j-l)}\right) & j \geq l \\ \\ \dfrac{1}{i\hbar v_g}\left(e^{ik(l-j)} - e^{ik(2i_0-j-l)}\right) & j \leq l \end{cases} \tag{3.30}$$

Equation (3.30) can be written as:

$$g_{j,l} = \frac{1}{i\hbar v_g}\left(e^{ik|j-l|} - e^{ik(2i_0-j-l)}\right) = g_{j,l}^{\infty} + \Psi_{j,l}^{i_0} \tag{3.31}$$

Here, $g_{j,l}^{\infty}$ is the Green's function of the infinite lattice and $\Psi_{j,l}^{i_0}$ is the mathematical representation of the boundary condition:

$$\Psi_{j,l}^{i_0} = -\frac{e^{ik(2i_0-j-l)}}{i\hbar v_g} \tag{3.32}$$

At the boundary, $j = l = i_0 - 1$, the Green's function will have the simple form:

$$g_{i_0-1,i_0-1} = -\frac{e^{ik}}{\gamma} \tag{3.33}$$

In case of decoupled leads, α = 0, the decoupled Green's function is given by:

$$g = \begin{pmatrix} -\frac{e^{ik}}{\gamma} & 0 \\ 0 & -\frac{e^{ik}}{\gamma} \end{pmatrix} = \begin{pmatrix} gL & 0 \\ 0 & gR \end{pmatrix} \tag{3.34}$$

In case of coupled leads, Dyson's equation is required to obtain the Green's function, *G*, of the system:

$$G^{-1} = (g^{-1} - V) \tag{3.35}$$

Here, *V* is the operator, which describes the interaction between leads:

$$V = \begin{pmatrix} 0 & V_c \\ V_c^{\dagger} & 0 \end{pmatrix} = \begin{pmatrix} 0 & -\alpha \\ -\alpha^* & 0 \end{pmatrix} \tag{3.36}$$

The solution of Dyson's equation (equation (3.35)) is:

$$G = \frac{1}{\alpha^2 - \gamma^2 e^{-2ik}} \begin{pmatrix} \gamma e^{-ik} & -\alpha \\ -\alpha^* & \gamma e^{-ik} \end{pmatrix} \tag{3.37}$$

Utilizing the Fisher-Lee relation [3, 7], the transmission, $t$, and reflection, $r$, amplitudes can be calculated from equation (3.37). It is obvious that the Green's function contains information of the transmission and reflection, since $G_{j,l}$ is a response at point $j$ to a source at point $l$. When point $j$ is in the right lead and point $l$ is in the left lead, the source emits two waves travelling outwards, one away from the scatterer and one towards the scatterer with amplitudes A and B respectively. For the right-going wave to affect point $j$, it has to travel through the scatterer. Therefore, the Green's function contains information on two waves; a left moving wave $\left(Ae^{-ik|j-l|} + Bre^{ik|j-l|}\right)$ and the transmitted right-moving wave $\left(Bte^{ik|j-l|}\right)$. The transmission, $t$, and reflection, $r$ coefficients are introduced.

The points before and after the scatterer, $i_0 - 1$ and $i_0$ respectively, which act as boundary states, are defined in equation (3.33). Since $A = B = \frac{1}{i\hbar v_g}$, the equation (3.38) gives a definition for the transmission and reflection coefficients, equations (3.39) and (3.40):

$$G_{1,1} = \frac{1}{i\hbar v_g}(1 + r)$$

$$G_{2,1} = \frac{1}{i\hbar v_g} te^{ik} \quad (3.38)$$

$$1 + r = i\hbar v_g G_{1,1} \quad (3.39)$$

$$t = i\hbar v_g G_{2,1} te^{ik} \quad (3.40)$$

These amplitudes correspond to particles incident from the left. The same expressions could be used for the transmission, $t'$, and reflection, $r'$, amplitudes for the particles are travelling from the right.

Based on these coefficients, the probability is defined: $\mathcal{T} = tt^*,\ R = rr^*$. Consequently, the transmission probability for this system can be written as:

$$\mathcal{T} = \frac{\sigma^2}{(\gamma^2 - \alpha^2)^2 + \sigma^2} \tag{3.41}$$

Here, $\sigma = 2\gamma\alpha sink$. Now, if $\alpha = \gamma$ the equation becomes unity, $\mathcal{T} = 1$, which is expected, and when $\alpha$ is greater or smaller than $\gamma$, it creates a scattering region, which result in a transmission probability, $\mathcal{T} \leq 1$.

The full scattering matrix has been obtained, and therefore the Landauer formula (equation (3.4)) can be used to calculate the zero bias conductance. The procedures by which this analytical solution for the conductance of a one-dimensional scatterer was found can be generalized for more complex systems, and can be considered as a base for the analytical formula of the quantum circuit rule as shown in chapter 4.

# 3.4. Generalization of the Scattering Formalism

This section, based on Lambert's derivation [8, 9] shows a generalized method to perform transport calculations. This is analogous to the former approach. It involves three parts, first the surface Green's function of crystalline leads is computed. The second one is represented by using of decimation technique to lower the dimensionality of the scattering region. Finally, by using a generalization of the Fisher-Lee relation the scattering amplitudes will be calculated.

### 3.4.1 Hamiltonian and Green's Function of the Leads

The general semi-infinite crystalline electrode of arbitrary complexity will be studied here. Because the leads are crystalline, the structure of the Hamiltonian is a generalization of the one-dimensional electrode Hamiltonian in equation (3.2).

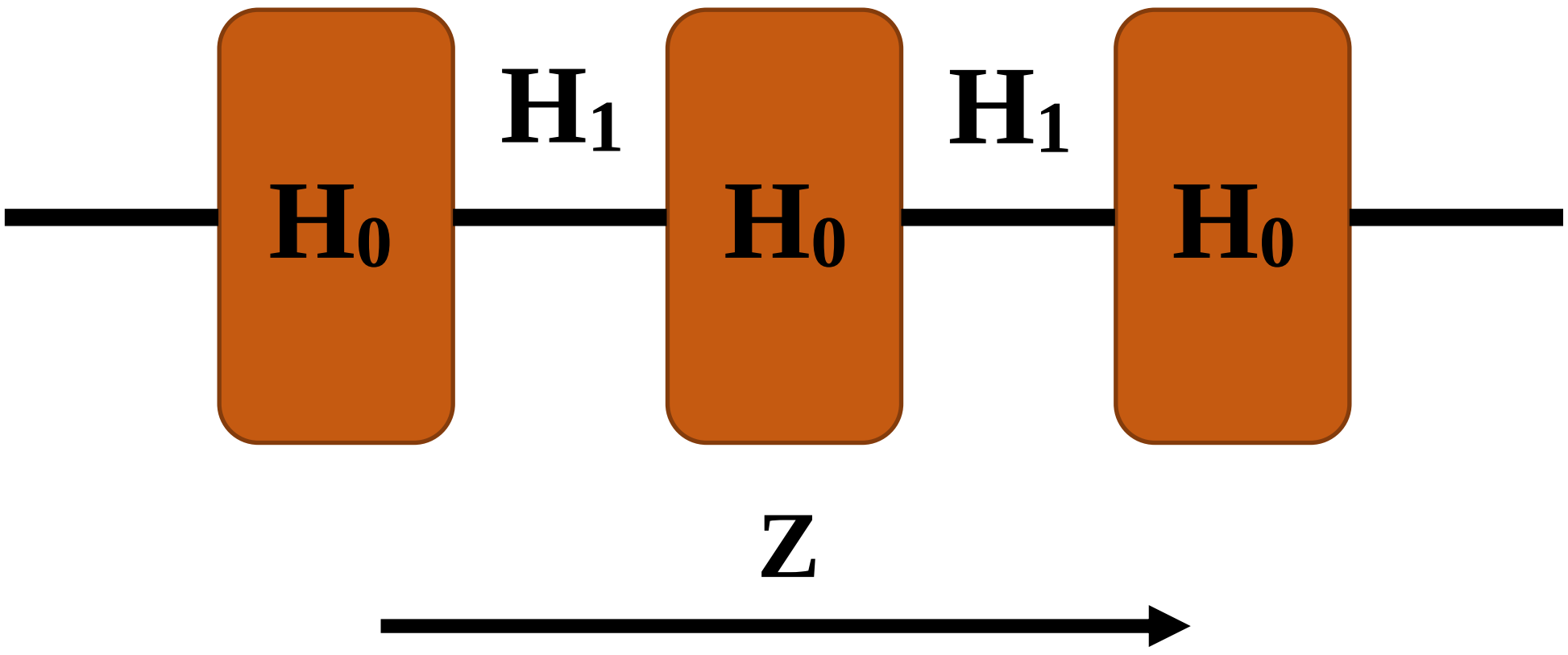

*Figure 3.5: Schematic representation of a semi-infinite generalized lead. States described by the Hamiltonian $H_0$ are connected via generalized hopping $H_1$. The direction Z is defined to be parallel to the axis of the chain.*

Figure 3.5, shows the general system topology. Instead of single site energies, the Hamiltonians for each repeating layer of the bulk electrode are described by $H_0$, and a coupling matrix to describe the coupling between these layers is denoted $H_1$. The total Hamiltonian for this system is:

$$H = \begin{pmatrix} \ddots & H_1 & 0 & 0 \\ H_1^\dagger & H_0 & H_1 & 0 \\ 0 & H_1^\dagger & H_0 & H_1 \\ 0 & 0 & H_1^\dagger & \ddots \end{pmatrix} \tag{3.42}$$

Here, $H_0$ and $H_1$ are in general complex matrices and only the restriction is that the full Hamiltonian, $H$, should be Hermitian. The main aim of this section is to calculate the Green's function of such a lead for a general $H_1$and $H_0$. Solving the Schrödinger equation is the way to calculate the spectrum of the Hamiltonian, and therefore the Green's function can be calculated.

$$E\Psi_{\mathcal{Z}} = H_0\Psi_{\mathcal{Z}} + H_1\Psi_{\mathcal{Z}+1} + H_1^{\dagger}\Psi_{\mathcal{Z}-1} \tag{3.42}$$

Here, $\Psi_{\mathcal{Z}}$ is the wave function describing layer $\mathcal{Z}$, where $\mathcal{Z}$ is an integer measured in units of inter-layer distance. The assumption here, that the system is infinitely periodic in the $\mathcal{Z}$ direction only, and therefore the on-site wavefunction, $\Psi_{\mathcal{Z}}$, can represented in Bloch form; consisting of a product of a propagating plane wave and a wavefunction, $\phi_k$, which is perpendicular to the transport direction, $\mathcal{Z}$. If the layer Hamiltonian, $H_0$, has dimensions $M \times M$ (or in other words consists of $M$ site energies and their respective hopping elements), the perpendicular wavefunction, $\phi_k$, will have $M$ degrees of freedom and take the form of a $1\times M$ dimensional vector. Therefore, the wavefunction, $\Psi_{\mathcal{Z}}$:

$$\Psi_{\mathcal{Z}} = \sqrt{n_k}\, e^{ik\mathcal{Z}}\phi_k \tag{3.43}$$

Here, $n_k$ is an arbitrary normalization parameter. Substituting this into the Schrödinger equation (3.42) gives:

$$\left(H_0 + e^{ik\mathcal{Z}}H_1 + e^{-ik\mathcal{Z}}H_1^{\dagger} - E\right)\phi_k = 0 \tag{3.44}$$

To obtain the band structure for such a problem, the $k$ values should be selected and then calculate the eigenvalues at that point, $E = E_1(k)$, where $l$ = 1,….$M$. Here, $l$ denotes the band index. For each value of $k$, there will be $M$ solutions to the eigenproblem, and so $M$ energy values. In this way, by selecting multiple values for $k$, it is relatively simple to build up a band structure.

In the scattering problem, the values of $k$ have been obtained at a given $E$, instead of finding the $E$ values at a given $k$, and this is approaching the problem from the opposite direction. A root-finding method is typically used to perform this, and therefore a huge numerical effort is required, because the wave numbers are in general complex. Alternatively, an eigenvalue problem can be written down in which the energy is the input and the wave numbers are the results.

$$\vartheta_k = e^{-ikZ}\phi_k \tag{3.45}$$

The combining of equations (3.45) and (3.44) gives:

$$\begin{pmatrix} H_1^{-1}(E - H_0) & -H_1^{-1}H_1^{\dagger} \\ I & 0 \end{pmatrix}\begin{pmatrix} \phi_k \\ \vartheta_k \end{pmatrix} = e^{ikZ}\begin{pmatrix} \phi_k \\ \vartheta_k \end{pmatrix} \tag{3.46}$$

The layer Hamiltonian, $H_0$, of size $M \times M$, from equation (3.46) yields $2M$ eigenvalues, $e^{ik_lZ}$, and eigenvectors $\phi_k$, of size $M$. These states can be sorted into four categories according to whether they are propagating or decaying and whether they are left going or right going. A state is propagating if it has a real wave number, $k_l$, and is decaying if it has an imaginary part. If the imaginary part of the wave number is positive then it is a left decaying state, while if the imaginary part is negative, then it is a right decaying state.

The propagating states are sorted according to the group velocity of the state:

$$\vartheta_{k_l} = \frac{1}{\hbar}\frac{\partial E_{k,l}}{\partial k} \tag{3.47}$$

If the group velocity, $\vartheta_{k_l}$, of the state is positive, there would be a right propagating state. In contrast, if the group velocity is negative, a left propagating state would be obtained.

*Table 3.1: Classification of the eigenstates into left and right propagating or decaying states according to the wave number and group velocity.*

| **Category** | **Left** | **Right** |
|---|---|---|
| **Decaying** | $Im\ (k_l) > 0$ | $Im\ (k_l) < 0$ |
| **Propagating** | $Im\ (k_l) = 0, \vartheta_{k_l} < 0$ | $Im\ (k_l) = 0, \vartheta_{k_l} > 0$ |

For convenience, from now on the $\overline{k_l}$ refers to the numbers which belong to the left propagating/decaying set, and $k_l$ indicates the numbers which belong to the right propagating/decaying set, and therefore $\phi_{k_l}$ is a wave function associated to a right state, while $\phi_{\bar{k}_l}$ is associated to a left state. If $H_1$ is invertible, there must be exactly the same number, *M*, of left and right going states. On the other hand, if $H_1$ is singular, the matrix in equation (3.46) cannot be constructed, since it relies of the inversion of $H_1$. However, various methods can be used to overcome this problem. The first one [10] uses the decimation method to create an effective non-singular $H_1$.

Another solution might be to populate a singular $H_1$ with small random numbers, hence introducing an explicit numerical error. This method is reasonable as the introduced numerical error can be as small as the numerical error introduced by decimation.

Another solution is to re-write equation (3.46), such that $H_1$ need not be inverted:

$$\begin{pmatrix} E - H_0 & -H_1^{\dagger} \\ I & 0 \end{pmatrix} \begin{pmatrix} \phi_k \\ \vartheta_k \end{pmatrix} = e^{ikZ} \begin{pmatrix} H_1 & 0 \\ 0 & I \end{pmatrix} \begin{pmatrix} \phi_k \\ \vartheta_k \end{pmatrix} \qquad (3.48)$$

However, solving this generalized eigenproblem is more computationally expensive and a singular $H_1$ continues to disrupt the mathematics further in the theory. The solution to the eigenproblem (equation (3.44)) at a given energy, *E*, will not generally form an orthogonal set of states. This is crucial, due to the dealing with non-orthogonality when constructing the Green's function, and therefore, it is necessary to introduce the duals to $\phi_{k_l}$ and $\phi_{\bar{k}_l}$ in such a way that obey:

$$\phi_{k_i}\tilde{\phi}_{k_j}^{\dagger} = \phi_{\bar{k}_i}\tilde{\phi}_{\bar{k}_i}^{\dagger} = \delta_{ij} \tag{3.49}$$

This gives the generalized completeness relation:

$$\sum_{l=1}^{M} \tilde{\phi}_{k_l}^{\dagger}\,\phi_{k_l} = \sum_{l=1}^{M} \tilde{\phi}_{\bar{k}_l}^{\dagger}\,\phi_{\bar{k}_l} = I \tag{3.50}$$

Now, the Green's function can be calculated first for the infinite system and then, by satisfying the appropriate boundary conditions, for the semi-infinite leads at their surface. Since the Green's function satisfies the Schrödinger equation when $\mathcal{Z} \neq \mathcal{Z}'$, the Green's function can be built up from the mixture of the eigenstates $\phi_{k_l}$ and $\phi_{\bar{k}_l}$.

$$g(\mathcal{Z},\mathcal{Z}') = \begin{cases} \sum_{l=1}^{M} \phi_{k_l} e^{ik_l(\mathcal{Z}-\mathcal{Z}')}\mathcal{W}_{k_l}^{\dagger} & \mathcal{Z} \geq \mathcal{Z}' \\ \\ \sum_{l=1}^{M} \phi_{\bar{k}_l} e^{i\bar{k}_l(\mathcal{Z}-\mathcal{Z}')}\mathcal{W}_{\bar{k}_l}^{\dagger} & \mathcal{Z} \leq \mathcal{Z}' \end{cases} \tag{3.51}$$

Here the *M*-component vectors $\mathcal{W}_{k_l}$ and $\mathcal{W}_{\bar{k}_l}$ are to be determined. It is important to note the structural similarities between this equation and equation (3.13), and also that all the degrees of freedom in the transverse direction are contained in the vectors $\phi_k$ and $\mathcal{W}_k$. The priority now is to obtain the $\mathcal{W}$ vectors. As in section (3.3.1), it is known that equation (3.51) must be continuous at $\mathcal{Z} \neq \mathcal{Z}'$ and should be achieved for the Green's function (equation (3.12)). The first condition is expressed as:

$$\sum_{l=1}^{M} \phi_{k_l} \mathcal{W}_{k_l}^{\dagger} = \sum_{l} \phi_{\bar{k}_l} \mathcal{W}_{\bar{k}_l}^{\dagger} \tag{3.52}$$

The second condition is expressed as:

$$\sum_{l=1}^{M} \left[ (E - H_0) \phi_{k_l} \mathcal{W}_{k_l}^{\dagger} + H_1 \phi_{k_l} e^{ik_l} \mathcal{W}_{k_l}^{\dagger} + H_1^{\dagger} \phi_{\bar{k}_l} e^{-i\bar{k}_l} \mathcal{W}_{\bar{k}_l}^{\dagger} \right] = I$$

$$\sum_{l=1}^{M} \left[ (E - H_0) \phi_{k_l} \mathcal{W}_{k_l}^{\dagger} + H_1 \phi_{k_l} e^{ik_l} \mathcal{W}_{k_l}^{\dagger} + H_1^{\dagger} \phi_{\bar{k}_l} e^{-i\bar{k}_l} \mathcal{W}_{\bar{k}_l}^{\dagger} + H_1^{\dagger} \phi_{k_l} e^{-ik_l} \mathcal{W}_{k_l}^{\dagger} + H_1^{\dagger} \phi_{k_l} e^{-ik_l} \mathcal{W}_{k_l}^{\dagger} \right] = I$$

$$\sum_{l=1}^{M} \left[ H_1^{\dagger} \phi_{-\bar{k}_l} e^{-i\bar{k}_l} \mathcal{W}_{\bar{k}_l}^{\dagger} + H_1^{\dagger} \phi_{k_l} e^{-ik_l} \mathcal{W}_{k_l}^{\dagger} \right] + \sum_{l=1}^{M} \left[ (E - H_0) + H_1 e^{ik_l} + H_1^{\dagger} e^{-ik_l} \right] \phi_{k_l} \mathcal{W}_{k_l}^{\dagger} = I \tag{3.53}$$

It is known from the Schrödinger equation (equation (3.44)) that:

$$\sum_{l=1}^{M} \left[ (E - H_0) + H_1 e^{ik_l} + H_1^{\dagger} e^{-ik_l} \right] = \phi_{k_l} = 0 \tag{3.54}$$

$$\sum_{l=1}^{M} H_1^{\dagger} \left( \phi_{\bar{k}_l} e^{-i\bar{k}_l} \mathcal{W}_{\bar{k}_l}^{\dagger} - \phi_{k_l} e^{-ik_l} \mathcal{W}_{k_l}^{\dagger} \right) = I \tag{3.55}$$

The using of dual vectors defined in equation (3.49) and multiplying equation (3.50) by $\tilde{\phi}_{k_p}$ gives:

$$\sum_{l=1}^{M} \tilde{\phi}_{k_p}^{\dagger} \phi_{\bar{k}_l} \mathcal{W}_{\bar{k}_l}^{\dagger} = \mathcal{W}_{k_p}^{\dagger} \tag{3.56}$$

and similarly multiplying by $\tilde{\phi}_{\bar{k}_l}$ yields:

$$\sum_{l=1}^{M} \tilde{\phi}_{\bar{k}_p}^{\dagger} \phi_{k_l} \mathcal{W}_{k_l}^{\dagger} = \mathcal{W}_{\bar{k}_p}^{\dagger} \tag{3.57}$$

Using the continuity equation (3.52) and equations (3.56 and 3.57), the Green's function equation (3.55) becomes:

$$\sum_{l=1}^{M} \sum_{p=1}^{M} H_1^{\dagger} \left( \phi_{k_l} e^{-ik_l} \tilde{\phi}_{k_l}^{\dagger} - \phi_{\bar{k}_l} e^{-i\bar{k}_l} \tilde{\phi}_{\bar{k}_l}^{\dagger} \right) \phi_{\bar{k}_p} \mathcal{W}_{\bar{k}_p}^{\dagger} = I \tag{3.58}$$

In what follows:

$$\sum_{l=1}^{M} \left[ H_1^{\dagger} \left( \phi_{k_l} e^{-ik_l} \tilde{\phi}_{k_l}^{\dagger} - \phi_{\bar{k}_l} e^{-i\bar{k}_l} \tilde{\phi}_{\bar{k}_l}^{\dagger} \right) \right]^{-1} = \sum_{p=1}^{M} \phi_{\bar{k}_p} \mathcal{W}_{\bar{k}_p}^{\dagger} = \sum_{p=1}^{M} \phi_{k_p} \mathcal{W}_{k_p}^{\dagger} \tag{3.59}$$

This immediately gives an expressions for $\mathcal{W}_k^{\dagger}$ :

$$\mathcal{W}_k^{\dagger} = \tilde{\phi}_k^{\dagger} \mathcal{V}^{-1} \tag{3.60}$$

Here, $\mathcal{V}$ is defined as:

$$\mathcal{V} = \sum_{l=1}^{M} H_1^{\dagger} \left( \phi_{k_l} e^{-ik_l} \tilde{\phi}_{k_l}^{\dagger} - \phi_{\bar{k}_l} e^{-i\bar{k}_l} \tilde{\phi}_{\bar{k}_l}^{\dagger} \right) \tag{3.61}$$

The wave number, $k$, in equation (3.60) indicates both left and right type of states. Substituting equation (3.60) into equation (3.51), gives the Green's function of an infinite system:

$$g^{\infty}_{\mathcal{Z},\mathcal{Z}'} = \begin{cases} \sum_{l=1}^{M} \phi_{k_l} e^{ik_l(\mathcal{Z}-\mathcal{Z}')} \tilde{\phi}^{\dagger}_{k} \mathcal{V}^{-1} & \mathcal{Z} \geq \mathcal{Z}' \\ \\ \sum_{l=1}^{M} \phi_{\bar{k}_l} e^{i\bar{k}_l(\mathcal{Z}-\mathcal{Z}')} \tilde{\phi}^{\dagger}_{\bar{k}} \mathcal{V}^{-1} & \mathcal{Z} \leq \mathcal{Z}' \end{cases} \tag{3.62}$$

In order to get the Green's function for a semi-infinite lead, a wave function should be added to the Green's function, in order to satisfy the boundary conditions at the edge of the lead, as with the one dimensional case. The boundary condition here is that the Green's function must vanish at a given place, $\mathcal{Z} = \mathcal{Z}_0$. In order to perform that $(g = g^{\infty} + \Delta)$ has been added to the Green's function (equation (3.62)).

$$\Delta = \sum_{l,p=1}^{M} \phi_{\bar{k}_l} e^{i\bar{k}_l(\mathcal{Z}-\mathcal{Z}_0)} \tilde{\phi}^{\dagger}_{\bar{k}_l} \phi_{k_p} e^{ik_p(\mathcal{Z}-\mathcal{Z}_0)} \tilde{\phi}^{\dagger}_{k_p} \mathcal{V}^{-1} \tag{3.63}$$

This gives the surface Green's function for a semi-infinite lead going left:

$$g_L = \left( I - \sum_{l,p} \phi_{\bar{k}_l} e^{-i\bar{k}_l} \tilde{\phi}^{\dagger}_{\bar{k}_l} \phi_{k_p} e^{ik_p} \tilde{\phi}^{\dagger}_{k_p} \right) \mathcal{V}^{-1} \tag{3.64}$$

and going right:

$$g_R = \left( I - \sum_{l,p} \phi_{\bar{k}_l} e^{i\bar{k}_l} \tilde{\phi}^{\dagger}_{\bar{k}_l} \phi_{k_p} e^{-ik_p} \tilde{\phi}^{\dagger}_{k_p} \right) \mathcal{V}^{-1} \tag{3.65}$$

## 3.4.2. Effective Hamiltonian of the Scattering Region

The coupling matrix between the surfaces of the semi-infinite leads has been shown in section (3.3.2), as well as the Dyson equation (3.36) can be used to calculate the Green's function of the scatterer. However, the scattering region is not generally described simply as a coupling matrix between surfaces. Therefore, it is useful to use the decimation method to reduce the Hamiltonian down to such a structure. Other methods have been developed [11, 12]. In this thesis, the decimation method has been used.

Consider again the Schrödinger equation:

$$\sum_{j} H_{ij}\Psi_j = E\Psi_i \qquad (3.66)$$

If we separate the *lth* degree of freedom in the system:

$$H_{il}\Psi_l + \sum_{j\neq l} H_{ij}\Psi_j = E\Psi_i \qquad i \neq l \qquad (3.67)$$

$$H_{ll}\Psi_l + \sum_{j\neq l} H_{lj}\Psi_j = E\Psi_l \qquad i = l \qquad (3.68)$$

Now, $\Psi_l$ can be expressed from equation (3.68) as:

$$\Psi_l = \sum_{j\neq l} \frac{H_{lj}\Psi_j}{E - H_{ll}} \qquad (3.69)$$

Substituting of equation (3.69) into equation (3.67) yields:

$$\sum_{j\neq l} \left[ H_{ij}\Psi_j + \frac{H_{il}H_{lj}\Psi_j}{E - H_{ll}} \right] = E\Psi_i \qquad i \neq l \qquad (3.70)$$

Equation (3.70) can be considered as an effective Schrödinger equation, where the number of degree of freedom is lowered by one compared to equation (3.66).

Hence, the new effective Hamiltonian, $H'$, is given by:

$$H'_{ij} = H_{ij} + \frac{H_{il}H_{lj}}{E - H_{ll}} \tag{3.71}$$

This Hamiltonian is the decimated Hamiltonian produced by simple Gaussian elimination. A notable feature of the decimated Hamiltonian is that it is energy dependent, which suits the method presented in former section very well. Without the decimation method, the Hamiltonian describing the system in general would take the form:

$$H = \begin{pmatrix} H_L & V_L & 0 \\ V_L^\dagger & H_{scatt} & V_R \\ 0 & V_R^\dagger & H_R \end{pmatrix} \tag{3.72}$$

Here, $H_L$ and $H_R$ denote the semi-infinite leads, $H_{scatt}$ denotes the Hamiltonian of the scatterer; $V_L$ and $V_R$ are the coupling Hamiltonians, which couple the original scattering region to the leads.

After decimation, an effectively equivalent Hamiltonian has been produced:

$$H = \begin{pmatrix} H_L & V_c \\ V_c^\dagger & H_R \end{pmatrix} \tag{3.73}$$

Here, $V_c$ denotes an effective coupling Hamiltonian, which now describes the whole scattering process. Now the same steps as with the one-dimensional case can be applied; using Dyson's equation (3.35). Hence, the Green's function for the whole system is described by the surface Green's function (equations (3.64 and 3.65)), and the effective coupling Hamiltonian from equation (3.73).

$$G = \begin{pmatrix} g_L^{-1} & V_c \\ V_c^\dagger & g_R^{-1} \end{pmatrix}^{-1} = \begin{pmatrix} G_{00} & G_{01} \\ G_{10} & G_{11} \end{pmatrix} \tag{3.74}$$

# Chapter 4

# A quantum circuit rule for interference effects in single-molecule electrical junctions

## 4.1. Introduction

Studies of the electrical conductance of single molecules attached to metallic electrodes not only probe the fundamentals of quantum transport, but also provide the knowledge needed to develop future molecular-scale devices and functioning circuits [1 – 9]. Owing to their small size (on the scale of angstroms) and the large energy gaps (on the scale of eV), transport through single molecules can remain phase coherent even at room temperature, and constructive or destructive quantum interference (QI) can be utilized to manipulate their room temperature electrical [10 – 13] and thermoelectrical [14 – 15] properties. In previous studies, it was reported theoretically and experimentally that the conductance of a phenyl ring with meta (m) connectivity is lower than the isomer with para (p) connectivity by several orders of magnitude [16 – 26]. This arises because partial de Broglie waves traversing different paths through the ring are perfectly out of phase leading to destructive QI in the case of meta coupling, while for para or ortho coupling they are perfectly in phase and exhibit constructive QI.

It is therefore natural to investigate how QI in molecules with multiple aromatic rings can be utilized in the design of more complicated networks of interference-controlled molecular units. The basic unit for studying QI in single molecules is the phenyl ring, with thiol [17, 21], methyl thioether [27], amine [17], or cyanide [19] anchors directly connecting the aromatic ring to gold electrodes. Recently, Arroyo *et al.* [28, 29] studied the effect of QI in a central phenyl ring by varying the coupling to various anchor groups, including two variants of thienyl anchors. However, the relative importance of QI in central rings compared with QI in anchor groups has not been studied systematically because the thienyl anchors of Arroyo *et al.* [28, 29] were five-membered rings, which exhibit only constructive interference.

To study the relative effect of QI in anchors, the molecules were examined with terminal groups formed from six-membered pyridyl rings for which constructive and destructive interference is possible. It has been reported [30] that pyridyl rings are excellent anchor groups for attaching single molecules to metallic electrodes because of their well-defined binding geometry, in which the nitrogen plays the role of the anchoring site. The ability to substitute pyridine either para (p), meta (m) or ortho (o) to the nitrogen offers the possibility of systematically investigating the relative importance of QI in this anchor group in molecules of the type X-Y-X, where X is a pyridyl ring and Y is a central phenyl ring. This question is rather subtle because charge transport from the electrode into the pyridyl rings takes place via the N atoms and also via metal–π coupling [31] and the interplay between these two mechanisms will determine the importance and robustness of QI in the terminal [32] rings.

In the present work, the aim is to compare the effects of QI in both the terminal rings X and the central ring Y of molecules of the type X-Y-X, and to study the relationship between their conductances. The results were found to satisfy the quantum circuit rule $G_{ppp}/G_{pmp} = G_{mpm}/G_{mmm}$, which demonstrates that the contribution to the conductance from the central ring is independent of the para versus meta nature of the anchor groups. Combinatorial rules for QI in molecules with other geometries including fused rings or rings connected in parallel are discussed in refs [33–36].

The pyridyl-terminated oligo(phenyleneethynylene) (OPE) derivatives shown in figure 4.1 are studied in this chapter, which possess a variety of connectivities of the central ring and locations of the nitrogen in the anchor units. In group 1, the central unit Y is a phenyl ring with para connectivity, whereas the central ring of group 2 has meta connectivity. The anchor units X are pyridyl rings with their nitrogens located in either meta, ortho or para positions.

This chapter presents all theoretical details and experimental conductance measurements as a part of a published paper. For more details regarding to experimental methods and synthesis details see *Manrique D. Z.; Huang C.; Baghernejad M.; Zhao X.; Oday A. Al-Owaedi, Sadeghi H.; Kaliginedi V.; Hong W.; Gulcur M.; Wandlowski T.; Bryce M. R.; Lambert C. J. A Quantum Circuit Rule for Interference Effects in Single-Molecule Electrical Junctions. Nat. Commun. 2015. 6(6389): p. 1-8.*

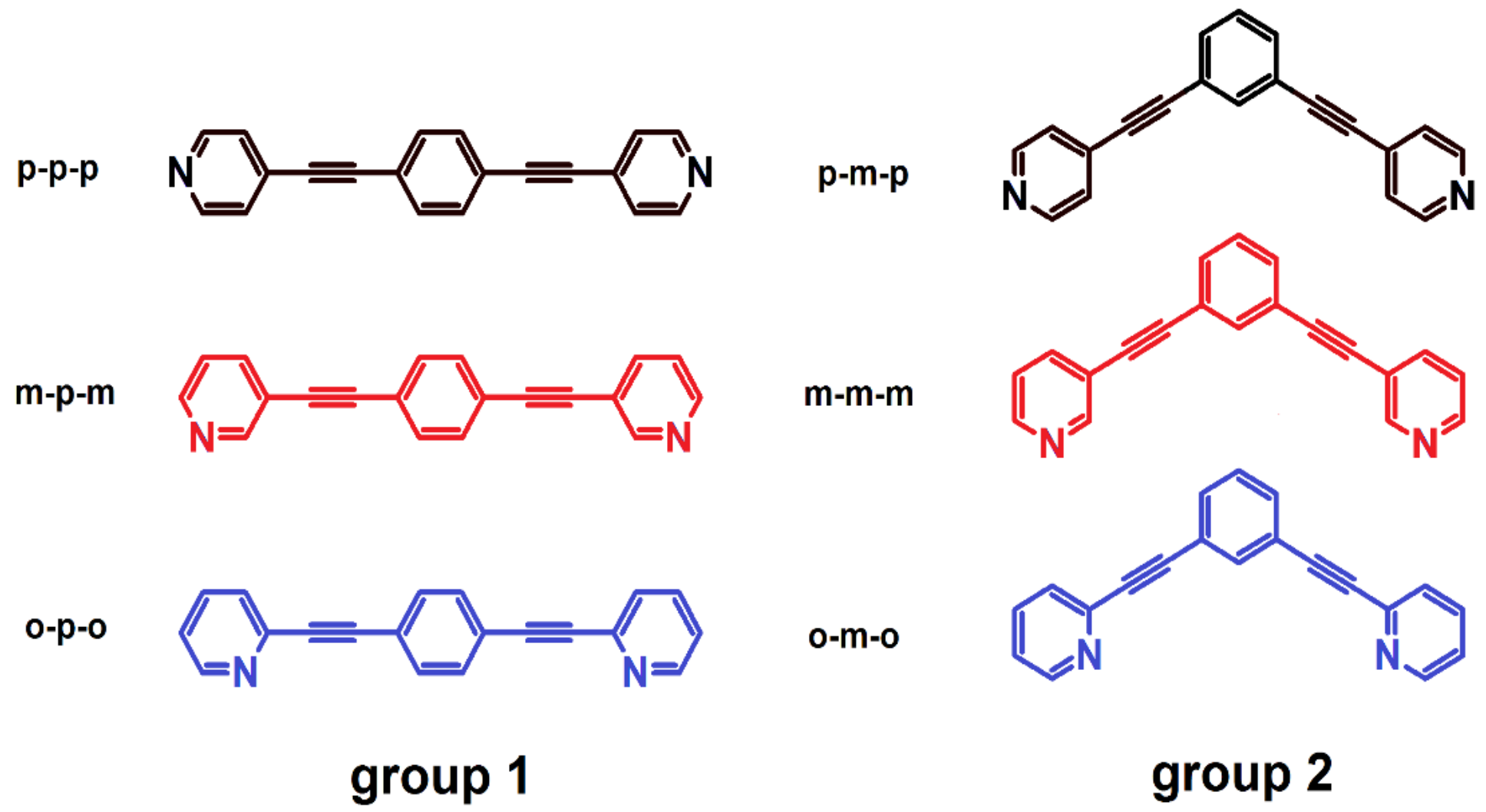

*Figure 4.1: The molecular structures studied in this work. These are divided into two groups, based on the presence of a para (group 1) or meta (group 2) central phenyl ring.*

# 4.2. Experimental and Theoretical Methods

## 4.2.1. Experimental Methods

The transport characteristics in single-molecule junctions were studied by mechanically controllable break junction (MCBJ) and scanning tunnelling microscopy break junction (STM-BJ) measurements in solution at room temperature. All details of experimental methods are presented in ref. [37].

### 4.2.2. Theoretical Methods

The DFT-Landauer approach used in the modeling assumes that on the time scale taken by an electron to traverse the molecule, inelastic scattering is negligible. This is known to be an accurate assumption for molecules up to several nanometers in length [26]. Geometrically optimizations were carried out using the DFT code SIESTA, with a generalized gradient approximation [38, 39] (PBE functional), double ζ polarized basis set, 0.01 eV/A force tolerance, a real-space grid with a plane wave cut-off energy of 250 Ry, zero bias voltage and 1 k-point.

All molecules in this study were initially geometrically relaxed in isolation to yield the geometries presented in figure 4.9. To investigate ideal junction geometries, a small four-atom gold pyramid was attached to the N atoms of the molecules, with Au–N–C angle being 120$^\circ$ and Au–N bond length being 2.1 Å, as shown in figure 4.2. Many of these junction geometries are unlikely to happen in break-junction (BJ) experiments, and during the junction elongation typically the gap between the electrodes is shorter than the molecular length. Another set of idealized junction geometries were constructed, where the gold pyramid is attached to rings from the side, perpendicular to the ring. These junction geometries are shown in figure 4.10. For transport calculations, the four-atom gold pyramids that are presented in figure 4.10 are extended to a 35-atom gold pyramid.

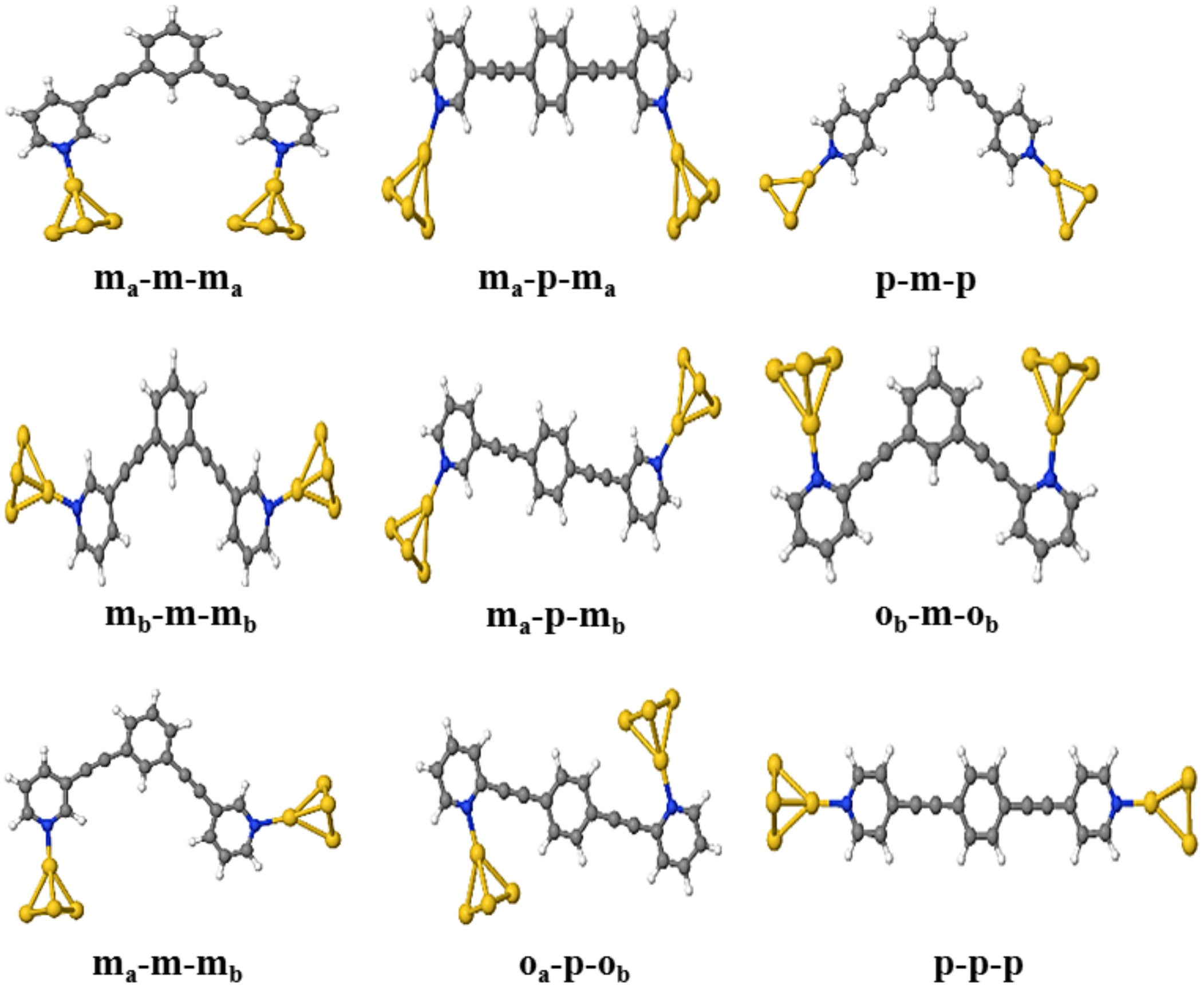

*Figure 4.2: Various idealized junctions with connected para, meta and ortho aromatic rings, illustrating the versatile planar conformations of the three-aromatic ring systems as possible components of molecular circuits.*

In break-junction (BJ) experiments, more complicated structures are expected. For this reason, BJ simulations have been performed. The geometrically optimized molecules were inserted between two opposing 35-atom (111) directed pyramids with four different tip separations (The tip separation is defined to be the centre to centre distance between the apex atoms of the two opposing pyramids).

In the initial geometries, the molecules were shifted slightly towards one of the pyramids and the initial Au–N distances were ~ 2.5 Å. Then the constructed structure was geometrically relaxed such that the base layers of the pyramids were kept fixed during the optimizations. The optimized junction geometries are shown in figures 4.3 and 4.4.

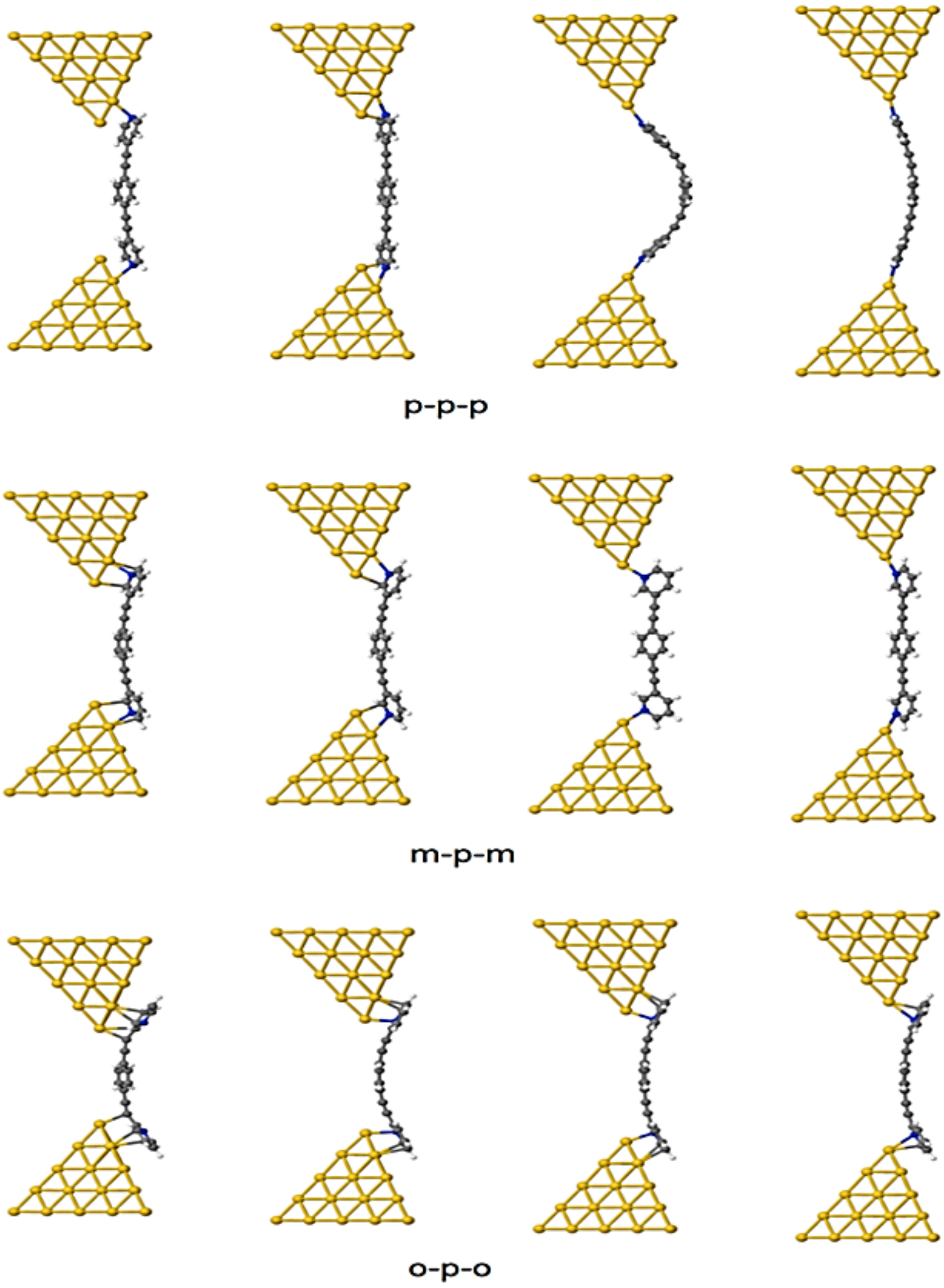

*Figure 4.3: Junction geometries for group 1 molecules in the BJ simulation. The electrode separation ($z_{the}$) increases from left to right. For each molecule there are four geometries that correspond to the four theoretical conductance points in the simulated trace in figure 4.13.*

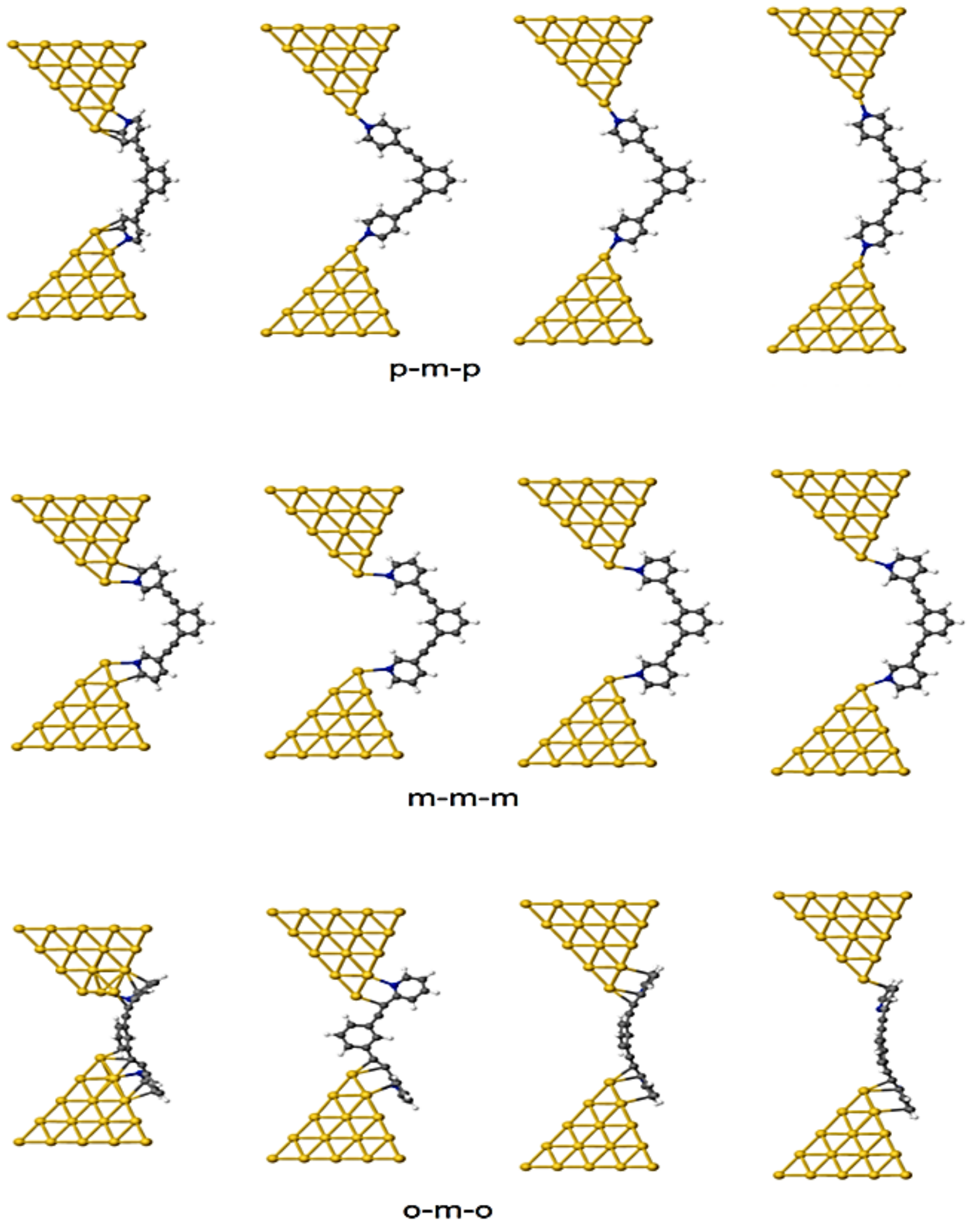

*Figure 4.4: Junction geometries for group 2 molecules in the BJ simulation. The electrode separation ($z_{the}$) increases from left to right. For each molecule there are four geometries that correspond to the four theoretical conductance points in the simulated trace in figure 4.13.*

For each relaxed junction geometry, the transmission coefficient, *T(E)*, describing the propagation of electrons of energy *E* from the left to the right electrode was calculated by first obtaining the corresponding Hamiltonian and overlap matrices using SIESTA and then using the GOLLUM code [40]. The transmission coefficient for all junction geometries in this study was obtained using wide-band electrodes with $\Gamma = 4.0$ eV. The wide-band electrodes were coupled to the two base layers of gold atoms of the 35-atom pyramids. A few additional transmission coefficient functions have been shown in figure 4.12. To produce the conductance-trace curves in figure 4.13, the conductance $G/G_0 = T(E_F)$ was obtained by evaluating *T(E)* at the Fermi energy $E_F$. To determine $E_F$, the predicted conductance values of all molecules have been compared with the experimental values and chose a single common value of $E_F$ which gave the closest overall agreement. This yielded a value of $E_F - E_F^{DFT} = -0.65$ eV, which is used in all theoretical results. This is commonly accepted procedure in molecular electronics DFT-based calculations [41].

To further demonstrate the generality of the product rule, DFT based transport calculations (with the exact same methodology that gave the result in figure 4.10) for two pyridyl ring systems with para and meta connections have been performed. The structures of three molecules studied are shown in figure 4.5. The theoretical derivation shown later implies that for two rings $G_{pp}\, G_{mm} = G^2_{pm}$, Where the $G_{pp}$ and $G_{mm}$ are the conductances of para-para and meta-meta pyridyl rings, and $G_{mp}$ is the conductance of the molecule for meta and para pyridyl rings. All transmission coefficient curves are shown in figure 4.14.

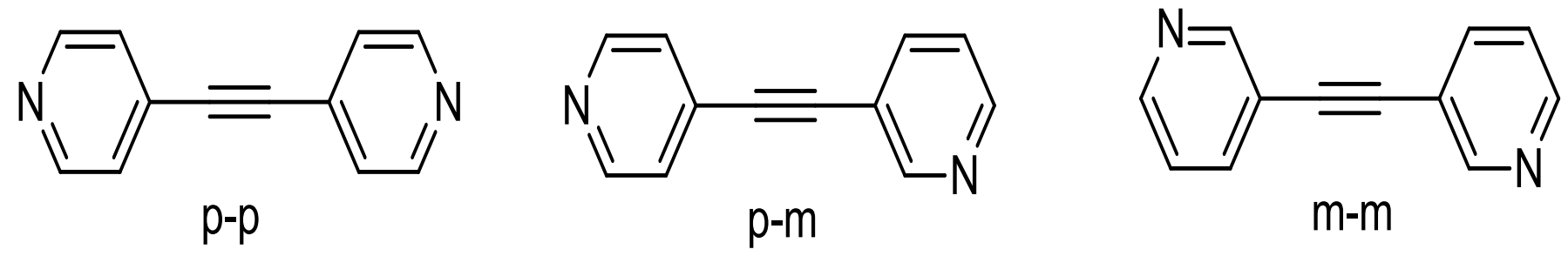

*Figure 4.5: Structures of two pyridyl rings with para and meta connections. These molecules have been studied theoretically.*

# 4.3. Results and Discussion

Nuclear magnetic resonance spectra for these molecules and details of synthesis are presented in ref. [37]. In this section experimental single molecule conductance results and theoretical, such as transmission coefficient results and the derivation of the quantum circuit rule for molecular conductances have been presented.

## 4.3.1. Experimental Results

Charge transport characteristics of single-molecule junctions formed from the molecules in figure 4.1 were investigated using both the mechanically controllable break junction (MCBJ) and scanning tunnelling microscopy break junction (STM-BJ) techniques, as reported elsewhere [6, 30, 42, 43].

Figure 4.6a displays typical traces of the conductance G (in units of the conductance quantum $G_0 = 2e^2/h$) versus the relative electrode displacement (ΔZ) from measurement of the molecule p-p-p (Corresponding results for other molecules are presented in figure 4.7). ΔZ is defined to be zero when $G = 0.7G_0$. It is related to the electrode separation $Z_{exp}$ by $Z_{exp} = \Delta Z + \Delta Z_{corr}$, where the correction $\Delta Z_{corr} = 0.5 \pm 0.1$ nm accounts for the snap-back of the electrodes upon breaking of the gold–gold atomic contact [44]. For this molecule, the $\log(G/G_0)$ versus ΔZ stretching traces possess well-defined plateaus in the range of $\log(G/G_0)$ around -4.5, which is attributed to the conductance of single-molecule junctions. Two-dimensional (2D) histograms of p-p-p in figure 4.6b show features of gold point contacts around $G \geq 1G_0$ followed by a second accumulation in the cloud-like scatter plot in the range ($10^{-5.0}G_0 < G < 10^{-3.6}G_0$), centred at $G = 10^{-4.5}G_0$.

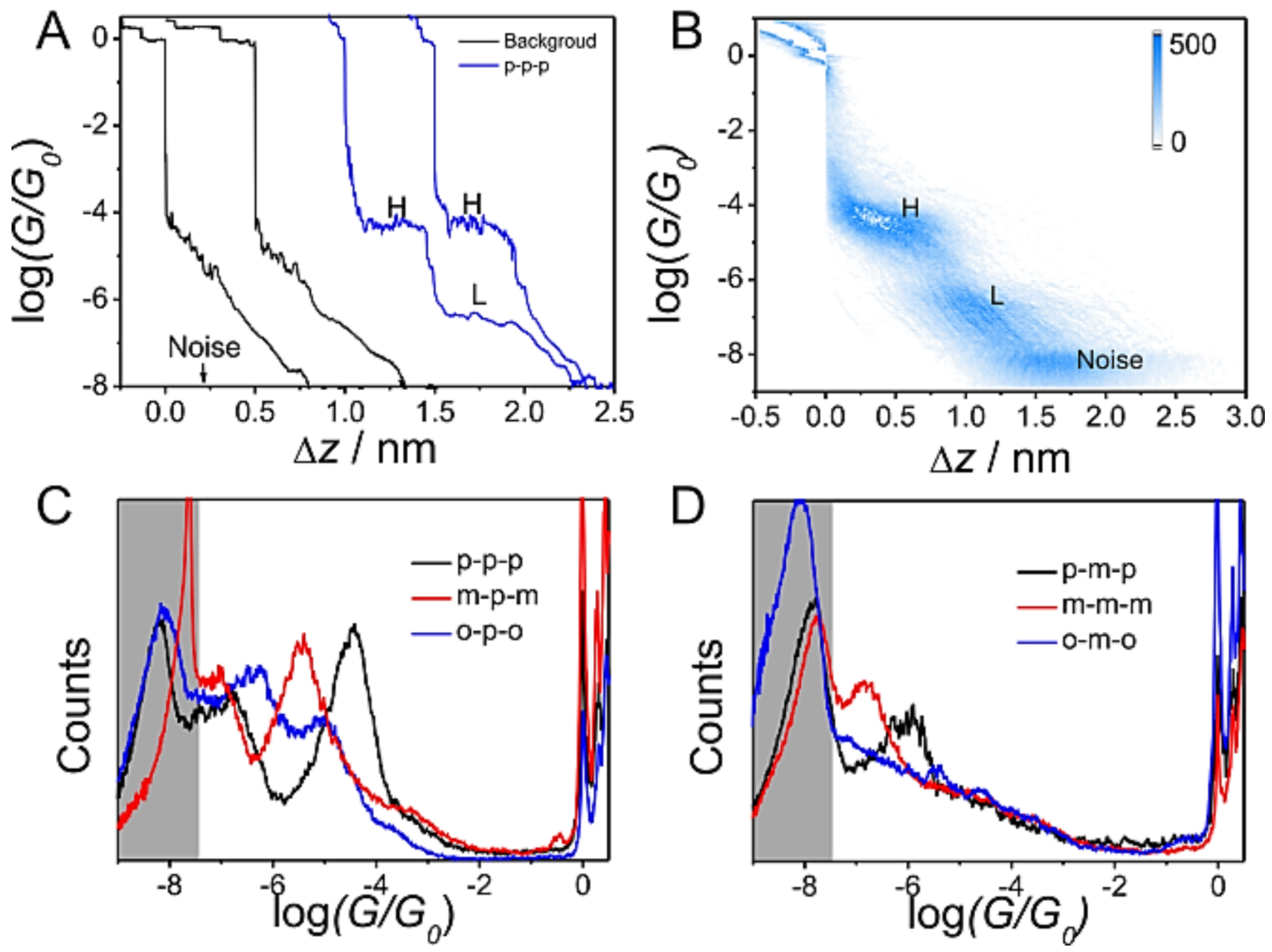

*Figure 4.6: (A) Typical individual conductance–distance traces of p-p-p (blue) and pure tunnelling traces (black). (B) All-data-point 2D conductance versus relative distance (ΔZ) of p-p-p. (C, D) All-data-point 1D conductance histograms constructed from 1000 MCBJ traces of molecules in group 1 (C) and in group 2 (D). The grey area is the noise level.*

The latter is attributed to the formation of single-molecule junctions. These clouds of conductance data lead to peaks in the corresponding one dimensional (1D) conductance histogram. The cloud-like pattern is observed in both MCBJ and STM-BJ measurements and the 1D histogram peaks are in good agreement with each other. Figure 4.6c,d displays the corresponding 1D conductance histograms of molecules belonging to groups 1 and 2 in a semi-logarithmic scale, constructed from 1000 experimental conductance–distance traces for each compound.

The sharp peaks ~ $G_0$ represent the conductance of a single-atom gold–gold contact. The prominent peaks between $10^{-7}G < G < 10^{-4}G_0$ represent molecular conductance features. The significant difference of single-molecule conductances is observed from the variety of connectivities of the central ring (figure 4.6c) and locations of the nitrogen in the anchor units (figure 4.6d), while para connection in both central and terminal rings shows the highest conductance in both cases. The statistically most-probable conductance values were obtained by fitting Gaussians to the maxima in the conductance histograms. The key results are summarized in table 4.1.

*Table 4.1: Most-probable experimental conductance, electrode separation $Z_H^*$ at the end of the high-conductance plateaus and junction formation probability (JFP) of pyridyl terminated OPE derivatives from MCBJ and STM-BJ. The origin of differences between MCBJ and STM-BJ results could be found in ref. [37]. Error bars were determined from the standard derivation in Gaussian fitting of conductance and $\Delta Z_H^*$ distribution. Comparison between theoretical lengths and most-probable end-of-plateau experimental electrode separations $Z_H^*$. The electrode separation $Z_H^*$ is closer to the theoretical N---N distance $L_{NN}$ than to the theoretical molecular length L.*

| Molecule | Conductance (Log (G/G0)) | | | JFP (%) | | $z_H^* = \Delta z_H^* + \Delta z_{corr}$ (nm) | | Theoretical Lengths | |
|---|---|---|---|---|---|---|---|---|---|
| | MCBJ | | STM-BJ | MCBJ | STM-BJ | MCBJ | STM-BJ | L (nm) | $L_{NN}$ (nm) |
| | High | Low | | | | | | | |
| p-p-p | -4.5±0.4 | -7.0±0.7 | -4.5±0.4 | 100 | 100 | 1.58±0.21 | 1.80±0.30 | 1.66 | 1.66 |
| m-p-m | -5.5±0.4 | -7.1±0.7 | -5.5±0.5 | 100 | 100 | 1.50±0.16 | 1.45±0.17 | 1.65 | 1.53 |
| o-p-o | -5.0±0.3 | -6.3±0.7 | -4.5±0.4 | 21 | 27 | 1.22±0.25 | 1.14±0.11 | 1.65 | 1.25 |
| p-m-p | -6.0±0.5 | - | -5.8±0.2 | 100 | 100 | 1.37±0.21 | 1.31±0.18 | 1.38 | 1.35 |
| m-m-m | -6.9±0.5 | - | < -6 | 100 | - | 1.36±0.14 | - | 1.40 | 1.40 |
| o-m-o | - | - | - | - | - | - | - | 1.40 | 1.17 |

In anticipation of the theoretical discussion below, it is interesting to note that to within experimental error, $\log(G_{ppp}/G_0)$ + $\log(G_{mmm}/G_0$; (that is, -4.5 to 6.9) is equal to $\log(G_{pmp}/G_0)$ + $\log(G_{mpm}/G_0$; (that is, -5.5 to 6.0), which suggests that the product of the conductances of p-p-p and m-m-m molecules is equal to product the conductances of p-m-p and m-p-m molecules, and the quantum circuit rule $G_{ppp}/G_{pmp} = G_{mpm}/G_{mmm}$ is satisfied.

Further statistical analysis of conductance versus ΔZ curves provides information about the junction formation probability (JFP) and allows us to determine the most-probable relative electrode displacement ($\Delta Z_H^*$) at the end of the high-conductance plateaus. For every $\log(G/G_0)$ versus ΔZ stretching trace, the relative electrode displacement at the end of the high-conductance plateau has been determined, $\Delta Z_H$, which is the largest ΔZ value within the range $-0.3 > \log(G/G_0) > \log(G_H^{end}/G_0)$, where $G_H^{end}$ is the end of a high-conductance feature. The most-probable values of $\Delta Z_H$ (denoted $\Delta Z_H^*$) are obtained by constructing a histogram and fitting a Gaussian function to the largest maxima. Taking into account the snap-back length, the most-probable electrode separations at the end of the high-conductance plateau are $Z_H^* = \Delta Z_H^* + \Delta Z_{corr}$. The representative $\Delta Z_H$ histograms with Gaussian fitting functions are shown in figure 4.7.

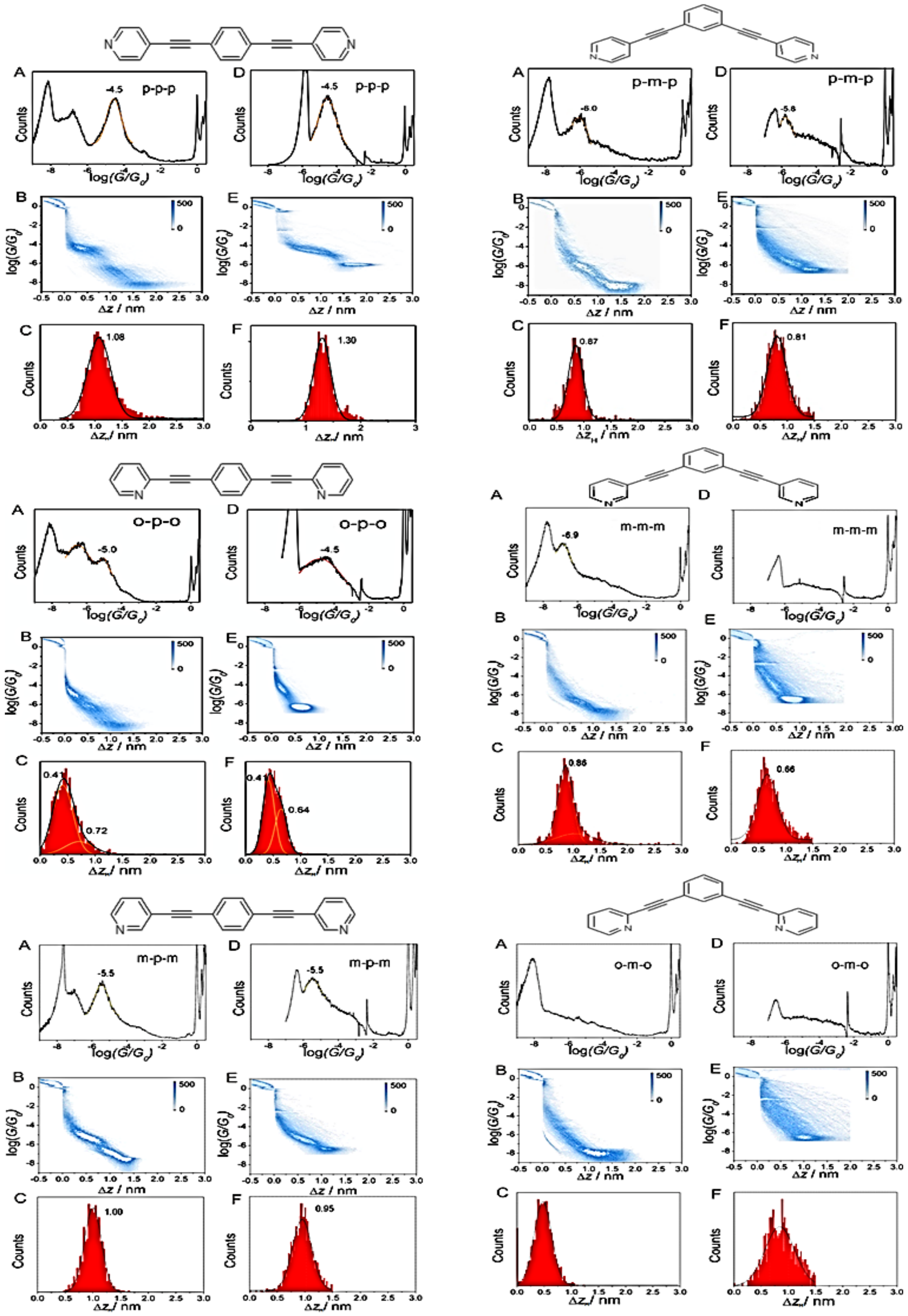

*Figure 4.7: 1D conductance histograms (A, D), 2D histograms (B, E), and distributions of the 'end-of-high-conductance-plateau displacements' $\Delta Z_H$ (C, F). The peaks of the latter are identified with $\Delta Z_H^*$. Columns (A-C) were obtained using MCBJ and (D-F) using STM-BJ.*

The JFP is calculated as the ratio of the area under the fitted Gaussian function and the total area of the $\Delta Z_H$ histogram. If no distinct peak is observed in the $\Delta Z_H$ histogram, then the JFP is considered to be zero. Table 4.1 also summaries the various distances obtained from both MCBJ and STM-BJ measurements. As shown in table 4.1, the JFP approaches 100% for molecules p-p-p, p-m-p, m-p-m and m-m-m. For molecule o-p-o, the JFP decreased sharply to 21% (MCBJ)/27% (STM-BJ) because the N in the terminal ortho pyridyl is partially hidden from the electrode surfaces and therefore it is difficult to form a bridge between the two gold electrodes [45]. The most-probable end-of-plateau electrode separations $Z_H^*$ follow the trends p-p-p > m-p-m > o-p-o, and p-m-p ≈ m-m-m that correlate with the molecular N…N distance, demonstrating that the gold-anchor link is primarily controlled by the gold–nitrogen bonds. Therefore, it is clear that changes in the position of the N atom within the anchors affects both the plateau length and the JFP, as well as the conductance.

According to the quantum circuit rule, the conductance of the o-m-o molecule is expected to be $10^{-6.5}G_0$, which is within the sensitivity limit ($\sim 10^{-8}G_0$) of MCBJ set up. However, as shown in figure 4.7, there is no ability to determine the molecular conductance feature of o-m-o molecular junction. To confirm this experimental finding, the conductance measurements have been repeated more than ten times and the same results have been obtained. The further analysis reveals that the absence of conductance feature in the break junction measurements is due to the combined effect of the relative short N…N length and low conductance, which resulted to the phenomenon that the direct tunnelling conductance between the two gold electrodes dominated on the single-molecule conductance.

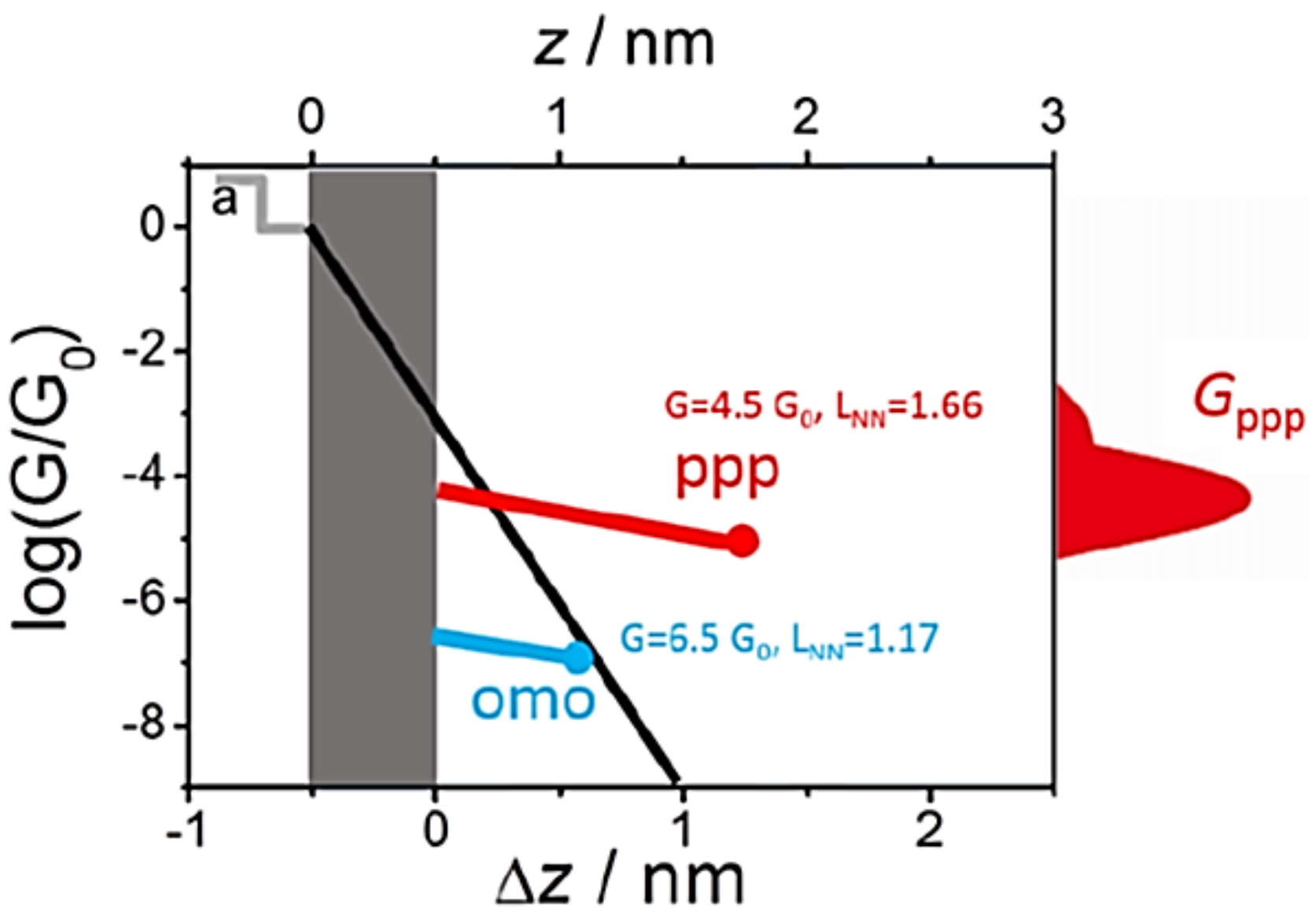

*Figure 4.8: A simple schematic to describe the parallel tunnelling pathway. ΔZ in the bottom axis describes the relative distance scale from the break junction measurement and Z in the upper axis describes the absolute distance scale between the two gold electrodes. The black line describes the direct tunnelling between gold electrodes in the absence of the molecule. The red and blue lines describe the expected conductance plateaus of p-p-p and o-m-o molecular junctions respectively. Parameters are estimated from table 4.1 and figure 4.7.*

Figure 4.8 shows a simple schematic of typical G – ΔZ traces in break junction. After breaking of the gold-gold contact (shown as step "a" at $G_0$), the tunnelling conductance decreases exponentially as the black line and the slope was estimated from the experimental data. The grey area indicates to the tunnelling conductance changes within the snapback area, which cannot be monitored from the break junction measurements. When the molecule is trapped between two gold electrodes, the tunnelling conductance through the molecular junction could be higher than that through the solvent.

For instance, in case of p-p-p molecules there is an ability to determine a clear molecular conductance plateau at around $10^{-4.5}G_0$ (red curve, the parameter is estimated from data shown in table 4.1), which contributed to a clear conductance peak in the conductance histogram after statistical analysis. In contrast, for o-m-o, the direct tunnelling conductance with the associated N…N distance of the o-m-o molecule (which is calculated to be 1.17 nm as shown in table 4.1) is ~$10^{-6}G_0$, which is slightly higher than the expected conductance of the o-m-o single-molecule junction ($10^{-6.5}G_0$). In this case, the expected molecular conductance plateau (blue line) would be completely dominated by direct tunnelling conductance, to the extent that the conductance peak cannot appear in the conductance histogram. A similar result was also observed and discussed in ref. [46].

## 4.3.2. Theoretical Results

### 4.3.2.1. Density Functional Theory (DFT) Results

To gain a deeper insight into the electronic characteristics of these compounds and the electrical behaviour of the junctions, the DFT based methods described in chapter 2 have been used. Before calculating the transport properties, the gas-phase electronic structures of all molecules were investigated to explore the distribution and composition of the frontier molecular orbitals. Plots of the HOMOs and LUMOs are given in figure 4.9. This figure shows that the HOMOs and LUMOs are extended across the backbone for each molecule as expected from previous studies of OPE-derivatives [47 – 49, 50].

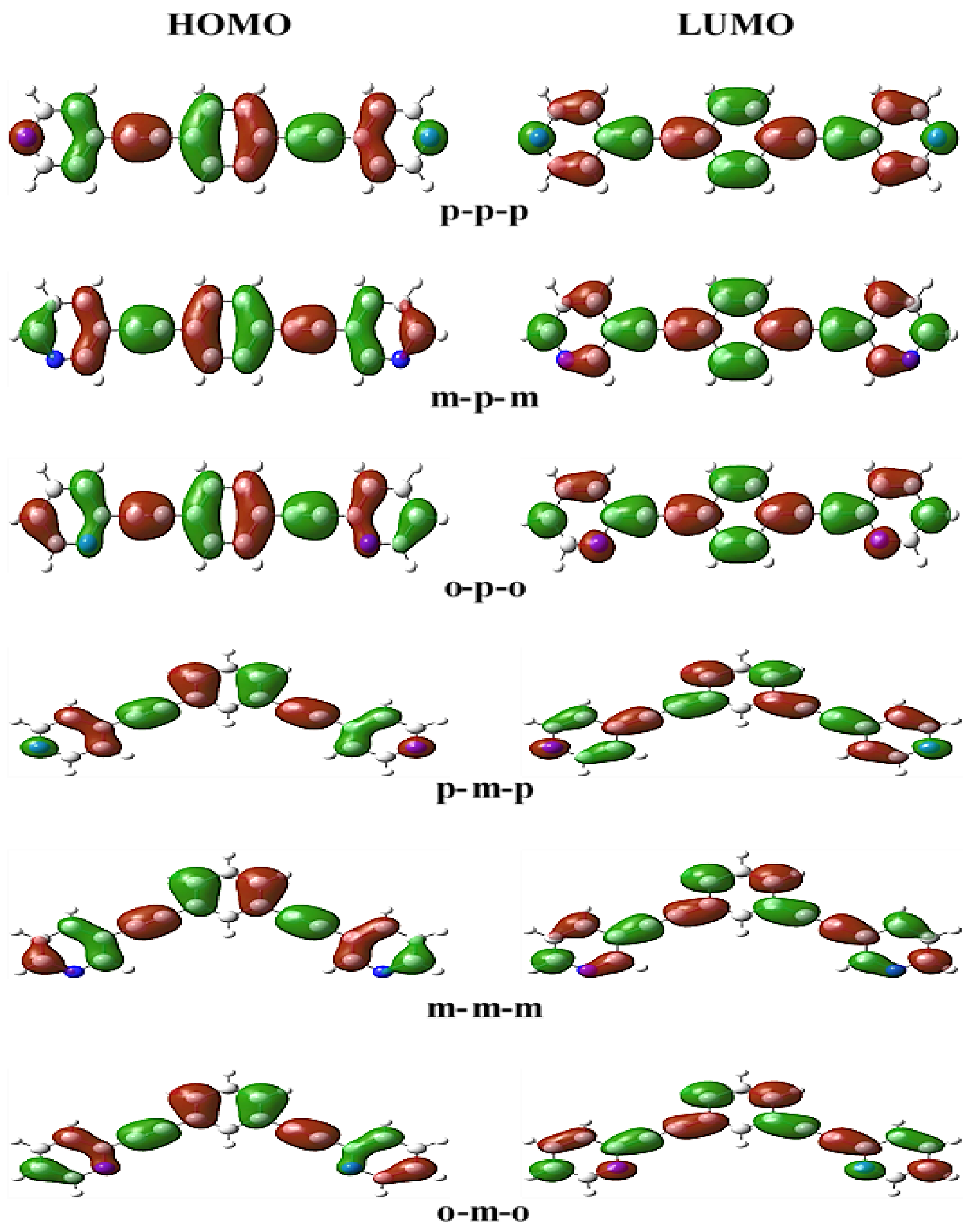

*Figure 4.9: The relaxed geometries and iso-surfaces of the HOMOs and LUMOs for all molecules.*

To elucidate the measured conductance trends, the density functional theory (DFT)-based optimization has been performed for each molecule using SIESTA and then used to carry out electron transport calculations.

Unlike the p-p-p molecule, the other molecules possess anchoring nitrogen atoms located at meta or ortho positions within the terminal rings that do not naturally bind to planar electrode surfaces.

To study the transport properties of these molecules, the anchor nitrogen atoms were attached to the apexes of pyramidal gold electrodes, as shown in figures 4.10 and 4.2. The molecules shown in these figures have been geometrically relaxed using SIESTA and it is clear that many of these geometries, such as the ma-m-ma conformation in the first column of figure 4.2 would not tend to form a bridge between two opposing pyramids. It could be noted also that molecular lengths (L), defined as the distance between the centres of the furthest non-hydrogen atoms in the molecule, is the same distance for these different conformations, whereas the N…N distance, which is the distance between the centres of the N atoms, does vary. Therefore, in table 4.1 the largest N…N distance ($L_{NN}$) of each molecule is calculated and compared with the most-probable electrode separation at the end of the high-conductance plateau, that is, $Z_H^*$. Figure 4.10 demonstrates that for a wide range of Fermi energy choice, the theoretical and experimental conductances of the X-p-X molecules of group 1 are distinctly higher than those of the molecules in group 2 (figure 4.1), as expected from previous studies [27, 28, 32,]. Figure 4.10 also shows that for a wide range of energies in the gap between the highest occupied molecular orbital (HOMO) and lowest unoccupied molecular orbital (LUMO), the ordering of the transmission coefficients follows the experimental conductance ordering. To demonstrate the resilience of the circuit rule, dotted lines are plots of (log $T_{mmm}$ + log $T_{ppp}$)/2 and (log $T_{mpm}$ + log $T_{pmp}$)/2. The similarity of these two curves shows that the product rule is satisfied over a wide range of energies within the HOMO-LUMO gap.

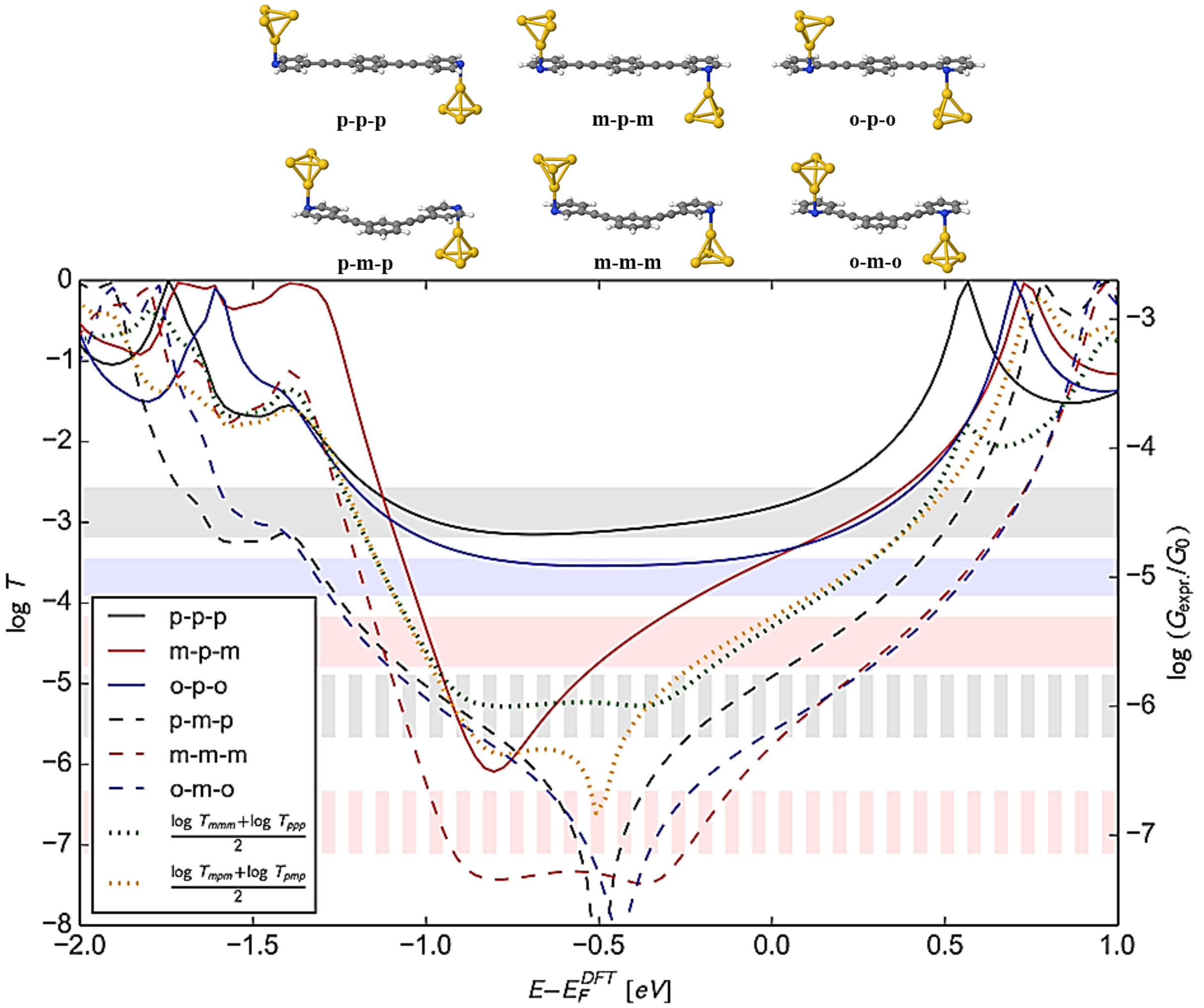

*Figure 4.10: Systematic junction geometries and transmission coefficient functions. The top figures show idealized junction geometries for all six molecules with the gold tip attached perpendicular to the pyridine ring. The bottom graphs show corresponding transmission coefficient curves. For comparison, the coloured and patterned horizontal shaded bars show the experimentally measured MCBJ values of log ($G/G_0$) (right-hand scale). The thickness of the bars corresponds to the error bar of the experimental log ($G/G_0$). These show that both the transmission coefficients around the $E_F$ and the experimental conductance values are separable into two groups corresponding to X-p-X and X-m-X connectivities. The green (yellow) dotted line is the average of log $T_{mmm}$ and log $T_{ppp}$ (log $T_{mpm}$ and log $T_{pmp}$). It should be noted that over most of the energy range from -0.8 to -0.4 eV, the m-m-m results have the lowest transmission coefficient. At these low transmissions, the contribution from the σ channel becomes comparable with that of the π channel and therefore in this energy range, the circuit rule is violated.*

To demonstrate that QI effects associated with variations in the positions of the N atoms are suppressed owing to the presence of a parallel conductance path associated with the electrode-ring overlap, the transport calculations have been performed, in which every Hamiltonian and overlap matrix element that couples carbon and hydrogen atoms to gold atoms are artificially set to zero, leaving couplings between the nitrogen atoms and gold as the only possible transport path. The resulting transmission coefficients are shown as dashed lines in figure 4.11 and demonstrate that without metal-ring interactions the meta link in the terminal ring of the m-p-m molecule reduces the conductance by orders of magnitude, which is comparable with the effect of a meta link in the central ring, whereas in the presence of metal-ring interactions meta coupling in the terminal rings has a much smaller effect.

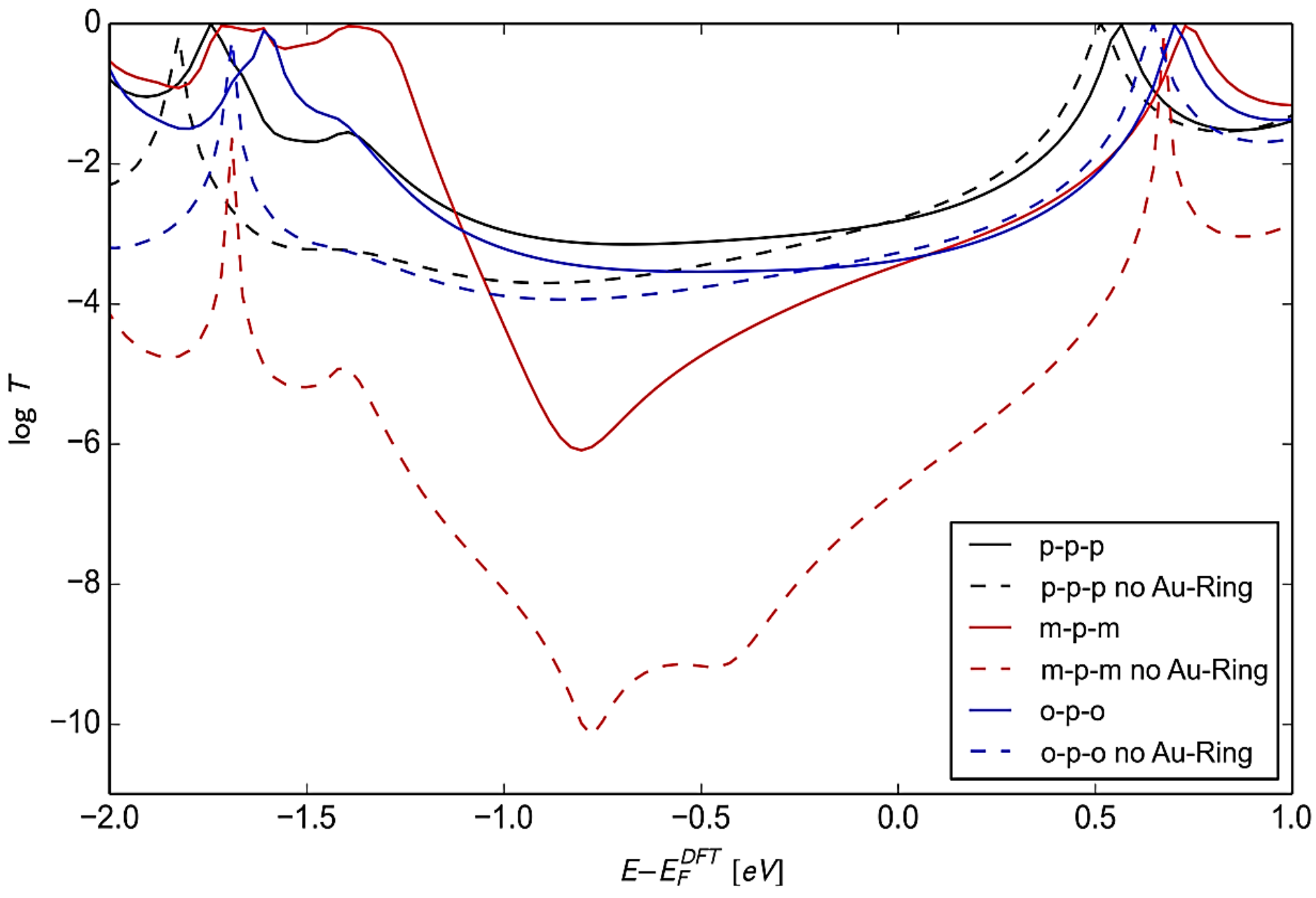

*Figure 4.11: Transmission curves with and without metal-ring coupling. The corresponding junction geometries are shown in figure 4.10. Dashed lines are without metal-ring coupling, the continuous curves are with metal-ring coupling.*

The central focus of this section is to understand the relative contribution to the conductance from QI in the terminal and central aromatic rings, and to demonstrate that the quantum circuit rule is satisfied at the level of DFT. The largest value of $Z_H^*$ among all the measured molecules (table 4.1) was found to be $Z_H^*$ = 1.80 nm for the p-p-p. The value of $Z_H^*$ for other molecules was typically shorter than this value. Therefore, the typical conductances occur when the molecules are not fully stretched and there is a possibility of metal-ring overlap between the gold electrode and the pyridyl rings. To analyse such events, the junction geometries with two 35-atom gold (111)-directed pyramids have been systematically constructed and attached to the N atoms perpendicular to the pyridyl rings, as shown in figure 4.10, and then computed the transmission coefficient curves *T(E)* for these geometries. Alternative junction geometries are investigated as shown in figures 4.3 and 4.4.

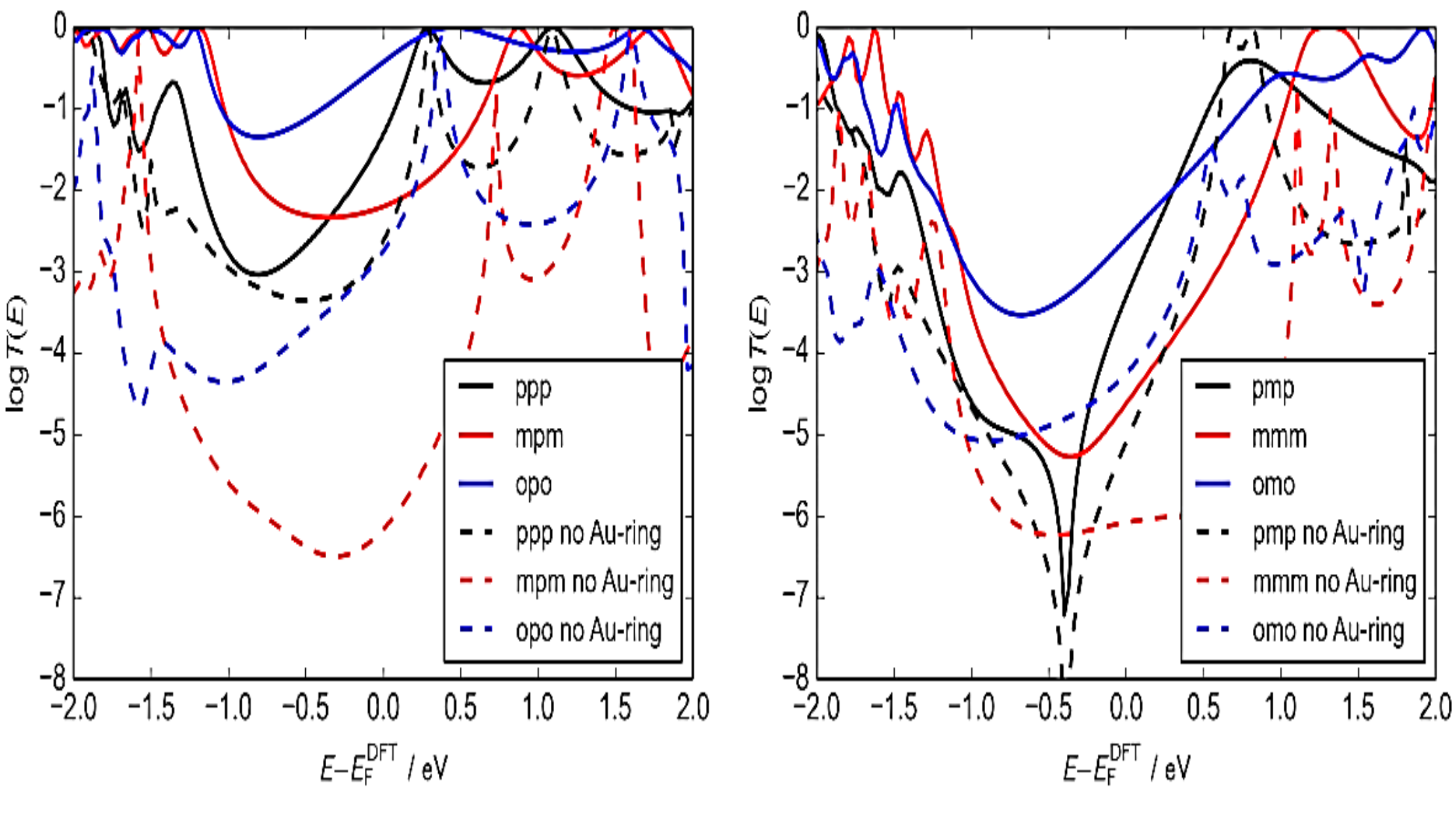

*Figure 4.12: Transmission coefficient functions for junction geometries shown at most left hand side in figures 4.3 and 4.4 for molecules in group 1 (left pane) and 2 (right pane), with ring-gold (continuous) and without ring-gold (dashed) coupling in the Hamiltonian matrix.*

Figure 4.12 shows a comparison between the transmission coefficients of the systems with and without metal-ring interactions. Removing the ring-gold coupling generally reduces the conductance as the dominant resonances become narrower. The primary signature of destructive interference in the cases of meta terminal and central rings is the low value of the conductance. The anti-resonance in the $\pi$ channel is not always evident, because it may be masked by parallel sigma-like channels [51, 52].

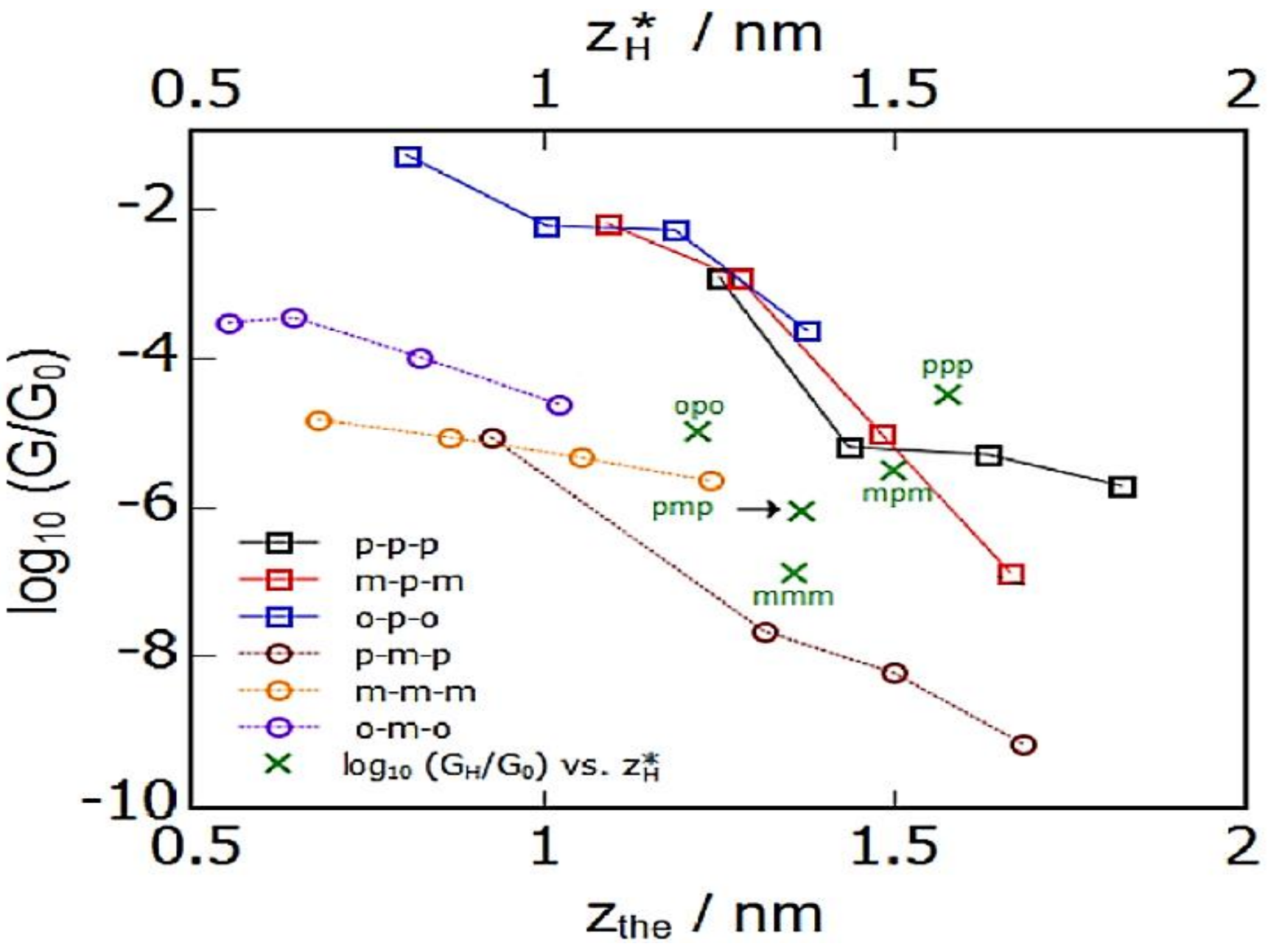

*Figure 4.13: Computed conductance vs electrode separation. Simulated trace curves for the first and second groups of molecules, marked with rectangle and circle markers, respectively. The figure shows the logarithm of conductance as a function of the theoretical electrode separation $Z_{the}$, defined as $Z_{the}=Z_{Au\text{-}Au}-0.25$ nm, where $Z_{Au\text{-}Au}$ is the tip separation in the relaxed structure, and 0.25 nm is the value of $Z_{Au\text{-}Au}$ when the conductance through the two contacting pyramids (in the absence of a molecule) is $G_0$. Beyond the last points of the simulated trace curves, the simulated junction (i.e. the Au-N bond) is broken. Junction configurations for each stage during the stretching can be seen in figures 4.3 and 4.4. The green crosses are the experimental most-probable high conductance values plotted against the experimental values of $Z_H^*$. The simulated trace curves and the experimental conductance values demonstrate a distinct separation between groups 1 and 2 (see figure 4.1).*

Figure 4.13 shows that for any given value of $Z_{the}$ or $Z_H^*$, both the theoretical and experimental conductances of the molecules in group 1 are distinctly higher than those of the molecules in group 2. Figure 4.13 also shows that for many values of $Z_{the}$ (in the electrode separation range 1.0 nm < $Z_{the}$ < 1.4 nm) the theoretical conductances of molecules within group 1 are rather similar and therefore it could be concluded that QI in the central ring is more important than QI in the anchors. For molecules in group 2, the theoretical conductance for m-m-m is larger than for p-m-p, which is in the reverse order of the experimental most probable high conductance values. The latter artefact is attributed to unrealistically-large metal-ring coupling in the m-m-m simulated optimized junction geometries that will rarely occur in room temperature environments.

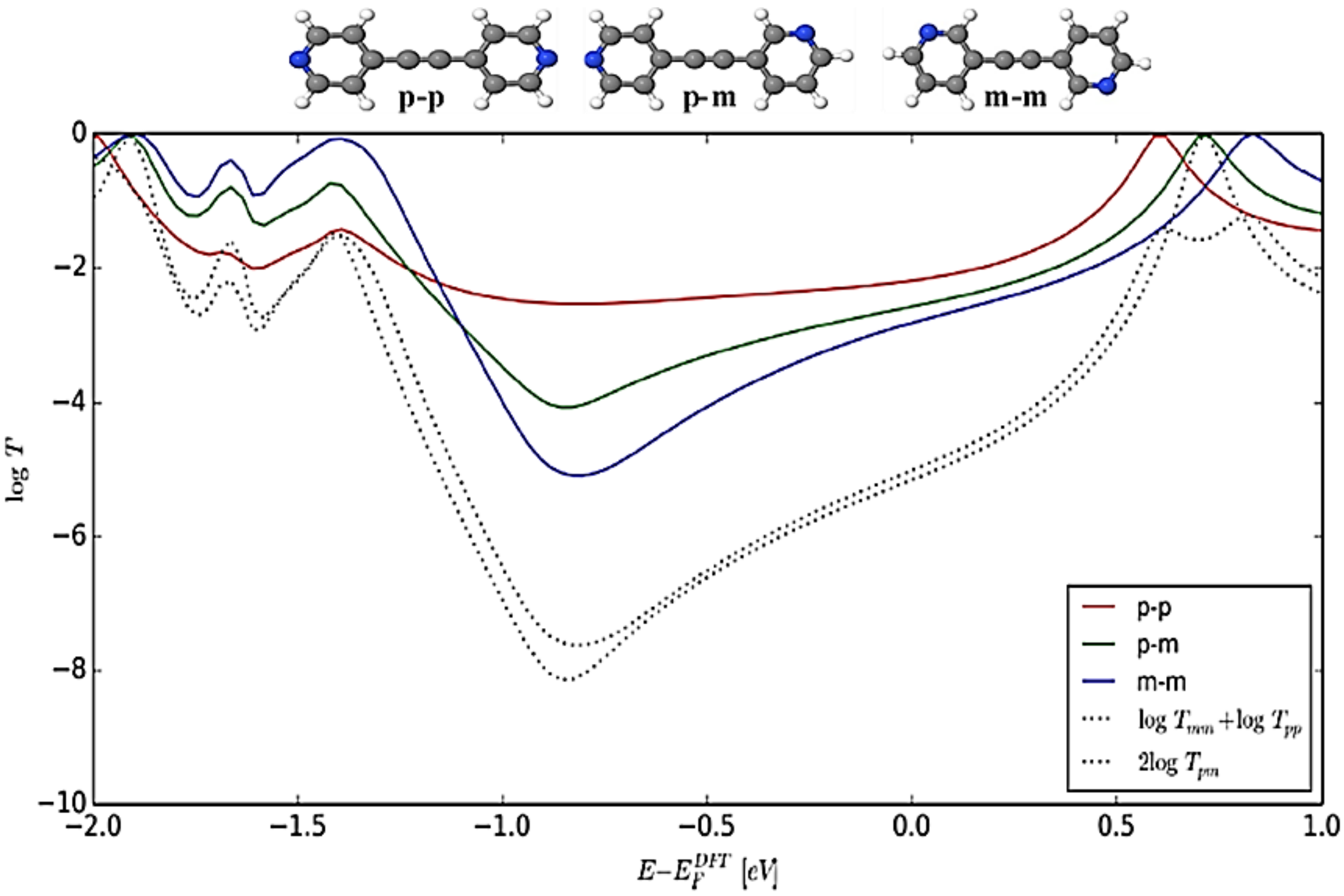

*Figure 4.14: Systematic gas-phase geometries and transmission coefficient functions. The top figures show idealized geometries for all three molecules. The bottom graphs show corresponding transmission coefficient curves.*

To further demonstrate the generality of the product rule the DFT based transport calculations have been performed (with the exact same methodology as gave the result in figure 4.10) for two pyridyl ring systems with para and meta connections. The structures of three molecules studied are shown in figure 4.14.

The theoretical derivation shown later implies that for two rings $G_{pp}G_{mm} = G_{pm}^2$, Where the $G_{pp}$ and $G_{mm}$ are the conductances of para-para and meta-meta pyridyl rings, and $G_{mp}$ is the conductance of the molecule for meta and para pyridyl rings. Figure 4.14 demonstrates that the product rule is satisfied for a wide range of energies (the dotted curves match).

#### 4.3.2.1.1. Phase averaging in an ensemble of measurements

Since the experimental conductance values are of statistical origin, a product rule for ensemble averaged conductances can arise due to inter-ring phase averaging. To illustrate this point, consider two quantum scatterers (labelled 1 and 2) in series, whose transmission and reflection coefficients are T1, T2 and R1, R2 respectively. It can be shown that the total transmission coefficient for the scatterers in series is

$T = \frac{T_1T_2}{1-2\sqrt{R_1R_2}cos\varphi+R_1R_2}$ where φ is a quantum phase arising from QI between the scatterers [9, 26].

It could be noted that the experimental quoted conductance is identified with the most probable value of $log_{10}\frac{G}{G_0}$ and if this possesses a Gaussian distribution, then it equates to the ensemble average of $log_{10}\frac{G}{G_0}$ which it is denoted by $\overline{log_{10}T}$ . From the above expression for $T$, $\overline{log_{10}T} = \overline{log_{10}T_1} + \overline{log_{10}T_2} - \overline{log_{10}(1 - 2\sqrt{R_1R_2}\, cos\varphi + R_1R_2}$.

Although each individual molecule is phase coherent, if the ensemble average involves an average over the phase, uniformly distributed between 0 and 2π, then the third term on the right hand side averages to zero, because of the mathematical identity

$\int_0^{2\pi} d\varphi\, \overline{log_{10}\left(1 - 2\sqrt{R_1R_2}\right)cos\varphi + R_1R_2} = 0$. Hence in the ensemble average, all information about the inter-scatterer quantum phase is lost and

$\overline{log_{10}T} = \overline{log_{10}T_1} + \overline{log_{10}T_2}$, or equivalently $\frac{G_{Total}}{G_0} = \frac{G_1}{G_0}\frac{G_2}{G_0}$.

For the three-ring OPEs of interest, if inter-ring QI is similarly absent in the ensemble average, then one would expect $\frac{G_{Tot}}{G_0} = \frac{G_t}{G_0}\frac{G_c}{G_0}\frac{G_t}{G_0}$ where $G_t$ is the conductance of the terminal pyridyl ring and $G_c$ is the conductance of the central phenyl ring. In the absence of inter-ring QI, the above expression implies that the 'quantum circuit rule' is satisfied.

It may seem surprising that phase averaging can lead to decoherence, even though the theory describing transport is fully phase coherent. Decoherence is perfectly reconciled with the calculations in figures 4.10 and 4.11, because although individual electrons remain phase coherent over the whole molecule, a restricted form of decoherence can arise in the ensemble averages of transport properties.

This is important, because all break-junction experiments involve averages over large ensembles in a single measurement and the typical conductance associated with a compound is the most occurred value of $\log(G/G_0)$ for several measurements. For a complete discussion, it has been shown that the possibility of the conductance product rule could arise from not only the intrinsic electronic properties of the individual molecules, but from the statistical origin of the experimental conductance values. Perhaps the discussion can be made clearer through an analogy. In the NMR literature, the phrase 'decoherence' is often used to describe two distinct phenomena, which are assigned T1 and T2 coherence times. The latter does not require inelastic scattering and arises when spins within an ensemble process in the plane perpendicular to the applied magnetic field with slightly different frequencies.

Consequently, even in the absence of inelastic process and even though individual spins remain coherent, the ensemble of spins eventually loses coherence. In contrast, T1 processes involve spin flips. These are inelastic processes and provide a second source of decoherence. In this work, all transport is elastic and individual electrons remain perfectly coherent. Nevertheless, for an ensemble of measurements on such electrons, decoherence in average values can arise from random elastic scattering, just as decoherence in NMR experiments arises from T2 processes.

#### 4.3.2.2. Derivation of the quantum circuit rule for molecular conductance

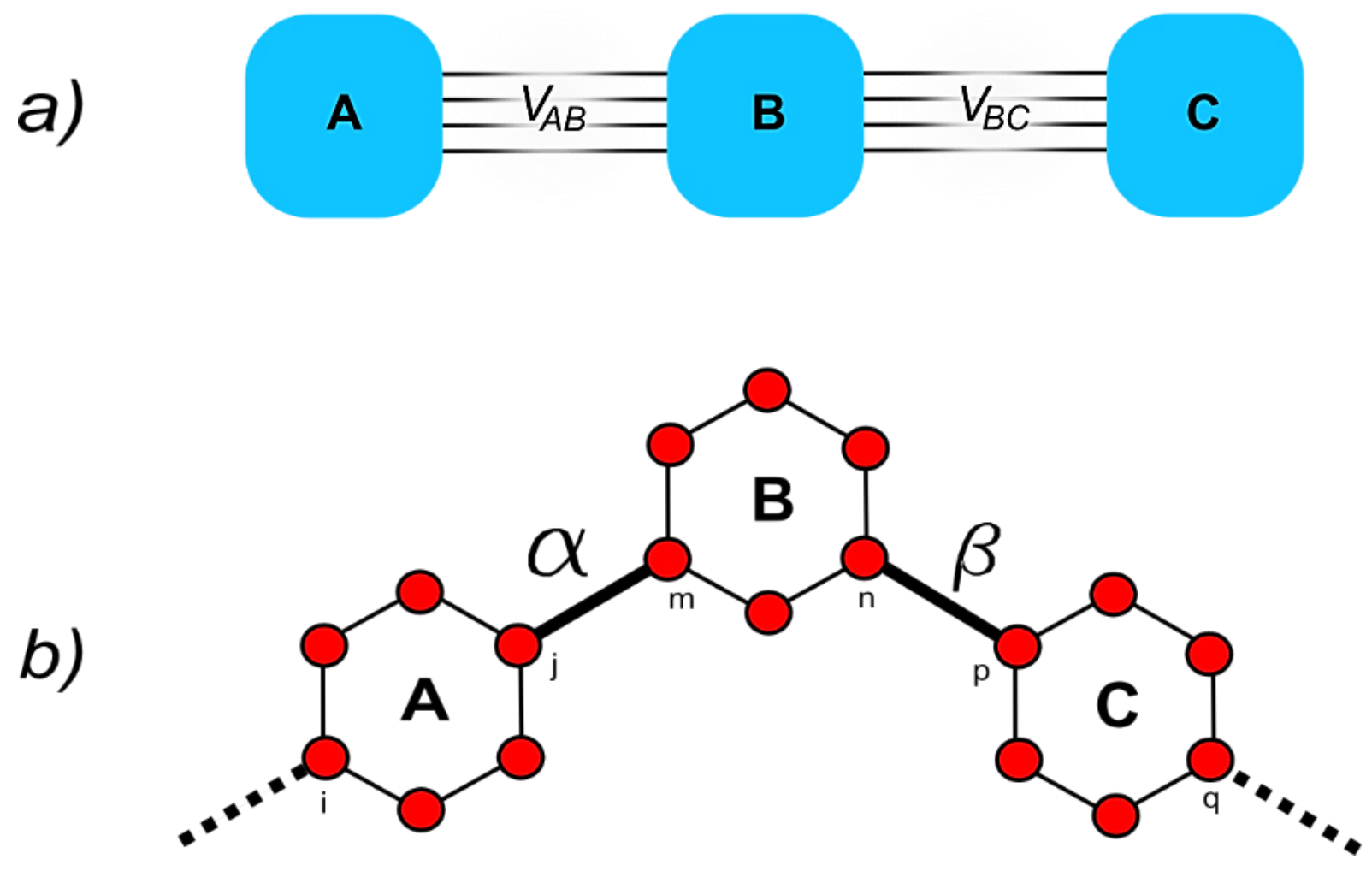

*Figure 4.15: Abstract models. (a) An abstract depiction of three coupled quantum subsystems. (b) A simple tight-binding model of the three ring system. (In the figure p-m-p connection is shown to illustrate the connecting sites i, j, m, n, p and q).*

Having demonstrated that the effect of QI in a given ring depends on the position of the ring within the molecule (that is, whether it is a terminal or a central ring).

The following quantum circuit rule describing the conductances of different ring combinations:

$$G_{ppp} \times G_{mmm} = G_{pmp} \times G_{mpm} \tag{4.1}$$

This captures the property that the contribution to the conductance from the ring Y in molecules of type p-Y-p and m-Y-m are identical.

To derive the circuit rule, consider the simple tight-binding model of figure 4.15b, in which three rings (labelled A, B and C) are coupled by nearest neighbour couplings α and β. To calculate the transmission coefficient arising when 1D electrodes are connected to site $i$ of ring A and $q$ of ring C, the Green's function $G_{qi}$ has been calculated of the whole structure in the absence of the electrodes. This structure is an example of three arbitrary-coupled quantum objects shown schematically in figure 4.15a, whose Green's function is given by

$$\begin{bmatrix} E-H_A & -V_{AB} & 0 \\ -V_{AB}^{\dagger} & E-H_B & -V_{BC} \\ 0 & -V_{BC}^{\dagger} & E-H_C \end{bmatrix} \begin{bmatrix} G_{AA} & G_{AB} & G_{AC} \\ G_{BA} & G_{BB} & G_{BC} \\ G_{CA} & G_{CB} & G_{CC} \end{bmatrix} = I \tag{4.2}$$

Electron propagation from subsystem A to subsystem C is described by the block $G_{CA}$, for which the above equation yields

$$G_{CA} = g_C V_C^{\dagger} G_{BB} V_A^{\dagger} g_A \tag{4.3}$$

Here,

$g_X = (E - H_X)\ (X = A\ or\ C)$ and $G_{BB} = \left(E - H_B - V_{ABgA}^{\dagger} V_{AB} - V_{BCgC} V_{BC}^{\dagger}\right)^{-1}$

As an example, for the tight-binding model of figure 4.15b, this becomes

$$G_{qi} = (g_C)_{qp} B_{nm} (g_A)_{ji} \tag{4.4}$$

Here, $B_{nm} = \beta \frac{(g_B)_{nm}}{(1-\sigma)} \alpha,$ and $\sigma = \alpha^2 (g_A)_{jj} (g_B)_{mm} + \beta^2 (g_C)_{pp} (g_B)_{nn} + \alpha^2 \beta^2 (g_A)_{jj} (g_C)_{pp} [(g_B)_{nm} (g_B)_{mn} - (g_B)_{nn} (g_B)_{mm}]$.

The crucial point is that $B_{nm}$ does not depend on the choice of *p*, *q*, *i* and *j*, due to the rotational symmetry, $(g_A)_{jj}$ and $(g_C)_{pp}$ are independent of *j* and *p* for any choice of *j* and *p*. This means that $B_{nm}$ does not depend on the choice of meta, ortho or para coupling for rings A and C and only depends on the connectivity of ring B. When the rings A and C are coupled to the electrodes, this rotational symmetry is broken slightly. However, provided $E_F$ lies in the HOMO-LUMO gap of A and C and the coupling to the electrodes is small (as is usually the case for molecules attached to gold electrodes, where the level broadening is much less than the level spacing), then the symmetry breaking is weak [26]. Since the electrical conductance is proportional to $|G_{qi}|^2$, this means that the electrical conductance $G_{XYZ}/G_0$ of molecules of the type X-Y-Z is proportional to a product of the form $a_X \times b_Y \times a_Z$ and hence the quantum circuit rule is satisfied. The factors $a_X$, $b_Y$, $a_Z$ are contributions to the overall conductance from the separate rings, but it should be noted that they are not themselves individually measureable conductances. The above analysis can be easily applied to a molecule with two rings yield the circuit rule $G_{pp}G_{mm} = G_{pm}^2$, where the $G_{pp}$ and $G_{mm}$ are the conductances of the molecule with para-para and meta-meta connected rings, and $G_{pm}$ is the conductance of the molecule with meta and para rings (see figure 4.14).

To check the experimental validity of the circuit rule for the OPEs of figure 4.1, the measured conductance values presented in table 4.1 have been examined. For example, rearranging equation (1) into the form $G_{pmp} = \frac{G_{ppp}}{G_{mpm}} G_{mmm}$ and substituting into the right-hand side, the measured values for $G_{ppp}$, $G_{mmm}$ and $G_{mpm}$ yields $G_{pmp}=10^{-5.9}G_0$, which compares well with the measured value of $G_{pmp}=10^{-5.8}G_0$ shown in table 4.1.

## 4.4. Summary

In this chapter charge transport studies of pyridyl terminated OPE derivatives have been studied, using the MCBJ and STM-BJ techniques, DFT-based theory and analytic Green's functions, and have investigated the interplay between QI effects associated with central and terminal rings in molecules of the type X-Y-X.

The results demonstrated that the contribution to the conductance from the central ring is independent of the para or meta nature of the anchor groups and the combined conductances satisfy the quantum circuit rule $G_{ppp}/G_{pmp} = G_{mpm}/G_{mmm}$. For the simpler case of a two-ring molecule, the circuit rule $G_{pp}G_{mm} = G_{pm}^2$ is satisfied (see figure 4.14). It should be noted that the circuit rule does not imply that the conductance $G_{XYX}$ is a product of three measureable conductances associated with rings X, Y and X. Indeed, the latter property does not hold for a single molecule. On the other hand, provided sample to sample fluctuations lead to a broad distribution of phases within an ensemble of measurements, a product rule for ensemble averages of conductances can arise.

The qualitative relationship between the conductances agrees well with the simple QI picture of molecular conduction. It has been reported that destructive QI exists in benzene with the meta connectivity and is responsible for the observed reduction of conductance [16, 20, 28], whereas for para and ortho connectivities, constructive QI should be observed [44 – 45, 53 – 56].

The transmission coefficient calculations through junctions where the metal-ring connection is artificially blocked (figure 4.11) show that the artificially coupled pyridyl ring exhibits similar behaviour to the benzene ring, with destructive QI in the case of the meta coupling significantly reducing the conductance compared with para and ortho connectivities. The dashed curves in the bottom panel in figure 4.11 clearly demonstrate that when the conduction is through only the nitrogen atoms, the conductance of the meta isomer is much lower than in the para and ortho isomers. More realistically, however, in the presence of metal-ring overlap, the effect of varying the positions of the nitrogens in the anchors becomes much weaker, and as demonstrated by figure 4.10, the major changes in the molecular conductance are caused by the variations in the connectivity of the central ring. The dominant influence of the central ring is accounted for by the fact that the central ring is not in direct contact with electrodes and therefore no parallel conductance paths are present, which could bypass the ethynylene connections to the anchors.

In a sub nanometre scale molecular circuit, as in standard complementary metal-oxide-semiconductor (CMOS) circuitry, electrical insulation is of crucial importance. Destructive interference in a two-terminal device may not be desirable, because of the lower conductance. However, for a three-terminal device minimizing the conductance of the third terminal is highly desirable, because the third (gate) electrode should be placed as close to the molecule as possible, but at the same time, there should be no leakage current between the molecule and gate. One way of achieving this may be to use an anchor group with built-in destructive interference. Therefore, destructive QI may be a vital ingredient in the design of future three-terminal molecular devices and more complicated networks of interference-controlled molecular units.

# Chapter 5

# Solvent Dependence of the Single Molecule Conductance of Oligoyne-Based Molecular Wires

## 5.1. Introduction

Oligoynes hold particular interest in molecular electronics as the ultimate one-dimensional molecular wires formed from simple linear strings of sp-hybridized carbon atoms [1 – 6]. In contrast to an infinite one-dimensional carbon string, oligoynes are terminated at either end by protons (H) or organic, inorganic, or organometallic moieties and can be represented by general formula R – $(C \equiv C)_n$ – R. The terminating groups, R, can be chosen to aid the formation of metal – molecule – metal junctions, with examples including pyridyl, cyano, dihydrobenzo[b]Thienyl, and other anchoring groups [3, 4, 7, 8]. Oligoynes feature extensive electron delocalization along the sp-hybridized carbon backbone, with appreciable bond length alternation evidence in structural studies. The presence of delocalized states makes oligoynes attractive targets for both theoretical and experimental studies of ultrathin wires for molecular electronics. Single molecule wires of oligoynes have been experimentally investigated up to octa-1,3,5,7-tetaynes (n = 4) [3, 4], while theoretical evaluations have been made for hypothetical infinitely long chains as well as finite chains [1]. However, it would be wrong to assume that oligoynes offer rigid-rod linear geometries; on the contrary, they can be considerably curved and flexible offering a surprisingly low bending force [9]. Nevertheless, this structural flexibility does not detract from the impressive electronic properties of oligoynes, which offer considerable interest for their electronic, optoelectronic, and electrical charge transport properties [10].

Despite their apparent chemical simplicity, it is a challenge to synthesize long carbon chains due to the potential instability of $R–(C≡C)_x–H$ intermediates and also instability of longer oligoyne products for certain, particularly small R groups. This issue was circumvented by Bohlmann, who introduced bulky tertiary butyl ($^tBu$) end groups in an n = 7 oligomer ($^tBu – (C≡C)_7 – {}^tBu$) [11], and then also by Johnson and Walton, who extended the chain to $^tBu – (C≡C)_{12} – {}^tBu$ [12]. It has been shown that in addition to a wider range of other bulky organic capping groups, cyano, aryl, or organometallic terminal groups can stabilize oligoyne chains [13]. A recent approach to oligoyne stabilization has been “insulation” of the oligoyne by threading through the cavity of a macrocycle to give rotaxane [14 – 17]. Metal centres have also been widely used to stabilizing end-caps to give complexes with the stoichiometry $[\{L_mM\}\{\mu\text{-}(C≡C)_n\}\{ML_m\}]^{x+}$, which have provided interesting test beds for examination of their electronic, physical, and chemical properties as a function of metal oxidation state [18]. In particular, electronic and vibrational spectroscopies of a complex of this type, coupled with detailed computational investigations, have enabled electronic structures of the all-carbon chain bridged complexes to be assessed [19]. More direct assessment of the electrical properties of oligoynes can be achieved by wiring them into electrical junctions with metal contacts [3, 4, 20].

In particular the influence of the solvents on conductance and β-values has been investigated. The molecular wires chosen here represent a strong choice for experimental investigations of solvent effects since they would be expected to be less affected by the small amounts of water present in even ostensibly well dried organic solvents. In this respect, it is mentioned that they have no functional groups in the backbone to coordinate the water (cf., the gating by water of thiophene-based molecules [21]) and no thiol end groups to be hydrated and thereby adversely affect the conductance (cf., ref [22]).

The experimental measurements of molecular conductance are complemented by density functional theory (DFT) computations of charge transport through the molecular bridge. Using DFT, changes in the conductance and β values in response to the solvent medium are explained by shifts in the Fermi energy of the contact, which impacts both the transmission coefficient of the system and the β-value.

This chapter presents all theoretical details and experimental conductance measurements of novel molecular wires $Me_3Si - (C{\equiv}C)_n - SiMe_3$ (n = 2, 3, 4, or 5; Scheme 5.1), as a part of a published paper. For more details regarding the experimental methods and synthesis details see *Milan David C.; Oday A. Al-Owaedi, Oerthel Marie-Christine, Marques-Gonzalez Santiago, Brooke Richard J.; Bryce Martin R.; Cea, Pilar, Ferrer Jaime, Higgins Simon J.; Lambert Colin J.; Low Paul J.; Manrique David Zsolt, Martin Santiago, Nichols Richard J.; Schwarzacher Walther, Garcia-Suarez Victor M. Solvent Dependence of the Single Molecule Conductance of Oligoyne-Based Molecular Wires. J. Phys. Chem. C, 2016. 120(29): p. 15666–15674.*

# 5.2. Experimental and Theoretical Methods

## 5.2.1 Experimental Methods

The transport characteristics of novel molecular wires Me3Si – (C≡C)n – SiMe3 (n = 2, 3, 4, or 5) (scheme 5.1), were studied by scanning tunnelling microscopy, using the current-distance (*I*(*s*)) technique [24 – 26]. All details of experimental methods are presented in ref. [23].

## 5.2.2. Theoretical Details: Computational Methods

The DFT-Landauer approach used in the modeling assumes that on the time scale taken by an electron to traverse the molecule, inelastic scattering is negligible. This is known to be an accurate assumption for molecules up to several nanometers in length [28]. Geometrically optimization were carried out using the DFT code SIESTA, with a generalized gradient approximation [29, 30] (PBE functional), double ζ polarized basis set, 0.01 eV/A force tolerance, a real-space grid with a plane wave cut-off energy of 250 Ry, zero bias voltage and 1 k points. All molecules in this study were initially geometrically relaxed in isolation to yield the geometries presented in figure 5.6. For each structure, the transmission coefficient, *T(E)*, describing the propagation of electrons of energy *E* from the left to the right electrode was calculated by first obtaining the corresponding Hamiltonian and overlap matrices using SIESTA and then using the GOLLUM code [31] to compute *T(E)* via the relation:

$$T(E) = T_r\{\Gamma_R(E)G^R(E)\Gamma_L(E)G^{R\dagger}(E)\} \quad (5.1)$$

In this expression,

$$\Gamma_{L,R}(E) = i\left(\Sigma_{L,R}(E) - \Sigma_{L,R}^{\dagger}(E)\right) \quad (5.2)$$

$\Gamma_{L,R}$ describes the level broadening due to the coupling between left (L) and right (R) electrodes and the central scattering region, $\Sigma_{L,R}(E)$ are the retarded self-energies associated with this coupling.

$$G^R = (ES - H - \Sigma_L - \Sigma_R)^{-1} \quad (5.3)$$

$G^R$ is the retarded Green's function, where H is the Hamiltonian and S is the overlap matrix (both of them are obtained from SIESTA). Finally the room temperature electrical conductance, G, was computed from the formula:

$$G = G_0 \int_{-\infty}^{\infty} dE T(E)[-df(E)/dE] \quad (5.4)$$

$G_0$ is the quantum of conductance and is given by:

$$G_0 = (2e^2)/h \quad (5.5)$$

$$f(E) = \left[e^{\beta(E-E_F)} + 1\right]^{-1} \quad (5.6)$$

$$\beta = 1/K_BT \quad (5.7)$$

*h* is the Planck's constant, *e* is the electron charge, *f(E)* is the Fermi-Dirac function, $E_F$ is the Fermi energy, $K_B$ is the Boltzmann constant, *T* is the temperature.

Since the quantity *df(E)/dE* is the a probability distribution peaked at $E = E_F$ with a width on the order $K_BT$, the previous expression shows that $G/G_0$ is obtained by averaging $T(E)$ over an energy range on the order $K_BT$ in the vicinity of $E = E_F$. It is well known that the Fermi energy predicted by DFT is not reliable, and therefore, the plots $G/G_0$ have been shown later as a function of $E_F - E_F^{DFT}$. To determine $E_F$, the predicted conductance values of all molecules have been compared with the experimental values and chose a single common value of $E_F$ which gave the closest overall agreement. This yielded a value of $E_F - E_F^{DFT} = -0.725\ eV$, which is used in all theoretical results. This is commonly accepted procedure in molecular electronics DFT-based calculations (cf., ref [32]).

The binding energies between anchor group and gold electrode for the optimized configurations have been calculated. It is well known that SIESTA employs a localized basis set and therefore these calculations are subject to errors. Consequently, the counterpoise method has been used to obtain accurate energies [33]. This involves calculating the total DFT energies for the system ((Molecule plus gold ($E_S$)). The molecule alone ($E_M$), but with the same conformation adopted in the presence of gold electrode and in the absence of the molecule ($E_G$). For these energy calculations, the same basis functions as those generated for the system have been utilized. With these data the binding energies $E_b$ have been calculated according to the expression $E_b = (E_S - E_M - E_G)$ [34, 35].

# 5.3. Results and Discussion

The family of trimethylsilyl (TMS) end-capped oligoynes $Me_3Si – (C≡C)_n – SiMe_3$ (n = 2 – 5) were chosen as the platform upon which to base investigation of molecular conductance and solvent effects in wire-like oligoyne derivatives. These compounds are readily available syntheses of oligoynes $Me_3Si – (C≡C)_n – SiMe_3$ (n = 2 – 5) [24, 38] being well known. The synthesizing details are presented in ref. [23]. In this section the experimental single molecule conductance results and theoretical, such as binding energies between the gold cluster and the TMS terminal groups, the calculated conductance and the decay constant results have been presented.

## 5.3.1. Experimental Results

The single molecule conductance of the oligoynes 2 – 5 (see scheme 5.1) was explored in each of three organic solvents (see figure 5.1) of different polarity, namely, mesitylene (MES), trichlorobenzene (TCB), and propylene carbonate (PC). Mesitylene is a nonpolar solvent with zero dipole moment, and it is commonly used in STM based single molecule electrical measurements because of its high boiling point and relatively low vapor pressure [33 – 35]. TCB is also frequently used in STM based single molecule electrical measurements, and like mesitylene it is a high boiling point and low vapor pressure organic solvent. However, it is a slightly polar solvent with a dipole moment of 1.35 D [36]. By contrast, propylene carbonate is a strongly dipolar solvent with a dipole moment of 4.9 D.

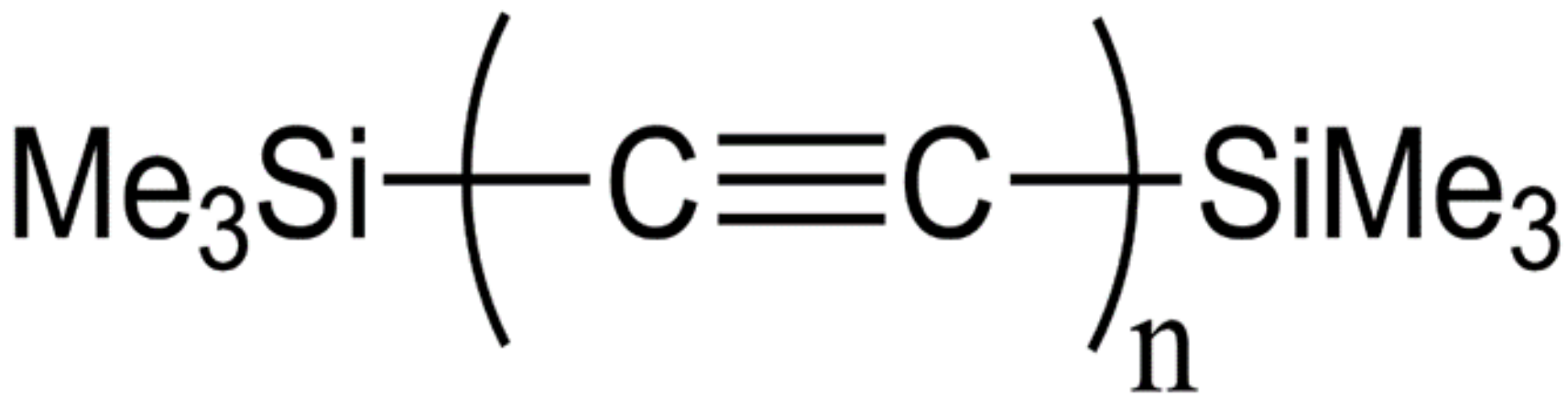

**2** (n = 2) **4** (n = 4)

**3** (n = 3) **5** (n = 5)

*Scheme 5.1: The molecules studied in this work.*

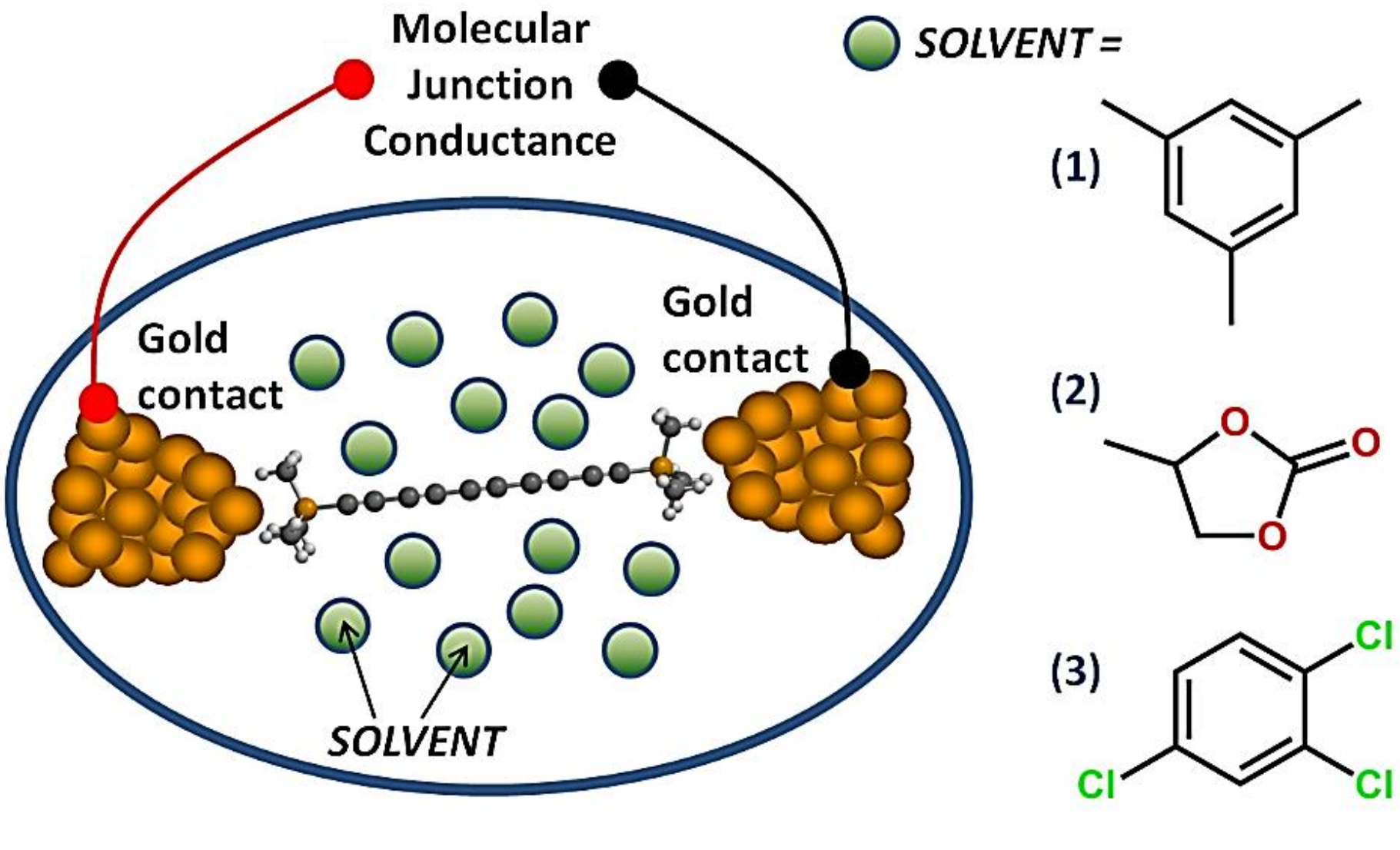

*Figure 5.1: A pictorial representation of single molecule conductance measurements, in (1) MES; (2) PC and (3) TCB solvents.*

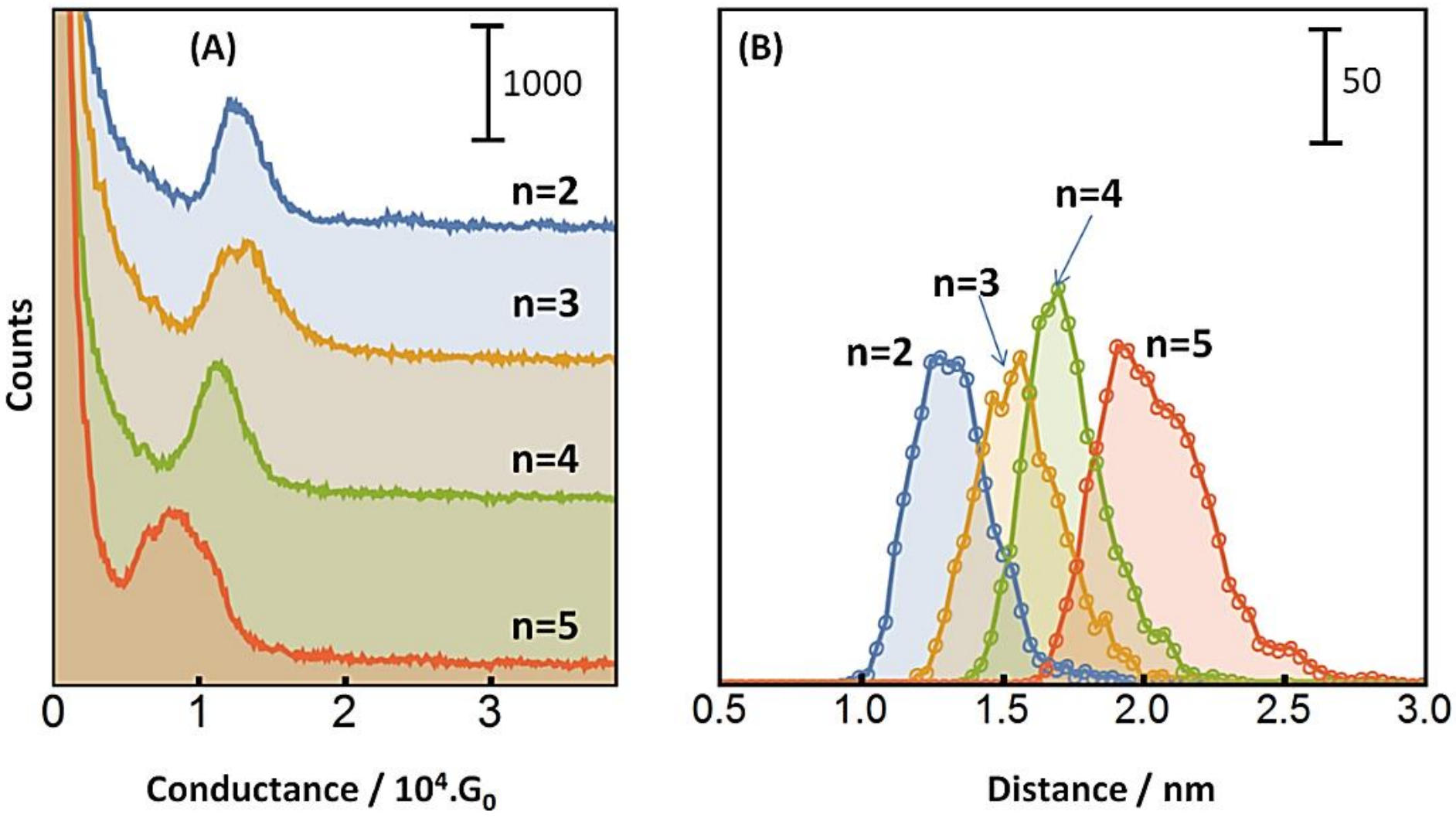

*Figure 5.2: I(s) conductance (A) and break-off distance (B) histograms recorded for series of oligoynes 2, 3, 4 and 5 in PC. Conductance histograms have been offset vertically for clarity.*

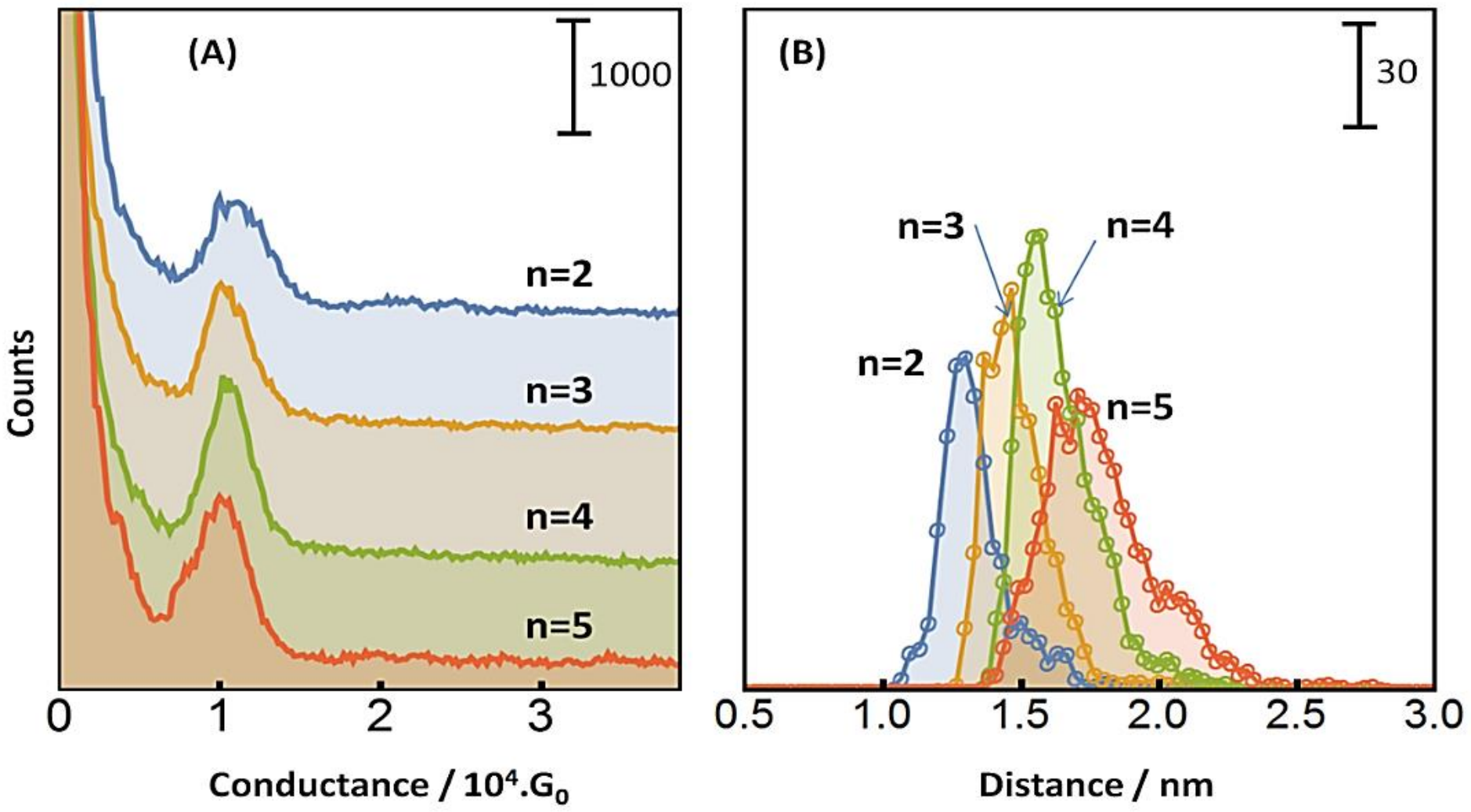

*Figure 5.3: I(s) conductance (A) and break-off distance (B) histograms recorded for series of oligoynes 2, 3, 4 and 5 in TCB. Conductance histograms have been offset vertically for clarity.*

Figure 5.2 shows conductance histograms for 2 – 5 recorded in propylene carbonate, which had been degassed with argon. As can be seen from figure 5.2, the conductance values decrease with the molecular length, and the break-off distance histogram distributions shift to longer distance along the series. The diyne, 2 (n = 2), shows a mean break-off distance of 1.3 nm, and this increases to 0.2 nm for pentayne (n = 5). Using the Si…Si distances computed from molecular modeling and a silicon to gold contact distance of 0.31 nm, also estimated from molecular modeling, and gives a theoretical Au-to-Au junction separation of 1.4 nm for 2 and 2.1 nm for 5. Figure 5.3 shows data recoded in a similar manner in TCB, while figure 5.4 shows data recoded in mesitylene.

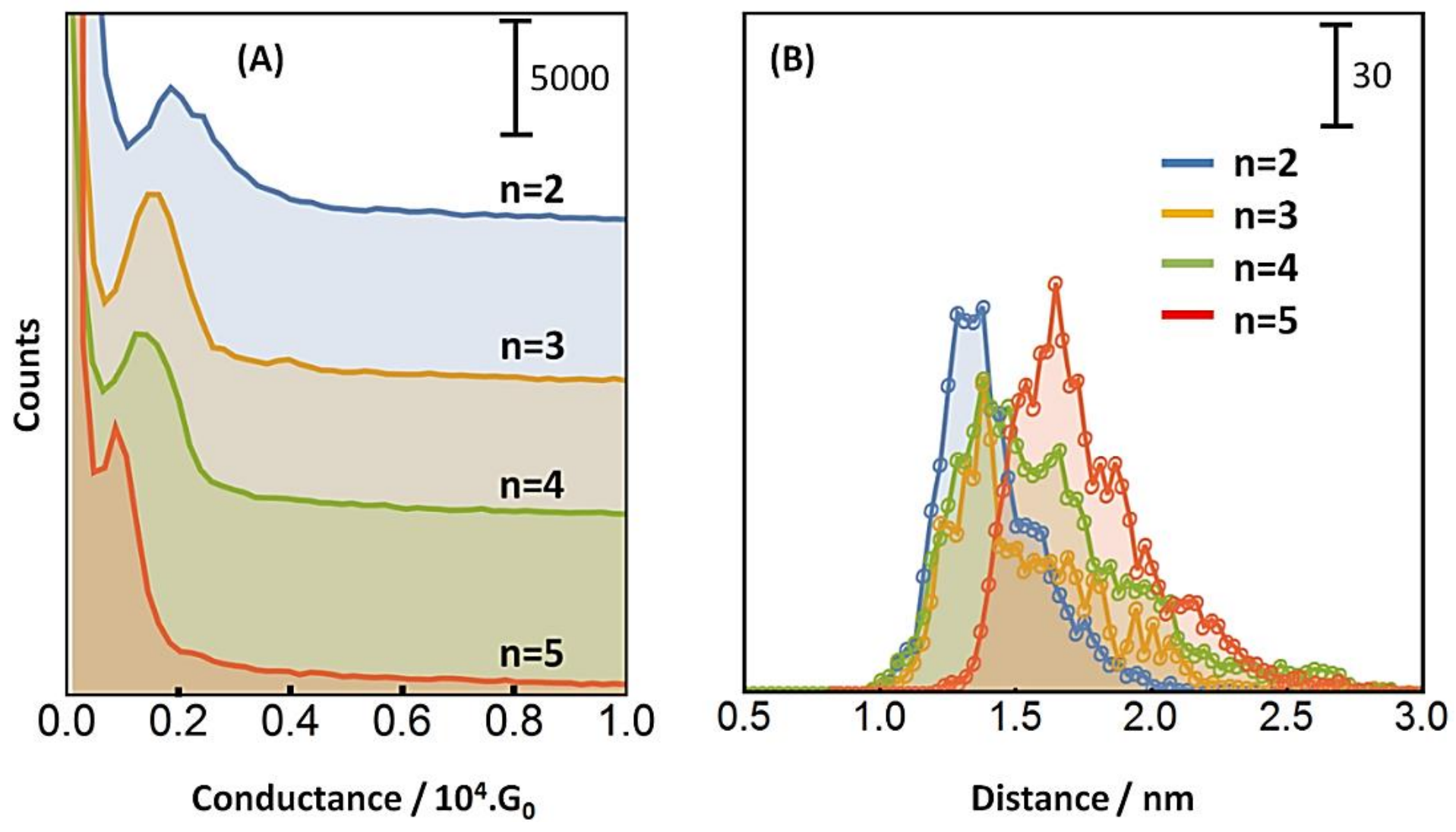

*Figure 5.4: I(s) conductance (A) and break-off distance (B) histograms recorded for series of oligoynes 2, 3, 4 and 5 in MES. Conductance histograms have been offset vertically for clarity.*

*Table 5.1: Theoretical and experimental conductances, and decay constant, in $nm^1(\beta(nm^{-1}))$ and per incremental –(C≡C)– unit, β (–(C≡C)–), in three different solvents (MES, TCB, and PC).*

| n | MES | | TCB | | PC | |
|---|---|---|---|---|---|---|
| | **theor G/G$_0$** | **expt G/G$_0$** | **theor G/G$_0$** | **expt G/G$_0$** | **theor G/G$_0$** | **expt G/G$_0$** |
| 2 | $2.03\times10^{-5}$ | $2.02\times10^{-5}$ | $10.3\times10^{-5}$ | $11\times10^{-5}$ | $14.2\times10^{-5}$ | $12.6\times10^{-5}$ |
| 3 | $1.35\times10^{-5}$ | $1.57\times10^{-5}$ | $10.2\times10^{-5}$ | $10.8\times10^{-5}$ | $13.7\times10^{-5}$ | $12.9\times10^{-5}$ |
| 4 | $1.16\times10^{-5}$ | $1.42\times10^{-5}$ | $9.60\times10^{-5}$ | $10.5\times10^{-5}$ | $12.6\times10^{-5}$ | $11.3\times10^{-5}$ |
| 5 | $0.94\times10^{-5}$ | $0.90\times10^{-5}$ | $6.16\times10^{-5}$ | $9.95\times10^{-5}$ | $10.6\times10^{-5}$ | $8.14\times10^{-5}$ |
| **β ($nm^{-1}$)** | 1.003 | 0.94 | 0.775 | 0.13 | 0.378 | 0.54 |
| **β (– (C≡C) –)** | 0.257 | 0.258 | 0.17 | 0.035 | 0.1 | 0.145 |

The single molecule conductance data are summarized in table 5.1. A plot of ln(conductance) versus junction is given in figure 5.5. The conductance values are larger in TCB and PC than those obtained in mesitylene solutions. From this, it is apparent that the same molecule can give conductance values that vary significantly with the solvent medium. Not only does the conductance change, but different length decays are also obtained across the series of oligoynes as quantified by the β values (figure 5.5).

The β value recorded in mesitylene (0.94 $nm^{-1}$) is substantially higher than those in TCB (0.13 $nm^{-1}$) and PC (0.54 $nm^{-1}$); the corresponding experimentally determined β values per incremental – (C≡C) – unit are 0.26 (MES), 0.035 (TCB), and 0.145 (PC).

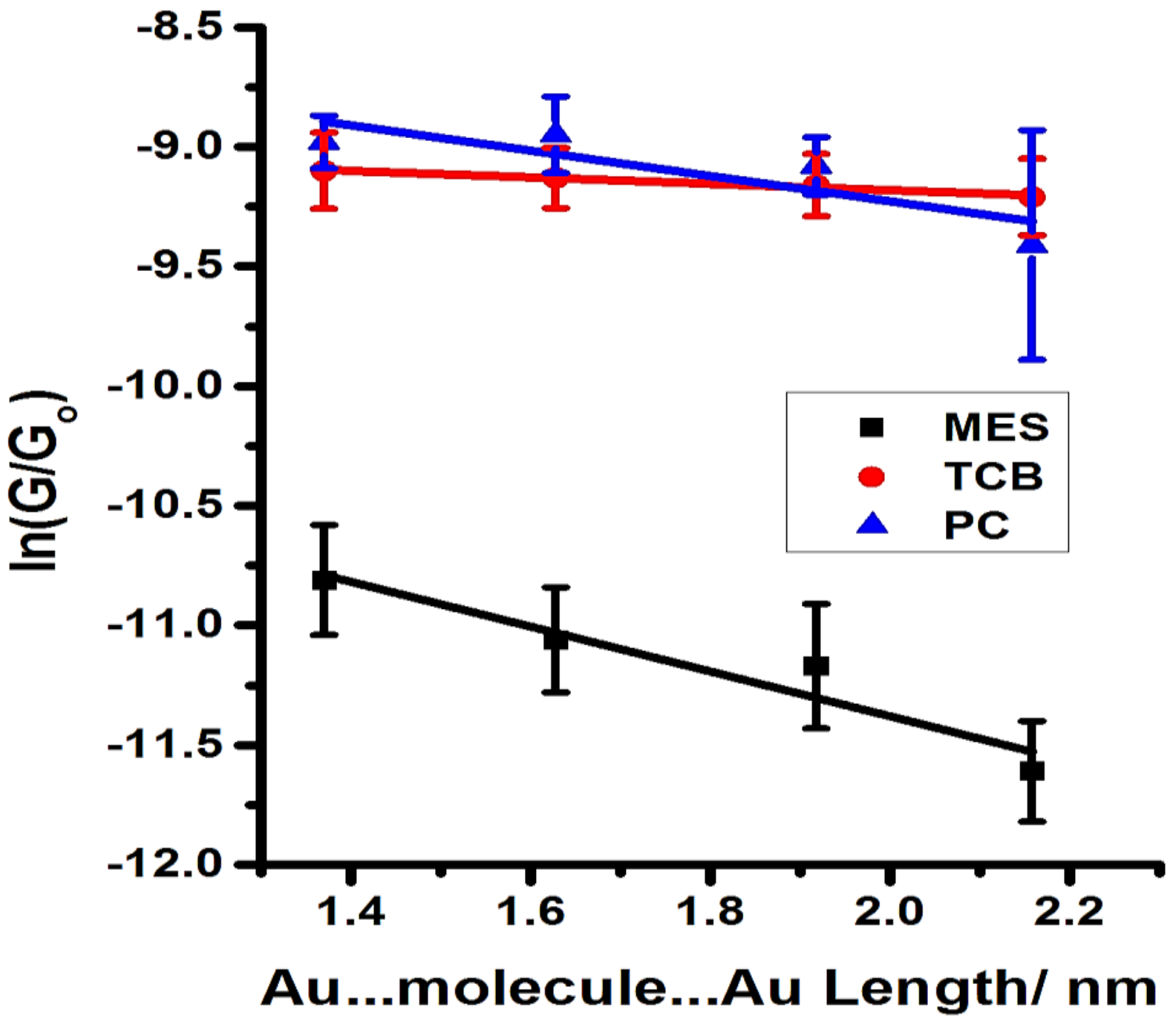

*Figure 5.5: Plots of ln(conductance) versus junction length for the series 2, 3, 4 and 5. Data recorded in mesitylene (MES), trichlorobenzene (TCB), and propylene carbonate (PC) as labeled. The linear fitting of each plot gives the β value of each solvent series.*

## 5.3.2. Theoretical Results

To gain a deeper insight into the role of the solvent medium and molecular length on molecular conductance, the computational modeling has been used. Before computing transport properties, all molecules were initially geometrically relaxed in isolation to yield the geometries shown in figure 5.6. This figure shows that the HOMOs and LUMOs are extended across the backbone for each molecule, with HOMOs showing large contributions from the carbon–carbon triple bonds, while the LUMOs have greater contributions from carbon–carbon single bonds, as expected from previous studies of oligoynes [3, 37, 38, 39].

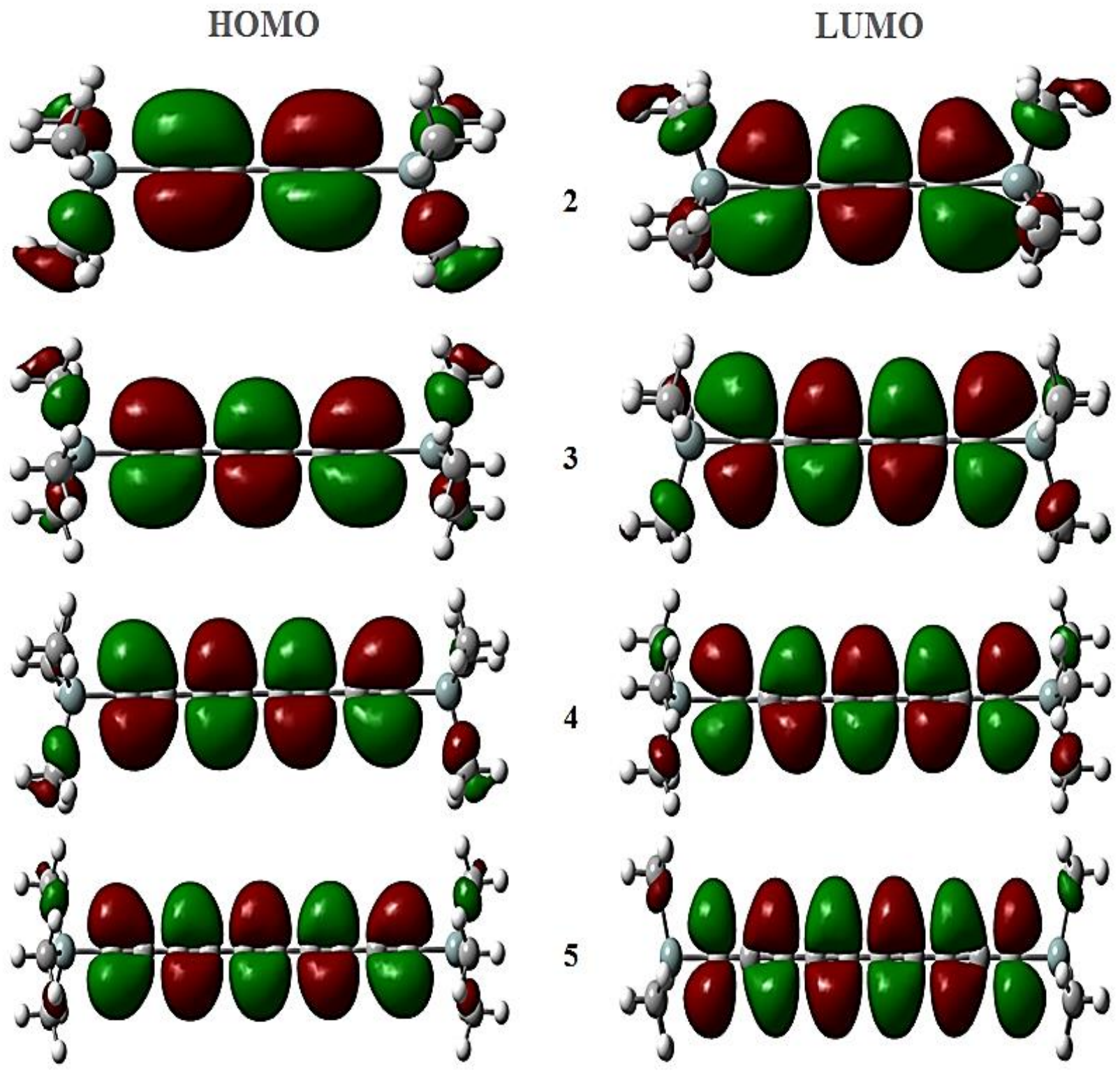

*Figure 5.6: The relaxed molecules in a gas phase and iso-surfaces of the HOMOs and LUMOs for 2 – 5.*

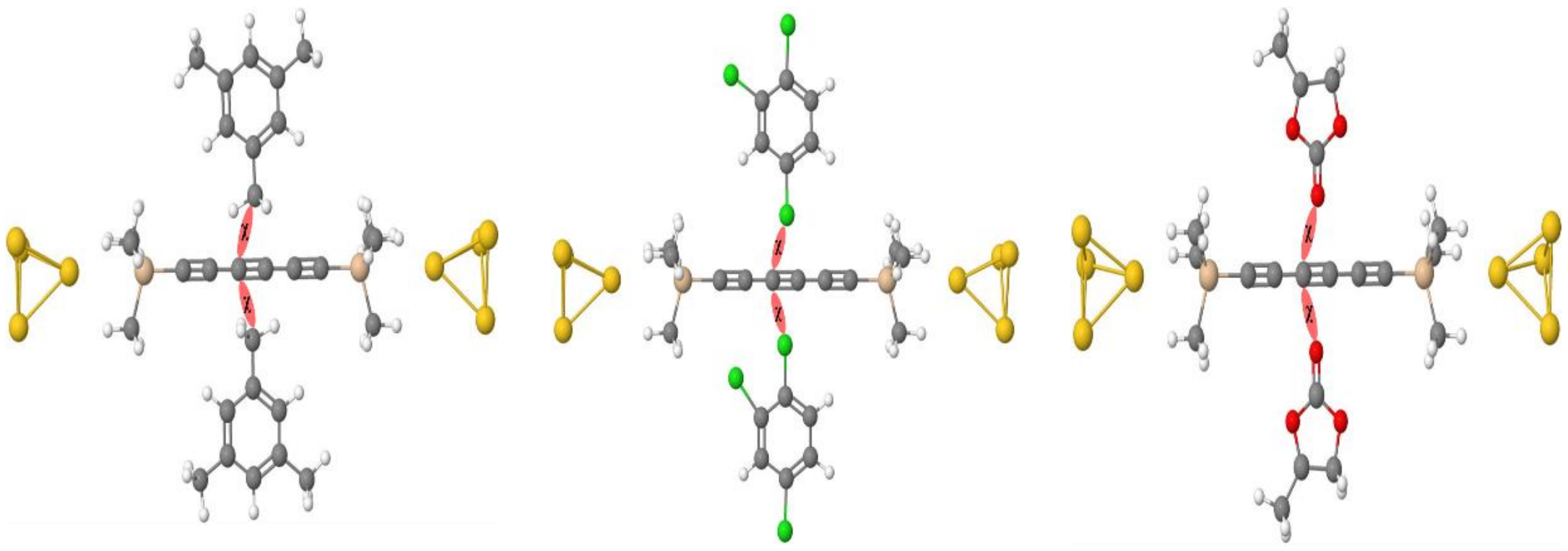

[3][MES] [3][TCB] [3][PC]

*Figure 5.7: Definition of the distance (χ) used to describe the various geometries representing solvent-oligoyne interaction with 3 by way of an example. χ is the distance between carbon (gray), chlorine (green), or oxygen (red) atoms of the solvent molecules (MES, TCB, or PC) and the nearest adjacent carbon atom of the backbone. For clarity, only two of the six solvent molecules employed to represent the first solvation shell are shown in these schematic representations.*

To explore how the solvent medium (MES, TCB, or PC) affects the conductance of these molecules, six molecules of each solvent were initially placed at a nearest distance χ from the oligoyne backbone within the model junction as shown in figure 5.7. In each simulation, the molecule plus solvent molecules together with few layers of gold at each pyramidal electrode were allowed to relax. These simulation were carried out with seven different initial distances χ for each solvent as shown in figure 5.9, resulting in seven different relaxed geometries. The relaxed structures are shown in figure 5.8 for the three different solvents.

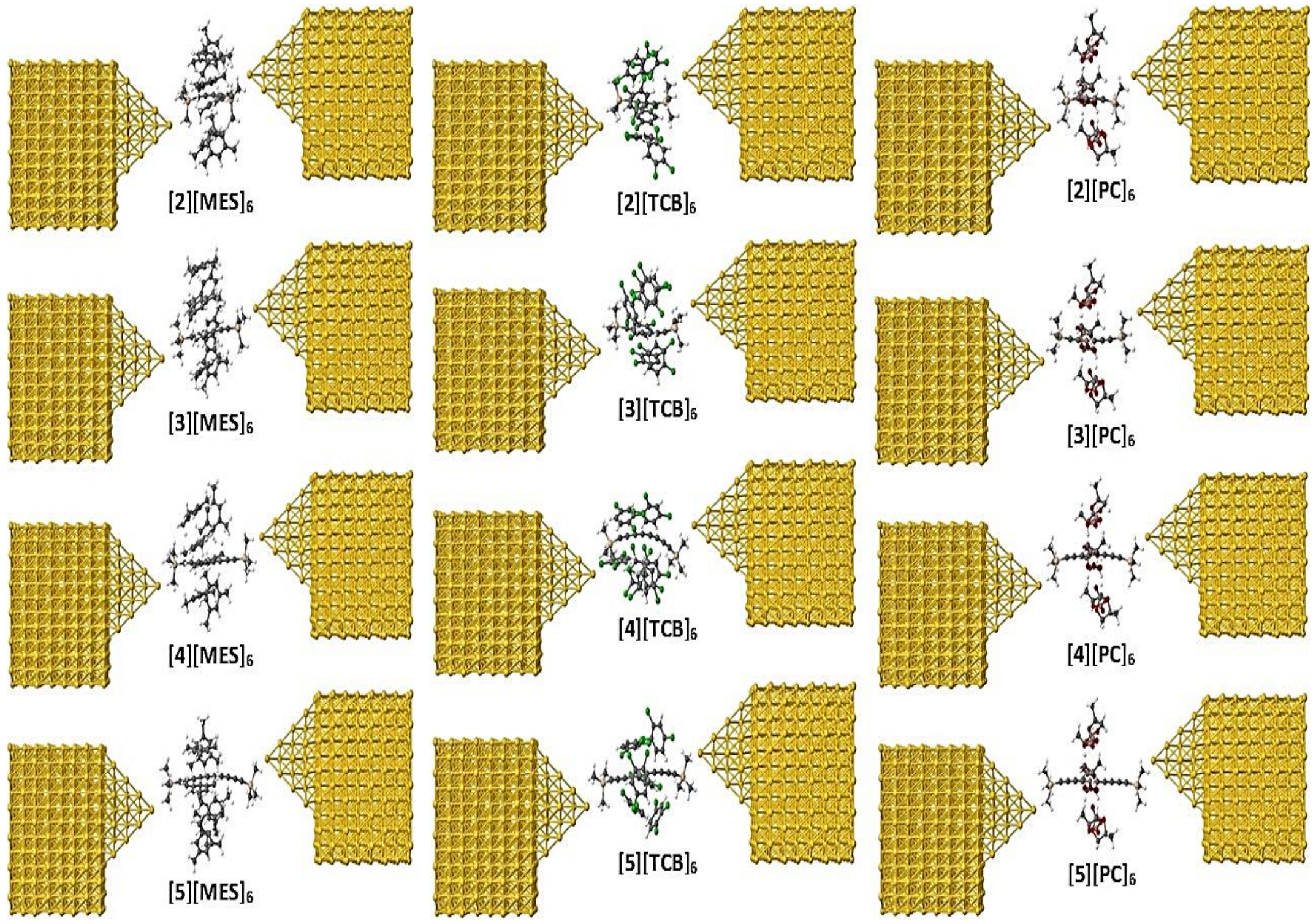

*Figure 5.8: Relaxed structures of junctions incorporating 2 – 5 in each three different solvents (MES, TCB, and PC).*

Figure 5.8 and table 5.2 show the relaxed structures and key optimized distances associated with the configurations. It could be noted that the interactions between the solvent molecules and the backbone affected the structural features of these molecules, since all bridges have been bended as shown in figure 5.8, and this leads to different molecular lengths as shown in table 5.2. The molecular length of 3 by way of an example differs from solvent to another (L = 0.993 nm in MES, 0.995 nm in TCB and 1.007 nm in PC).

*Table 5.2: The key optimized distances associated with the conformations; $d_{Au\text{-}Au}$ is the center-to-center distance of the apex atoms of the two opposing gold pyramids in the relaxed structures as shown in figure 5.8. L = $d_{Si\text{-}Si}$ is the molecular length, which is the optimized distance between the centres of silicon atoms. X is the bond length between the top gold atoms of the pyramids and the silicon atoms of the TMS terminal groups.*

| System | MES | | | TCB | | | PC | | |
|---|---|---|---|---|---|---|---|---|---|
| | $d_{Au\text{-}Au}$ (nm) | L = $d_{Si\text{-}Si}$ (nm) | X (nm) | $d_{Au\text{-}Au}$ (nm) | L = $d_{Si\text{-}Si}$ (nm) | X (nm) | $d_{Au\text{-}Au}$ (nm) | L = $d_{Si\text{-}Si}$ (nm) | X (nm) |
| 2 | 1.332 | 0.745 | 0.415 | 1.357 | 0.752 | 0.37 | 0.89 | 0.75 | 0.34 |
| 3 | 1.630 | 0.993 | 0.415 | 1.574 | 0.995 | 0.37 | 1.317 | 1.007 | 0.34 |
| 4 | 1.904 | 1.261 | 0.415 | 1.766 | 1.129 | 0.37 | 1.367 | 1.267 | 0.34 |
| 5 | 2.205 | 1.512 | 0.415 | 2.008 | 1.415 | 0.37 | 1.605 | 1.525 | 0.34 |

Before calculating the transport properties of these structures, the binding energies were computed for the contact formed between gold cluster and the TMS terminal groups for each of the different optimized distance $\chi$. Two points are note-worthy from the binding energy plots of figure 5.9. First, for the junctions modelled in figure 5.9 with compound 3 by way of an example, the optimized distance at which the contact binding energy reaches its maximum value is $\chi$ = 0.7 nm. Consequently, in what follows, the transport properties are calculated for the structures with the optimized distance $\chi$ = 0.7 nm. Second, the structures with TCB and PC solvation exhibit slightly stronger Au – TMS contact binding than for mesitylene solvation (-0.41 eV for the structures with MES and -0.44 eV for the structures with TCB and PC). However, the differences in binding energies are small and less than or on the order of $k_BT$ at room temperature.

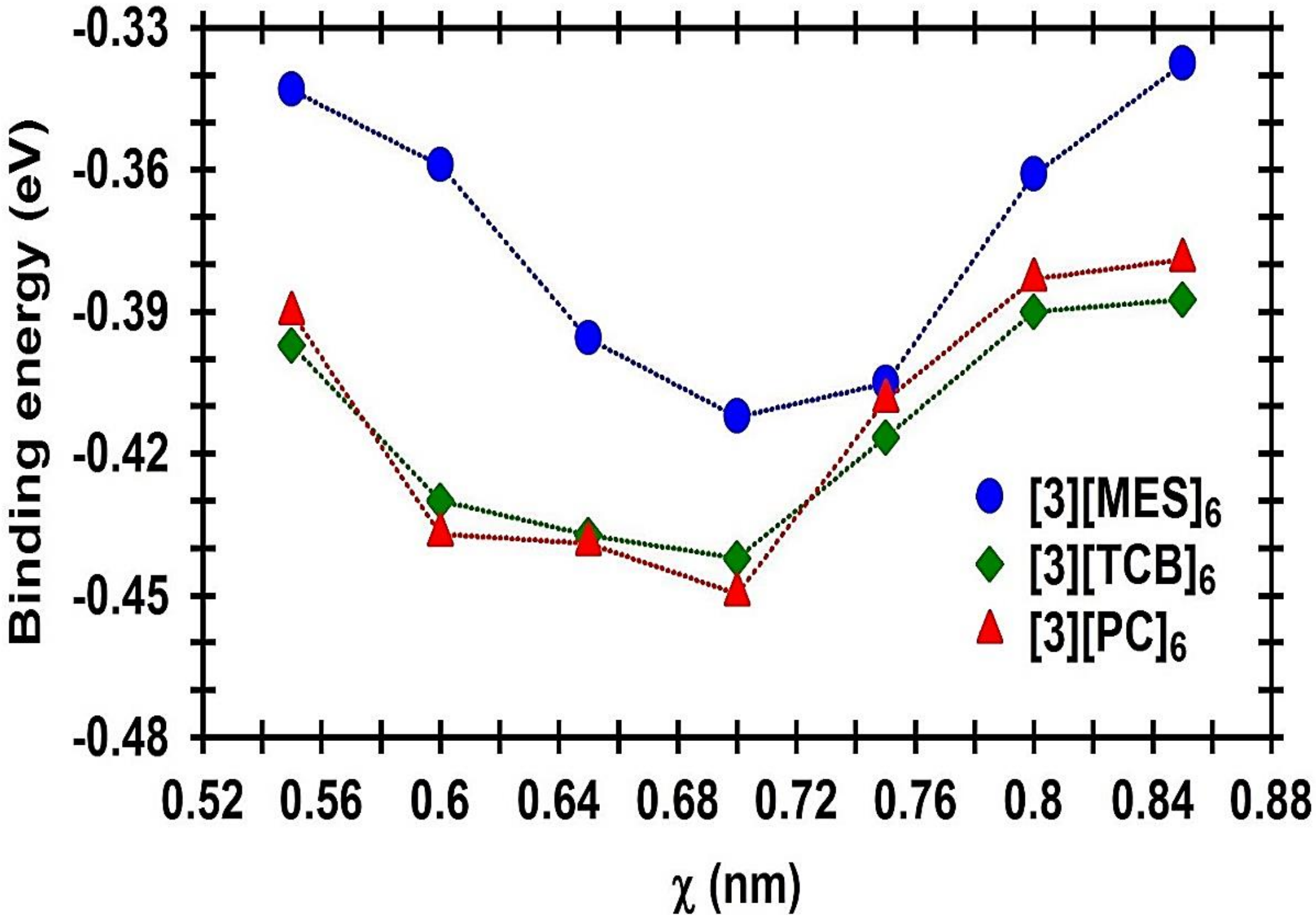

*Figure 5.9: The binding energy between the gold cluster and the TMS terminal group in 3 versus the distance between the solvent molecules and the backbone ($\chi$).*

To investigate the electronic properties of these molecules in three different environments, SIESTA code has been used, which employs norm-conserving pseudopotentials and linear combination of atomic orbitals (LCAO) to span the valance states. All systems were initially placed between two pyramidal gold electrodes, and then the oligoyne molecule plus solvent molecules and few layers of gold were allowed to relax to yield the structures. Then to calculate the transmission coefficient, $T(E)$ using the GOLLUM code, the resulting configurations were connected to bulk gold electrodes grown along the (111) direction as shown in figure 5.8 and described in detail elsewhere [31].

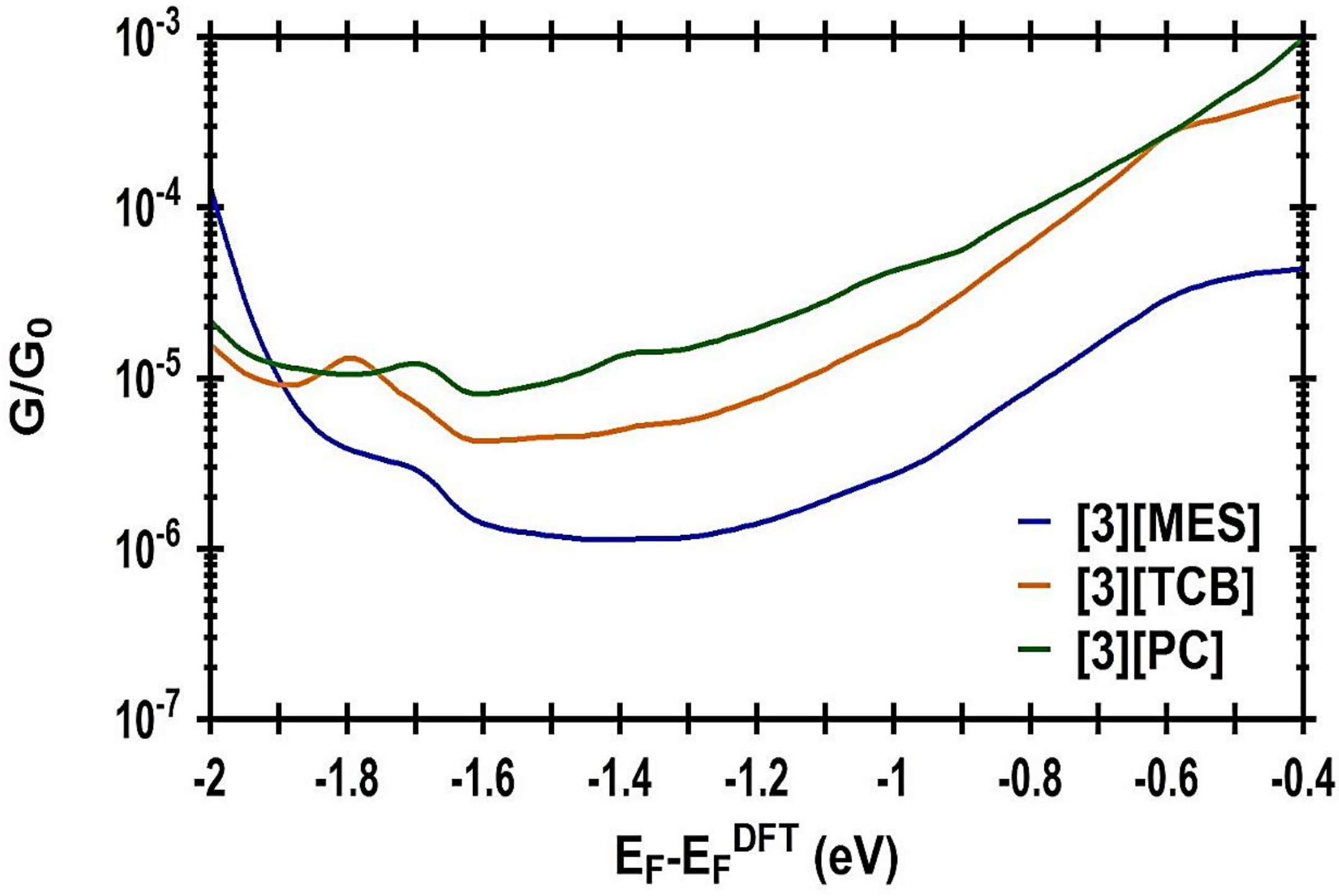

*Figure 5.10: Room-temperature conductance as a function of the Fermi energy for 3 in mesitylene (MES), trichlorobenzene (TCB), and propylene carbonate (PC).*

The theoretical and experimental data are summarized in figures 5.10 (illustrated for 3 by way of an example), 5.11, 5.12, and 5.13 and table 5.1. Figure 5.10 shows the calculated room temperature conductances (in $G/G_0$), plotted for energies in the HOMO – LUMO gap region, as a function of the Fermi energy ($E_F$) for 3, relative to the DFT-predicated Fermi energy ($E_F^{DFT}$). Figure 5.11 shows all conductance curves. It is worth to mention that DFT does not usually predict the correct value of the Fermi energy. Therefore, $E_F$ has been treated as a free parameter, which it has been determined by comparing the calculated conductance with experiment [32].

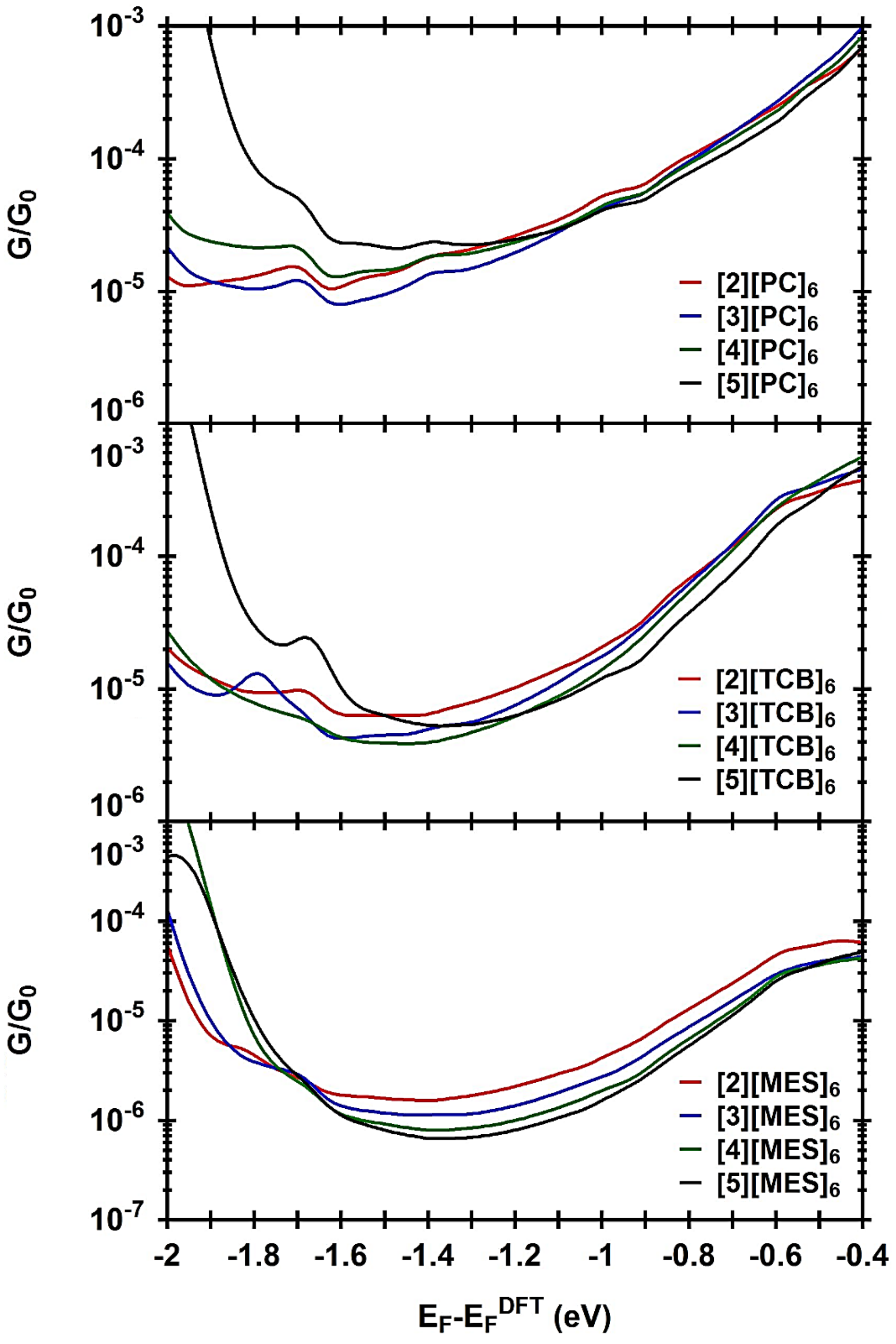

*Figure 5.11: The calculated conductances as a function of the Fermi energy for all molecules in mesitylene (MES), trichlorobenzene (TCB), and propylene carbonate (PC).*

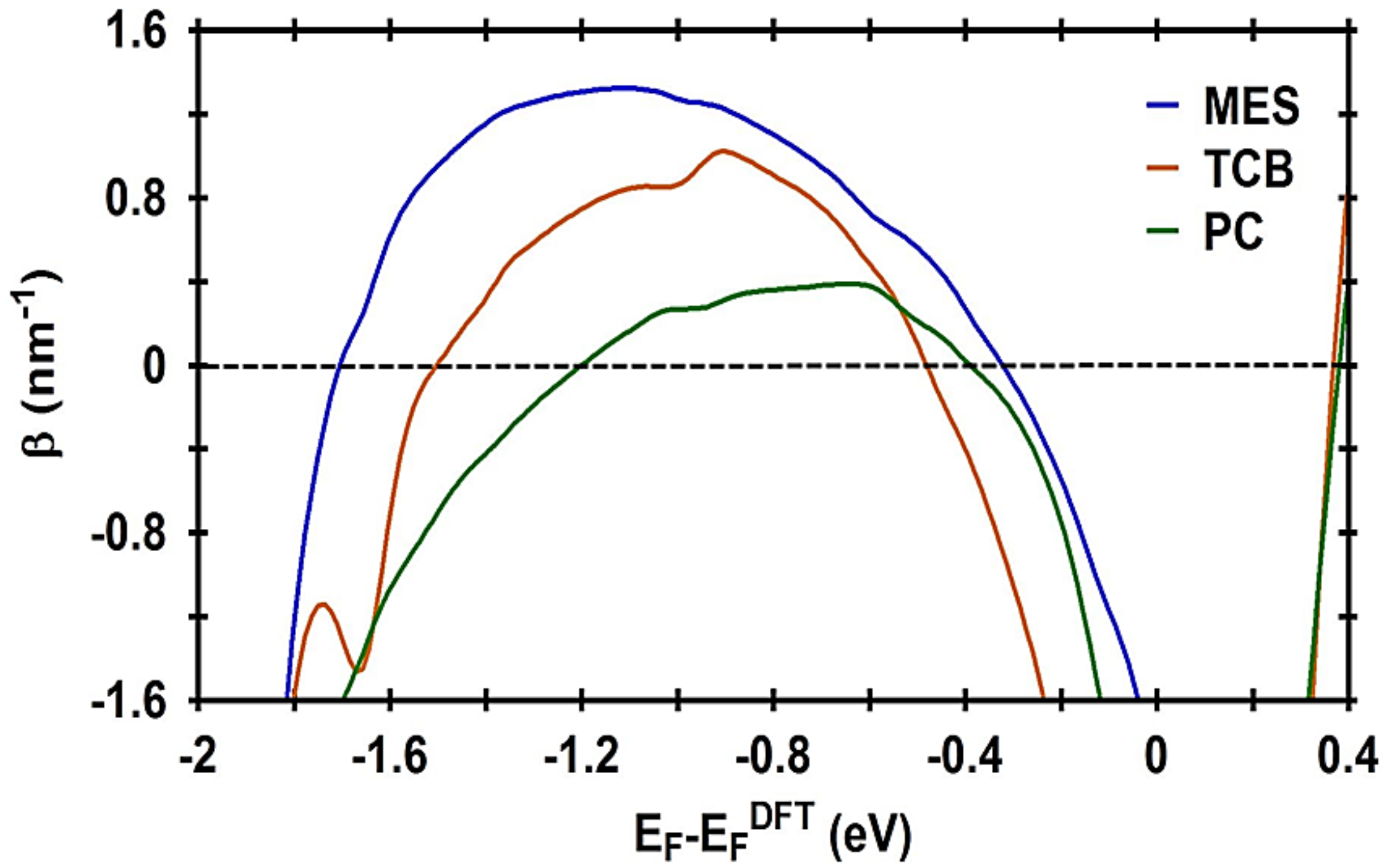

*Figure 5.12: Decay constant β (nm$^{-1}$) for the molecular series 2, 3, 4, and 5 as a function of the Fermi energy in three different solvents (MES, TCB, and PC).*

Figure 5.12 shows the Fermi energy dependence of the decay constant (β) for the molecular series 2, 3, 4, and 5 in MES, TCB and PC solutions, respectively. The best agreement with experimental data as shown in figure 5.13 is obtained at $E_F$ = - 0.72 eV, which is shifted toward the centre of the HOMO – LUMO gap, compared with DFT-predicted value. With this choice of $E_F$, both computational (figures 5.10 and 5.11), and experimental (figure 5.5) data sets show that the order of the conductance at the chosen Fermi energy is PC ≈ TCB > MES. At $E_F$ = - 0.72 eV, the computational β values (figure 5.12) follow the trend $\beta_{MES} > \beta_{TCB} > \beta_{PC}$, meaning that the experimentally observed β value recorded in TCB is not so well reproduced by theory. However, it can be seen from figure 5.12 that the β versus Fermi energy curve around the chosen $E_F$ value has a high slope. This sensitivity to precise Fermi energy value may explain this disparity.

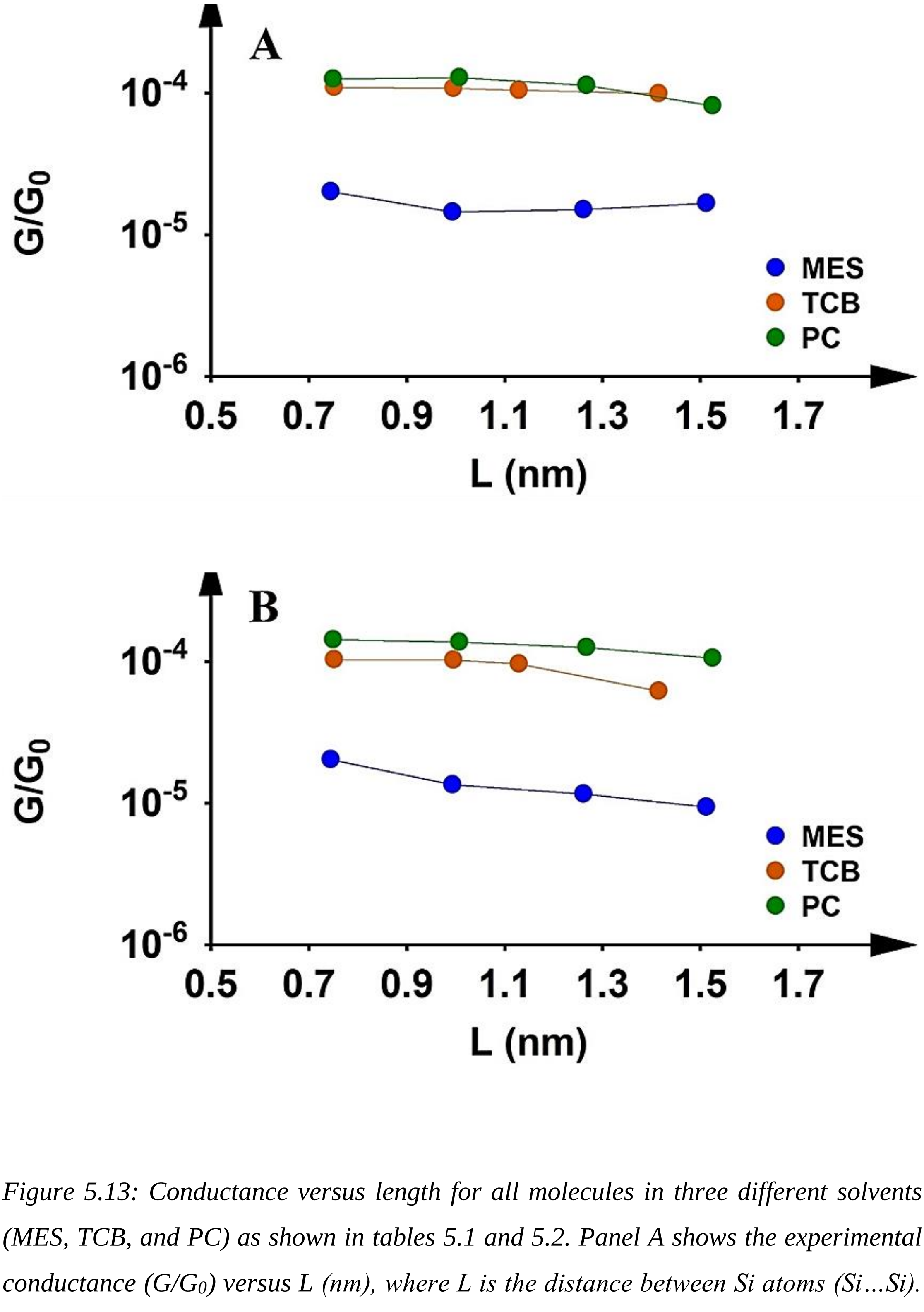

*Figure 5.13: Conductance versus length for all molecules in three different solvents (MES, TCB, and PC) as shown in tables 5.1 and 5.2. Panel A shows the experimental conductance ($G/G_0$) versus L (nm), where L is the distance between Si atoms (Si…Si). Panel B shows the calculated conductance ($G/G_0$) versus L.*

Figure 5.13 and table 5.1 summarize and compare the experimental and theoretical conductance values versus the molecular length (L). It could be noted here that for both experimental and computational data sets the conductance of the structures with PC and TCB is higher than that with MES.

Although oligoynes represent one of the most archetypical molecular wires, the effects of solvents on their single molecule conductance has not been considered before. The calculations described above show that the solvent surrounding the oligoyne molecule bridge has a strong impact on the computed electron transmission curves for the junctions. As a result different conductance and length dependence values are obtained across the homologous series. This is not the result of covalent interactions between the solvent and the bridge, but rather occurs from longer range electrostatic interactions. This could be described as a “solvent induced gating of the molecular junction electrical properties”.

Solvent effects have only been examined in detail in the literature for a relatively small number of single molecular junctions, which perhaps surprising given that most measurements of this kind are performed in a liquid or ambient environment. Li et al. found the conductance values of octanedithiol to be independent of solvent (toluene, dodecane, and water) [40]. This is perhaps not unexpected given by very large HOMO – LUMO gap for this far-off-resonance tunnelling system. On the other hand, Leary et al. have demonstrated large solvent dependence for the conductance of Oligothiophene-containing molecular wires [21]. This was attributed to water molecules directly interacting with the thiophene molecular rings and thereby shifting transport resonances with the effect of greatly increasing conductance.

Water dependence has also been seen in perylene tetracarboxylic diimide (PCTDI)-containing molecular bridges, with the measured conductance being temperature dependent in aqueous solvent but temperature independent in toluene [41]. In a theoretical study of this system, the water and temperature dependence was modelled through thermal effects on the hydrogen bonding network interacting primarily with the carbonyl moieties on PCTDI [42].

Other models have considered the effect of the solvent on the gold contact work function. In such a study, Fatemi et al. experimentally determined that solvents could increase the conductance of 1,4-benzenediamine (BDA) – gold molecular junctions by up to 50% [43 – 46]. This was attributed to shifts in the gold contact Fermi energies resulting from solvent binding, leading to better alignment to the HOMO of BDA and hence higher conductance. These studies collectively show the complexity of solvent effects in molecular junctions, which have, depending on the system, been modelled through electrostatic interactions between the solvent and molecular bridge, or solvent binding to gold contact atoms.

A present study demonstrates that the longer range solvent–molecular bridge interactions alone can describe the experimentally observed solvent effects on oligoyne junction conductance. In addition, this study rationalizes the previously unexpected differences observed between different studies of oligoyne molecular conductance [3, 4] which can be now attributed to solvent effects.

## 5.4. Summary

In this Chapter the electronic properties of oligoyne-based molecular wires in three different mediums have investigated experimentally and theoretically. It has been demonstrated that the changing of the solvent can lead to changes in both the conductance and the attenuation factor of oligoyne molecular bridges.

DFT computations shown that both the molecular junction conductance and the decay constant depend in a very sensitive manner on the position of the contact Fermi energies within the HOMO – LUMO gap.

In addition, it has been shown that the interactions between the solvent molecules and the oligoyne-bridges affected the structural features of these molecules, since all bridges have been bended, and that leads to different molecular lengths. By way of an example, the molecular length of 3 in MES-solvent is 0.993 nm, while in TCB and PC are 0.995 and 1.007 nm respectively.

Furthermore, it has been demonstrated that the structures with TCB and PC solvation exhibit slightly stronger Au – TMS contact binding than for mesitylene solvation (-0.41 eV for the structures with MES and -0.44 eV for the structures with TCB and PC).

Finally, these results shown that the solvent environment is an important variable to consider in interpreting conductance measurements and that the environment can give rise to dramatic changes in electronic properties of this kind of molecules.

# Chapter 6

# Effects of electrode-molecule binding and junction geometry on the single-molecule conductance of bis-2,2′:6′,2″-terpyridine based complexes

## 6.1. Introduction

The development of methods that permit the measurement of the electrical characteristics of single molecules under routine laboratory conditions [1, 2] coupled with the incentives for technological innovation arising from ever increasing challenges facing top-down miniaturisation of solid-state electronic devices, has seen a renaissance in the field of molecule electronics over the past decade [3 – 6]. In the context of developing molecular components for use in a hybrid solid-state/molecular electronics technology, many different molecular structures have been examined within molecular junctions, including oligophenylenes, [7] oligoaryleneethynylenes, [8] and oligoynes, [9] and arylene–ethynylene based molecular wires up to 8 nm in length [10 – 12]. However, whilst the majority of metal|molecule|metal junctions studied to date has been derived from organic molecules, the possibility that metal complexes may play a role in molecular electronics has been recognized [13], and inorganic and organometallic molecular components for electronics are now attracting increasing attention [14 – 18].

Various families of metal complexes have been explored for their wire-like properties and higher functionalities [19, 20], including porphyrin oligomers [21] and assemblies, [22] and metal alkynyl complexes [23 – 28]. Within the context of exploratory studies, bis-2,2′:6′,2″-terpyridine complexes are particularly attractive, being compatible with a broad cross-section of the metallic elements of the transition series, and thereby offering a wide range of metal d-electron configurations and charges, electro- and photo-chemical activity, and diverse synthetic approaches which include 'on surface' strategies that been used in the construction of quite complex surface bound mono-and multi-metallic [29, 30 – 32] films with impressive electrical characteristics [33 – 38]. Within single molecule junctions, the flexibility of the coordination bonds around the metal center has led to the opportunity for manipulation of transport properties through such Cardan-joint style metal complexes by mechanical stimulus [39, 40].

For assembly, as components in molecular electronics, metal complexes offer the potential for finer tuning of the frontier molecular orbitals in metal complexes to match the Fermi levels of the electrodes, [41] the possibilities of augmenting electronic characteristics through accessing available redox levels [42] and manipulating them through electrochemical gating, [43 – 45] the introduction of magnetic effects, [46, 47] and high thermoelectric efficiency [48]. These various factors are then expanded further by experimental and computational work in which multiple metal centres are introduced along a linear wire-like chain, either as an array of metal atoms [14, 49, 50] or in ligand-linked assemblies [24 – 27]. Whilst there is a body of experimental evidence, such as the observation of Kondo effects in transition metal complex based molecular junctions [51 – 53] and electrostatically gated spin-blockade effects, [46] which indicates that the metal center is involved directly in the transport mechanism, this is not always the case [41]. Recent studies have highlighted the potential role of metal centres as a structural element with the surrounding ligands providing the pathway for the through molecule current [54]. In such cases, the molecule-electrode contact and electronic structure of the ligand framework will play a more significant role in determining the overall transport properties of the molecule than the identity of the metal in the complex.

The important role of the molecule-electrode contact in determining transport properties of a molecular junction is now widely recognised, [55] and many different functional groups have been explored in this regard, with thiols, amines and pyridines being particularly widely used [56]. In addition to the chemical nature of the binding group, the electrode-molecule contact also depends on the structure of the electrode surface. For example, thiolate binds to a wide variety of sites on the gold surface including different points on flat terraces (atop surface atoms, in bridging or in hollow sites), adjacent step edges or ad-atoms [57, 58]. Each of these different contact types gives rise to a different conductance signature, which accounts for the appearance of multiple peaks in the conductance histograms of even simple thiolate contacted molecules [59]. One possible strategy to limit the range of these possible binding sites, and thereby simplifying the conductance profile of the molecular junction, would entail increasing the steric bulk around the surface coordinating atom.

Thioethers are beginning to attract attention in both studies of self-assembled monolayers (SAM) on gold, [60] and as a contact in molecular junctions where they often give rise to simpler conductance histograms than analogous thiolates [61 – 66]. Recently, the trimethylsilylethynyl moiety has been identified as a possible bulky anchoring moiety for use in single molecule electronics [28, 67 – 70].

Results from single molecule junctions indicate that the use of the trimethylsilylethynyl moiety as surface contact group leads to current histograms containing only a single conductance peak in the measureable current range, although junction formation probabilities are low (ca. 5%) [28, 67]. Detailed studies of SAM formed from trimethylsilylethynyl functionalised unsaturated hydrocarbons have indicated pit-etching features, consistent with a surprisingly strong Au-Si interaction [71]. A close registry of the silyl molecules with the underlying Au(111) surface and evidence for a degree of Si-Au interaction from synchrotron radiation photoelectron spectroscopy led to the suggestion of a local surface complex featuring a five-coordinate silicon atom in these self-assembled films [72, 73]. Later refinements to the model have shown the importance of lateral intermolecular van der Waals interactions in pre-organising the silyl head group in such a position as to promote the Si-Au interaction [74, 76, 77]. However, the nature of the Si-Au interaction in the case of the isolated molecules used in single molecule junction studies is an area for further investigation.

In what follows, this work seeks to extend these studies and arrive at a more detailed understanding of the role of the anchor unit and metal complex fragment on the behaviour of these junctions, by studying Fe(II), Ru(II) and Co(II) bis-2,2′:6′,2″-terpyridine complexes anchored by thiomethyl [60 – 63], and trimethylsilylethynyl [28, 67, 68 – 70] moieties within molecular junctions, supported by single molecule conductance measurements and quantum chemical models.

This chapter presents all theoretical details and experimental conductance measurements as a part of a published paper. For more details regarding the experimental methods and synthesis details see *Ross D.; Oday A. Al-Owaedi, David C. M.; Qiang Z.; Joanne T.; František, H.; Simon J. H.; Richard J. N.; Colin J. L.; Paul J. L. Effects of Electrode−Molecule Binding and Junction Geometry on the Single-Molecule Conductance of bis-2,2′:6′,2″-Terpyridine-based Complexes. Inorg. Chem. 2016. 55(6): p. 2691 – 2700.*

# 6.2. Experimental and Theoretical Methods

## 6.2.1. Experimental Methods

The transport characteristics in single-molecule junctions were studied by scanning tunnelling microscopy, using the current-distance (*I*(*s*)) technique. All details of experimental methods are presented in ref. [75].

## 6.2.2. Theoretical Details: Computational Methods

To avoid duplication, the same computational methods that have been previously described to relax the molecules (in a gas phase) in chapter 5, were utilized in this work as well.

To model the effect of an electrochemical environment, two tetrafluoroborate $[BF_4]^-$ or hexafluorophosphate $[PF_6]^-$ counterions were initially placed at different distances from the backbone (molecule). Two $[BF_4]^-$counterions have been used for the complexes containing a Co centre, while two $[PF6]^-$counterions have been used for the complexes containing Ru or Fe centres. Then the molecules plus counterions were allowed to relax. In this study, simulations were carried out with ten different initial distances $\chi$ between the fluorine atoms of the counterions and hydrogen atoms of the backbone, resulting in ten different relaxed distances $\chi$, as shown in figure 6.3 (illustrated by way of an example) and table 6.2.

In order to compute the electrical conductance of molecules in three different junction models, they were each placed between three distinct gold electrodes denoted Type I,

Type II and Type III. The complex cations and their associated counterions were allowed to relax, to yield the structures shown in figures 6.5, 6.6, and 6.7. For each structure, the transmission coefficient, $T(E)$, was calculated by first obtaining the corresponding Hamiltonian and overlap matrices using SIESTA [78] and then using the GOLLUM code [79].

To determine $E_F$, the predicted conductance values of all molecules have been compared with the experimental values and chose a single common value of $E_F$ which gave the closest overall agreement. This yielded a value of $E_F - E_F^{DFT} = -0.14\ eV$, which is used in all theoretical results. This is commonly accepted procedure in molecular electronics DFT-based calculations (cf., ref [80]).

Again, in order to prevent the repetition, the same theoretical method [81, 82] that has been employed to calculate the binding energies between anchor group and gold electrode for the optimized structures in chapter 5, was used in this work as well.

# 6.3. Results and Discussion

In this section, the experimental single molecule conductance results and theoretical, such as binding energies between the gold cluster and the TMS terminal groups, the calculated conductance and electrochemical computational results have been presented. Synthesis, spectroelectrochemistry and resonance Raman spectroscopy experimental results can be found in ref. [75].

## 6.3.1. Experimental Single Molecule Conductance Results

The molecular conductance of the bis-2,2′:6′,2″-terpyridine based complexes $[2\text{-M}](X)_2$ – $[3\text{-M}](X)_2$ (M = Fe, Ru, X = $PF_6^-$; M = Co, X = $BF_4^-$) on an Au(111) surface was investigated by scanning tunnelling microscopy (STM) under ambient conditions using the *I*(*s*) technique [1], as shown in figure 6.2 and table 6.1.

The conductance values, break-off distances and calculated molecular lengths are summarised in table 6.1, with conductance histograms obtained from the *I*(*s*) data shown in figure 6.2. To date, the few compounds featuring trimethylsilylethynyl based electrode contacts that have been studied in molecular junctions have been charge-neutral organic compounds [28, 68, 69, 70] or organometallic complexes [28]. The data in table 6.1 is consistent with literature studies of OPEs and oligoynes, whose tunnelling conductances decay with increasing numbers of phenyl rings and triple bonds respectively. One expects the conductances of $[2\text{-M}](X)_2$ to be lower than that of $[3\text{-M}](X)_2$, because each end of $[2\text{-M}](X)_2$ contains both a phenylene and a triple bond in series, where each end of $[3\text{-M}](X)_2$ contain only a single phenylene spacer or a single triple bond respectively. There are several competing factors here, including the increased molecular length (β for polyphenylene chains is said to be 0.6 $A^{-1}$) [83] and the decreased conjugation brought about by the twisting of the phenylene ring relative to the plane of the tpy π-system [84, 85]. Whilst for each series $[2\text{-M}]^{2+}$ and $[3\text{-M}]^{2+}$ (figure 6.1) the overall span of values is not more than a factor of three (table 6.1, figure 6.1), the apparent order of conductance is Ru ≥ Fe > Co for the trimethylsilylethynyl contacted complexes $[2\text{-M}]^{2+}$, whilst for the MeS derivatives $[3\text{-M}]^{2+}$ a trend of Co > Fe ≥ Ru is observed.

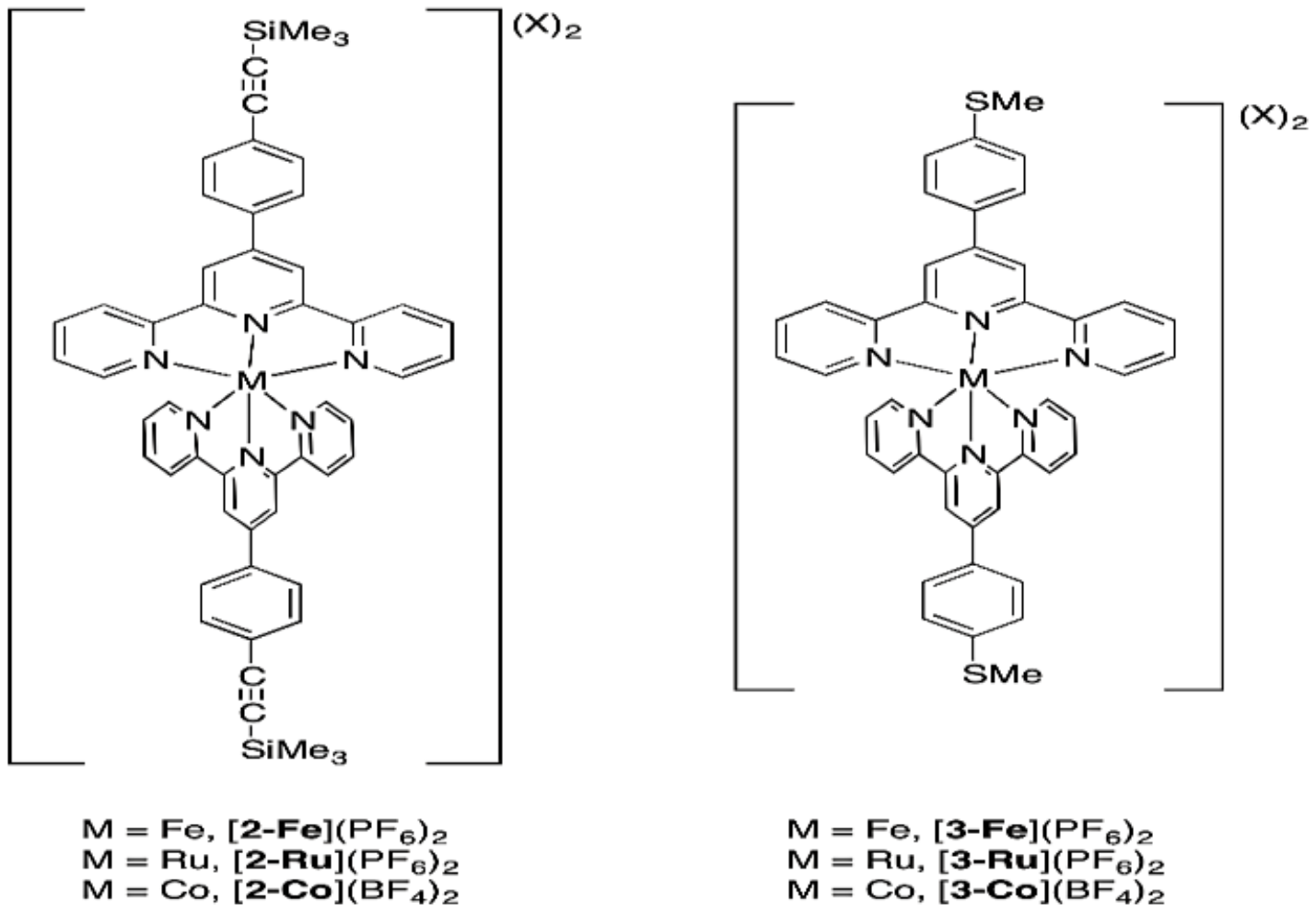

*Figure 6.1: Schematic representation of the series of complexes [2-M]$(X)_2$ and [3-M]$(X)_2$.*

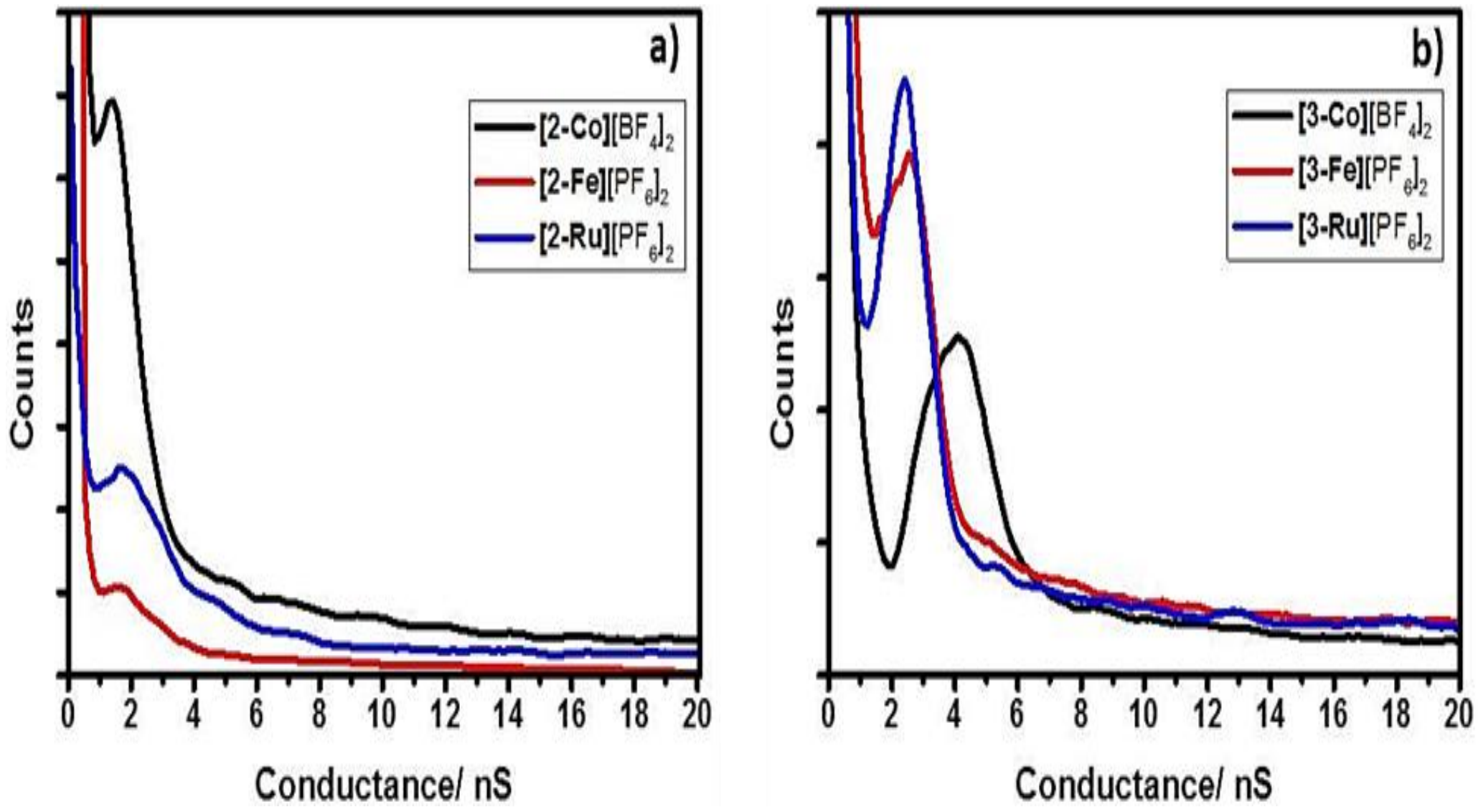

*Figure 6.2: Conductance histograms. (a) -C≡CSiMe$_3$ contacted complexes [2-Fe]$(PF_6)_2$, [2-Co]$(BF_4)_2$ and [2-Ru]$(PF_6)_2$. (b) -SMe contacted complexes [3-Fe]$(PF_6)_2$, [3-Co]$(BF_4)_2$ and [3-Ru]$(PF_6)_2$.*

*Table 6.1: The experimental (Exp. G) and calculated (Th. G) conductances, experimental break-off distances, and calculated geometric parameters from the type III junction geometries (vide infra), for complexes [2-M]$^{2+}$ and [3-M]$^{2+}$.*

*[a] Experimentally determined conductance G (nS). [b] Calculated conductance values Th. G (nS) at $E_F - E_F^{DFT} = -0.14$ eV. [c] experimental break-off distance $Z^*$ (nm). [d] The calculated electrode separation in a relaxed type III junction, $Z = d_{Au\text{-}Au} - 0.25$ nm, where 0.25 nm is the calculated center-to-center distance of the apex atoms of the two opposing gold pyramids when conductance = $G_0$ in the absence of a molecule. [e] $d_{Au\text{-}Au}$ is the calculated center-to-center distance of the apex atoms of the two opposing gold pyramids in the relaxed type III junctions (vide infra). [f] Distance between the centres of silicon atoms in the relaxed junction. [g] Distance between centres of sulfur atoms in the relaxed junction. [h] Bond length between the top gold atoms of the pyramids and the anchor atoms in the relaxed junctions.*

| Molecule | Exp. G /nS ($G_0$)[a] | Th. G /nS ($G_0$)[b] | Z* /nm[c] | Z /nm[d] | $d_{Au\text{-}Au}$ /nm[e] | d /nm[g,h] | X /nm[h] |
|---|---|---|---|---|---|---|---|
| [2-Fe]$(PF_6)_2$ | 1.9±0.7 (($2.5±0.9)\times10^{-5}$) | 2.67 ($3.45\times10^{-5}$) | 2.2 | 2.85 | 3.10 | 2.71[f] | 0.39 |
| [2-Ru]$(PF_6)_2$ | 2.0±0.7 (($2.6±0.9)\times10^{-5}$) | 2.77 ($3.58\times10^{-5}$) | 2.4 | 2.88 | 3.13 | 2.74[f] | 0.39 |
| [2-Co]$(BF_4)_2$ | 1.4±0.6 (($1.8±0.8)\times10^{-5}$) | 1.95 ($2.51\times10^{-5}$) | 2.7 | 2.83 | 3.08 | 2.69[f] | 0.39 |
| [3-Fe]$(PF_6)_2$ | 2.4±0.6 (($3.1±0.8)\times10^{-5}$) | 3.63 ($4.69\times10^{-5}$) | 2.4 | 2.21 | 2.46 | 2.15[g] | 0.30 |
| [3-Ru]$(PF_6)_2$ | 2.4±0.6 (($3.1±0.8)\times10^{-5}$) | 3.28 ($4.23\times10^{-5}$) | 2.4 | 2.24 | 2.49 | 2.19[g] | 0.30 |
| [3-Co]$(BF_4)_2$ | 4.1±1.0 (($5.3±1.3)\times10^{-5}$) | 5.60 ($7.23\times10^{-5}$) | 2.0 | 2.20 | 2.45 | 2.14[g] | 0.30 |

## 6.3.2. Theoretical Results

In seeking to explore the effect of surrounding environment, DFT-electrochemical calculations have been performed by using two kinds of counter ions tetrafluoroborate $[BF_4]^-$ and hexafluorophosphate $[PF_6]^-$ counterions. Two $[BF_4]^-$counterions were used for the molecules containing Co centres, while Two $[PF_6]^-$ were used for the complexes containing Ru or Fe centres. Initially these counter ions were placed at different distances from the backbone, where simulations carried out with ten different distances $\chi$ between the fluorine atoms of the counterion and hydrogen atoms of the backbone, resulting in ten different relaxed distances $\chi$, as shown in figure 6.3 (illustrated by a way of an example) and table 6.2. The calculated conductances of molecular junctions, denoted Type III in five different distances $\chi$ are shown in figures 6.16 and 6.17.

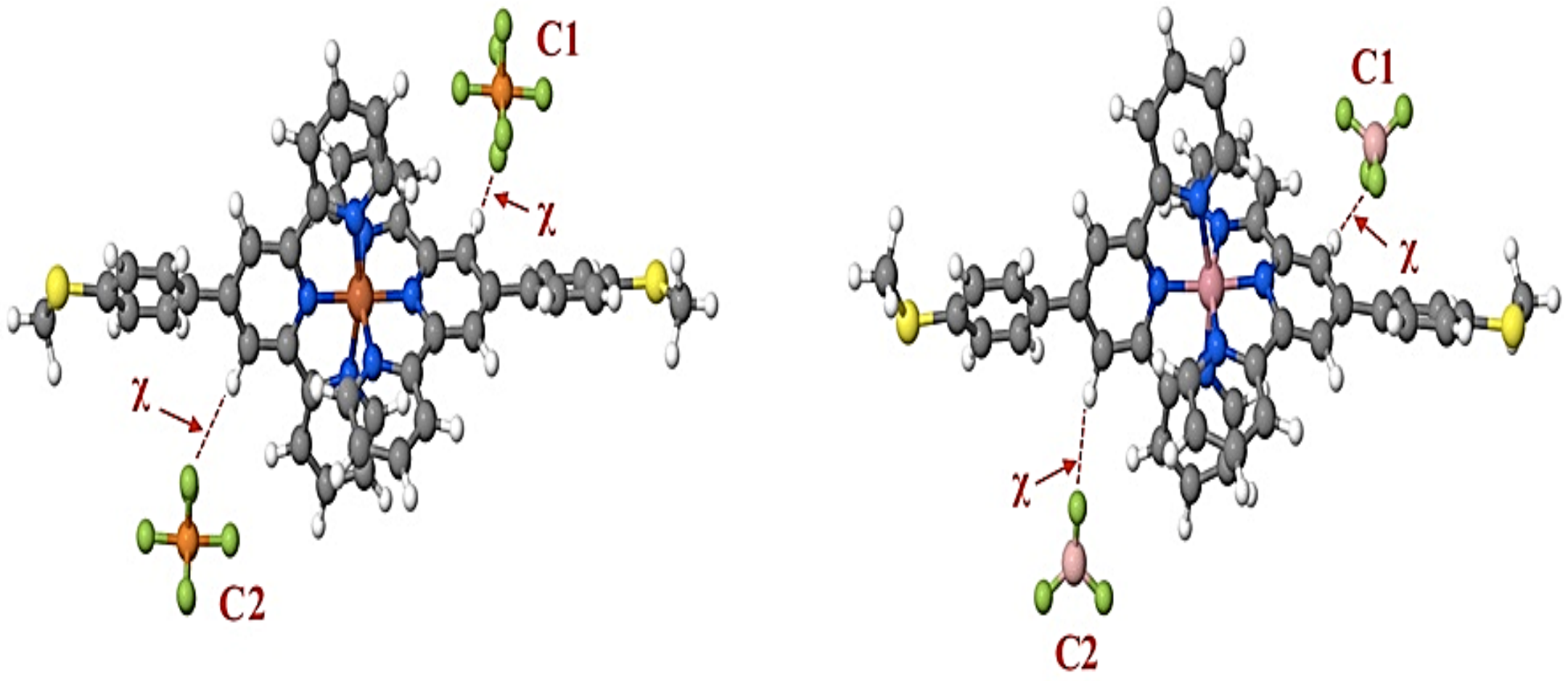

*Figure 6.3: The distance $\chi$ between fluorine atoms of counter ions and hydrogen atoms of the backbone. C1 is the counterion on one side of the backbone, and C2 is the counterion on the other side. By way of an example, this figure shows complexes continuing Fe and Co centres.*

Table 6.2 and figure 6.4 show that the number of the electrons (Γ) transferred from each molecule approaches zero for large χ and increases linearly with decreasing χ. In this region, the ordering of the electron transfer is $[Fe][PF_6]_2 \geq [Ru][PF_6]_2 > [Co][BF_4]_2$, which reflects the fact that the electrostatic field associated with $[PF_6]^-$ counterions is stronger than that of $[BF_4]^-$.

At smaller values of χ, the charge transfer Γ reaches a plateau between approximately χ = 3.4 A° and χ = 2.4 A°. At even smaller χ, the overlap between orbitals of the molecule and counter ions causes electrons to be shared between the two and reduces charge transfer at shorter distances. This chemical feature illustrates a crucial difference between electrochemical gating and the electrostatic gating.

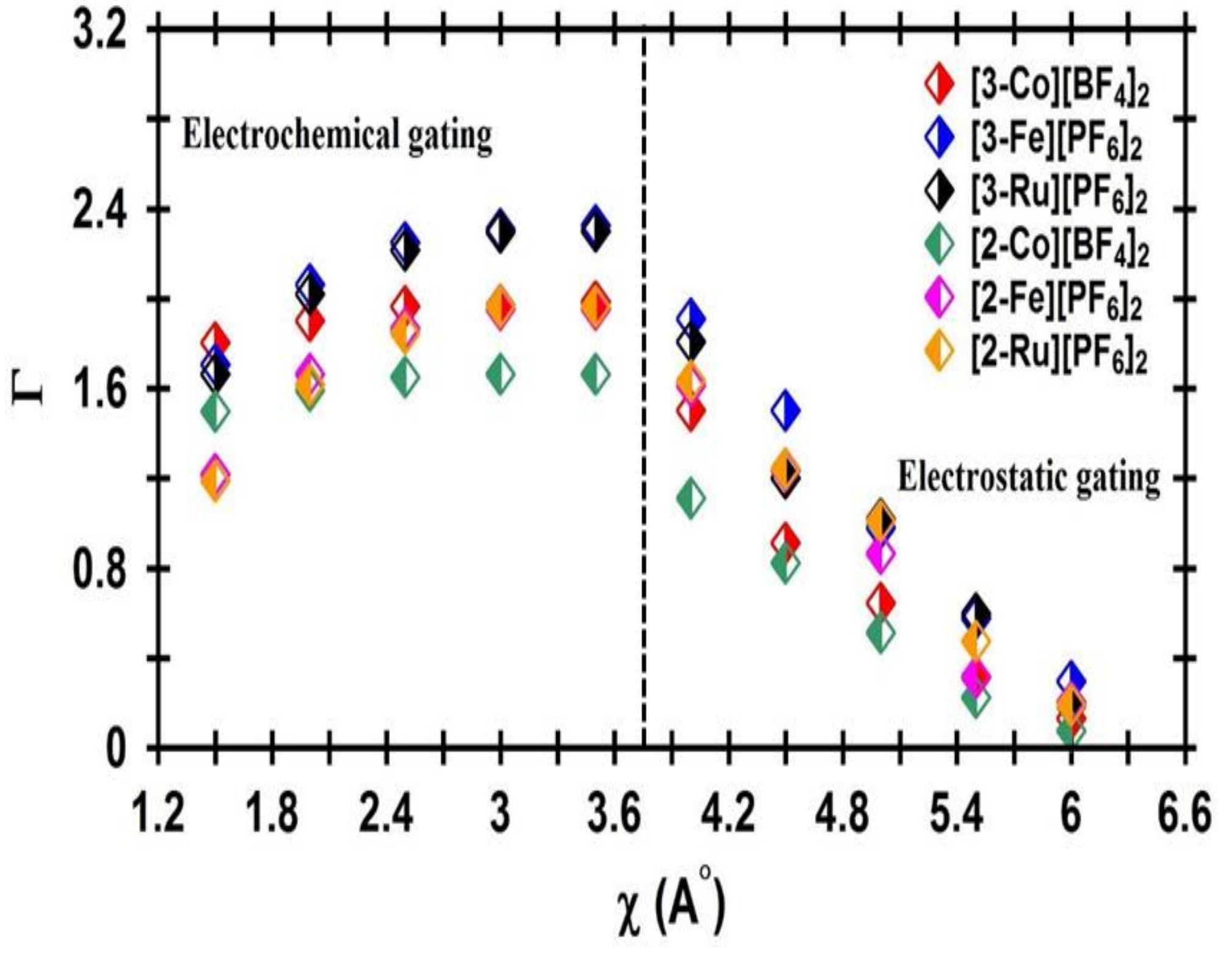

*Figure 6.4: The number of electrons (Γ) transferred from each molecule as a function of the distance (χ).*

*Table 6.2: Results of the charges on the backbone for ten different initial counterion distances $\chi$. $Q_C$ is the total charge on each backbone when it contacted to the gold electrodes in present of counterions. Q is the total charge on the backbone when it is contacted to the gold electrodes without counterions. $\Gamma = Q - Q_C$ is the number of electrons transferred from the backbone.*

| System | χ (A°) | $Q_C$ | Q | Γ |
|---|---|---|---|---|
| $[3\text{-Co}][BF_4]_2$-$\chi_1$ | 1.5 | 259.196 | 261.00 | 1.804 |
| $[3\text{-Co}][BF_4]_2$-$\chi_2$ | 2.0 | 259.099 | | 1.901 |
| $[3\text{-Co}][BF_4]_2$-$\chi_3$ | 2.5 | 259.033 | | 1.967 |
| $[3\text{-Co}][BF_4]_2$-$\chi_4$ | 3.0 | 259.032 | | 1.968 |
| $[3\text{-Co}][BF_4]_2$-$\chi_5$ | 3.5 | 259.015 | | 1.985 |
| $[3\text{-Co}][BF_4]_2$-$\chi_6$ | 4.00 | 259.999 | | 1.501 |
| $[3\text{-Co}][BF_4]_2$-$\chi_7$ | 4.50 | 260.187 | | 0.913 |
| $[3\text{-Co}][BF_4]_2$-$\chi_8$ | 5.00 | 260.355 | | 0.645 |
| $[3\text{-Co}][BF_4]_2$-$\chi_9$ | 5.50 | 260.685 | | 0.315 |
| $[3\text{-Co}][BF_4]_2$-$\chi_{10}$ | 6.00 | 260.867 | | 0.133 |
| | | | | |
| $[3\text{-Fe}][PF_6]_2$-$\chi_1$ | 1.5 | 258.292 | 260.00 | 1.708 |
| $[3\text{-Fe}][PF_6]_2$-$\chi_2$ | 2.0 | 257.936 | | 2.064 |
| $[3\text{-Fe}][PF_6]_2$-$\chi_3$ | 2.5 | 257.749 | | 2.251 |
| $[3\text{-Fe}][PF_6]_2$-$\chi_4$ | 3.0 | 257.692 | | 2.308 |
| $[3\text{-Fe}][PF_6]_2$-$\chi_5$ | 3.5 | 257.675 | | 2.325 |
| $[3\text{-Fe}][PF_6]_2$-$\chi_6$ | 4.00 | 258.289 | | 1.913 |
| $[3\text{-Fe}][PF_6]_2$-$\chi_7$ | 4.50 | 258.899 | | 1.501 |
| $[3\text{-Fe}][PF_6]_2$-$\chi_8$ | 5.00 | 259.118 | | 0.982 |
| $[3\text{-Fe}][PF_6]_2$-$\chi_9$ | 5.50 | 259.524 | | 0.576 |
| $[3\text{-Fe}][PF_6]_2$-$\chi_{10}$ | 6.00 | 259.705 | | 0.295 |
| | | | | |
| $[3\text{-Ru}][PF_6]_2$-$\chi_1$ | 1.5 | 258.338 | 260.00 | 1.662 |
| $[3\text{-Ru}][PF_6]_2$-$\chi_2$ | 2.0 | 257.981 | | 2.019 |
| $[3\text{-Ru}][PF_6]_2$-$\chi_3$ | 2.5 | 257.785 | | 2.215 |
| $[3\text{-Ru}][PF_6]_2$-$\chi_4$ | 3.0 | 257.705 | | 2.295 |
| $[3\text{-Ru}][PF_6]_2$-$\chi_5$ | 3.5 | 257.701 | | 2.299 |
| $[3\text{-Ru}][PF_6]_2$-$\chi_6$ | 4.00 | 258.191 | | 1.809 |
| $[3\text{-Ru}][PF_6]_2$-$\chi_7$ | 4.50 | 258.999 | | 1.201 |
| $[3\text{-Ru}][PF_6]_2$-$\chi_8$ | 5.00 | 259.181 | | 1.019 |
| $[3\text{-Ru}][PF_6]_2$-$\chi_9$ | 5.50 | 259.601 | | 0.599 |
| $[3\text{-Ru}][PF_6]_2$-$\chi_{10}$ | 6.00 | 259.812 | | 0.188 |
| [2-Co][BF$_4$]$_2$-$\chi_1$ | 1.5 | 299.501 | | 1.499 |
| [2-Co][BF$_4$]$_2$-$\chi_2$ | 2.0 | 299.408 | | 1.592 |
| [2-Co][BF$_4$]$_2$-$\chi_3$ | 2.5 | 299.347 | | 1.653 |
| [2-Co][BF$_4$]$_2$-$\chi_4$ | 3.0 | 299.337 | | 1.663 |
| [2-Co][BF$_4$]$_2$-$\chi_5$ | 3.5 | 299.336 | | 1.664 |
| [2-Co][BF$_4$]$_2$-$\chi_6$ | 4.00 | 299.889 | 301.00 | 1.111 |
| [2-Co][BF$_4$]$_2$-$\chi_7$ | 4.50 | 300.177 | | 0.823 |
| [2-Co][BF$_4$]$_2$-$\chi_8$ | 5.00 | 300.488 | | 0.512 |
| [2-Co][BF$_4$]$_2$-$\chi_9$ | 5.50 | 300.773 | | 0.227 |
| [2-Co][BF$_4$]$_2$-$\chi_{10}$ | 6.00 | 300.925 | | 0.075 |
| | | | | |
| [2-Fe][PF$_6$]$_2$-$\chi_1$ | 1.5 | 298.784 | | 1.216 |
| [2-Fe][PF$_6$]$_2$-$\chi_2$ | 2.0 | 298.338 | | 1.662 |
| [2-Fe][PF$_6$]$_2$-$\chi_3$ | 2.5 | 298.133 | | 1.867 |
| [2-Fe][PF$_6$]$_2$-$\chi_4$ | 3.0 | 298.049 | | 1.951 |
| [2-Fe][PF$_6$]$_2$-$\chi_5$ | 3.5 | 298.046 | 300.00 | 1.954 |
| [2-Fe][PF$_6$]$_2$-$\chi_6$ | 4.00 | 298.389 | | 1.611 |
| [2-Fe][PF$_6$]$_2$-$\chi_7$ | 4.50 | 298.767 | | 1.233 |
| [2-Fe][PF$_6$]$_2$-$\chi_8$ | 5.00 | 299.134 | | 0.866 |
| [2-Fe][PF$_6$]$_2$-$\chi_9$ | 5.50 | 299.681 | | 0.319 |
| [2-Fe][PF$_6$]$_2$-$\chi_{10}$ | 6.00 | 299.798 | | 0.202 |
| | | | | |
| [2-Ru][PF$_6$]$_2$-$\chi_1$ | 1.5 | 298.811 | | 1.189 |
| [2-Ru][PF$_6$]$_2$-$\chi_2$ | 2.0 | 298.384 | | 1.616 |
| [2-Ru][PF$_6$]$_2$-$\chi_3$ | 2.5 | 298.153 | | 1.847 |
| [2-Ru][PF$_6$]$_2$-$\chi_4$ | 3.0 | 298.034 | | 1.966 |
| [2-Ru][PF6]$_2$-$\chi_5$ | 3.5 | 298.033 | 300.00 | 1.967 |
| [2-Ru][PF$_6$]$_2$-$\chi_6$ | 4.00 | 298.365 | | 1.635 |
| [2-Ru][PF$_6$]$_2$-$\chi_7$ | 4.50 | 298.756 | | 1.244 |
| [2-Ru][PF$_6$]$_2$-$\chi_8$ | 5.00 | 298.991 | | 1.009 |
| [2-Ru][PF$_6$]$_2$-$\chi_9$ | 5.50 | 299.523 | | 0.477 |
| [2-Ru][PF$_6$]$_2$-$\chi_{10}$ | 6.00 | 299.807 | | 0.193 |

To further explore the properties of these molecular junctions, quantum chemical modeling of the junctions were undertaken to compare the electrical properties of the Fe, Ru and Co molecular pairs [2-M](X)$_2$ and [3-M](X)$_2$ (M = Fe, Ru, X = $PF_6^{-}$; M = Co, X = $BF_4^{-}$). To compute their transport properties the optimized structures (molecule and counterions) have been placed between gold electrodes grown along the (111) direction. The complexes and the first few layers of gold were allowed to relax, to yield the structures shown in figures 6.5, 6.6, and 6.7. In this study, the three distinct electrode geometries were explored, denoted Type I, Type II and Type III. The calculated conductances for each series [2-M](X)$_2$ and [3-M](X)$_2$ are shown in figures 6.11 and 6.12.

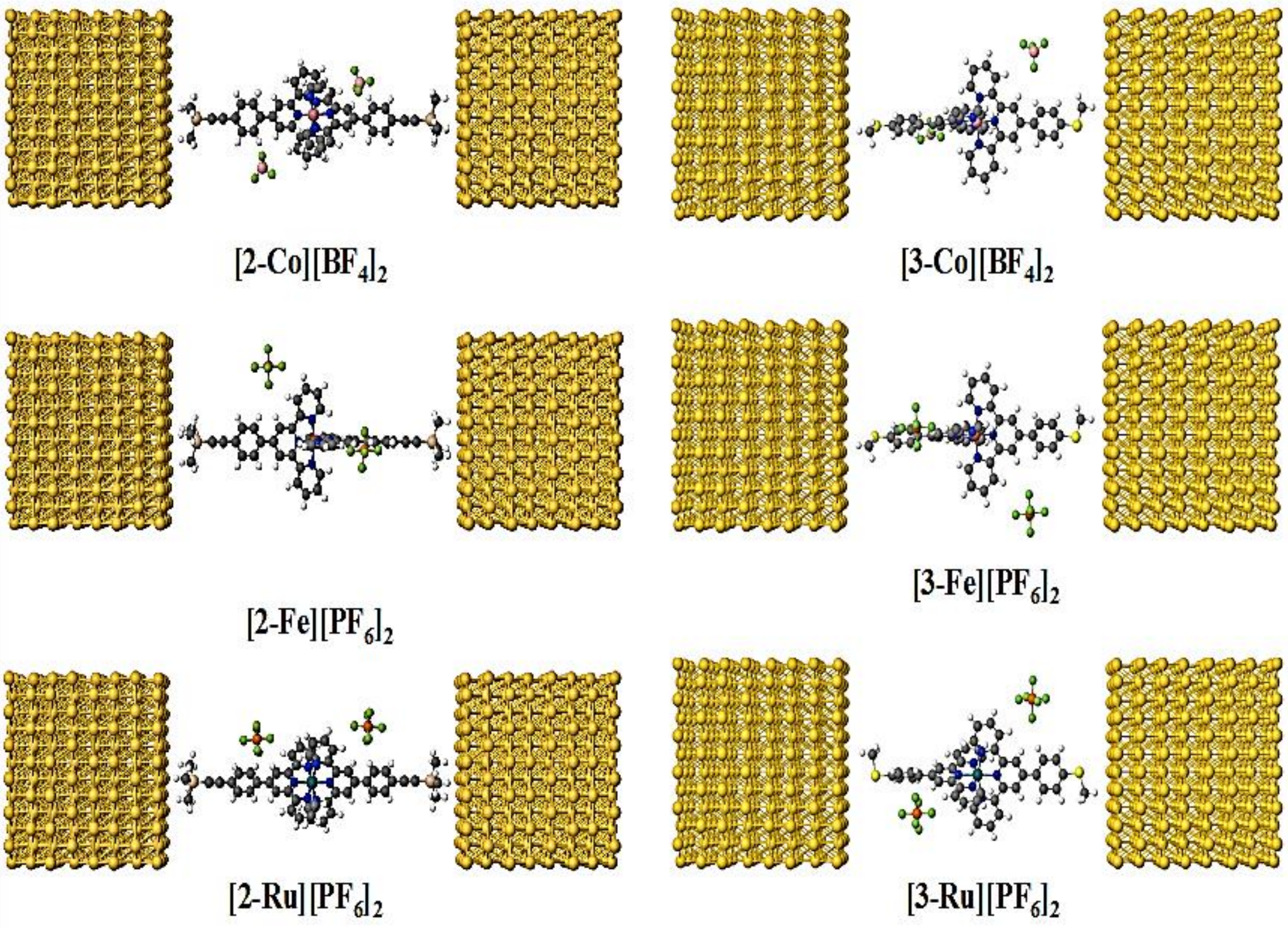

*Figure 6.5: The relaxed geometries, denoted Type I, illustrated for [2-M](X)$_2$ and [3-M](X)$_2$ (M = Fe, Ru, X = $PF_6^{-}$; M = Co, X = $BF_4^{-}$).*

Before calculating the electron transport properties, each member of the [2-M](X)$_2$ and [3-M](X)$_2$ series was optimized within the junctions. To explore the role of the electrode geometry, three electrode shapes were chosen to represent not only the idealized planar surface (Type I), but also surfaces containing a single add atom (Type II) and larger surface features modeled as gold pyramids (Type III).

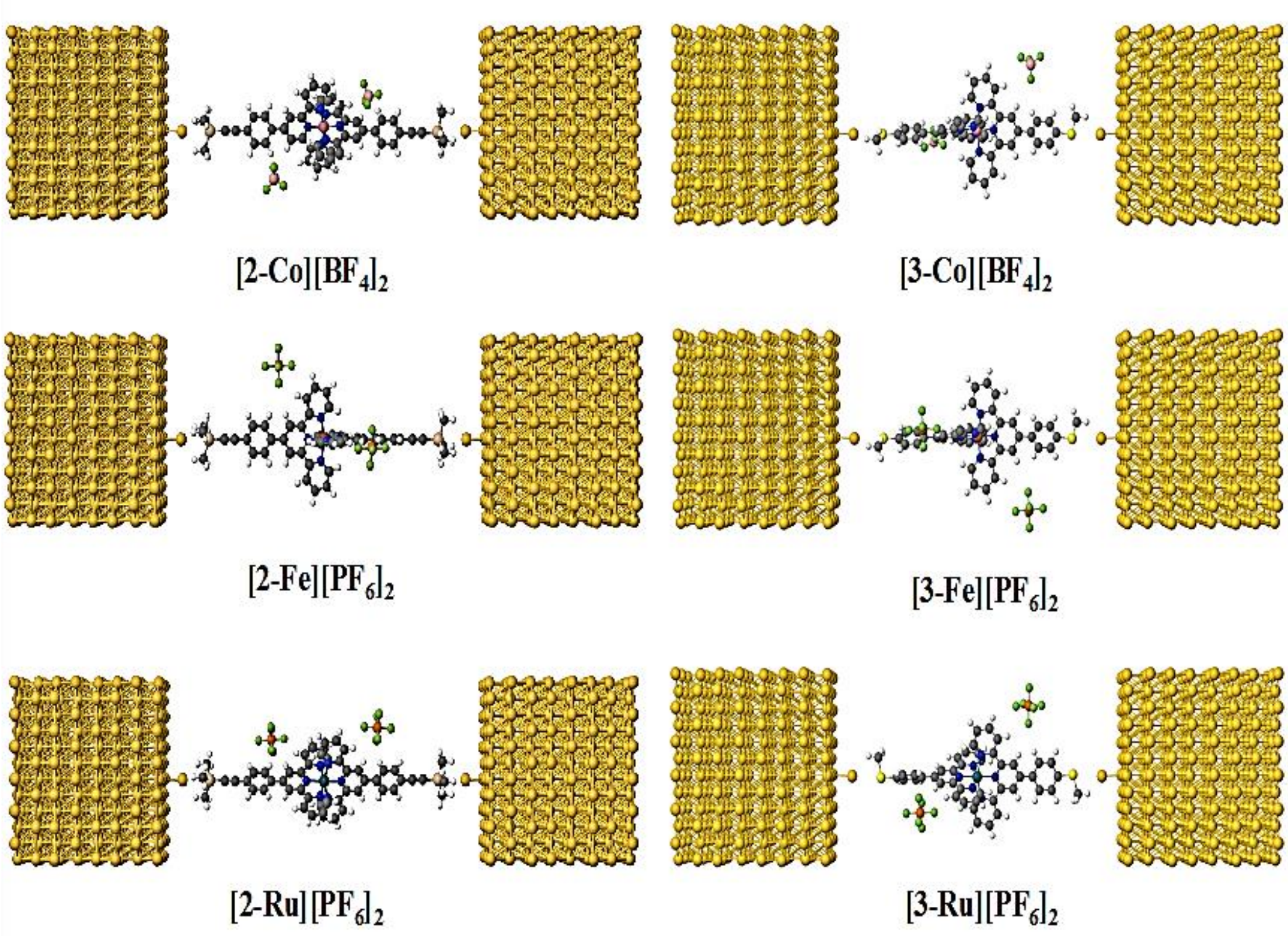

*Figure 6.6: The relaxed geometries, denoted Type II, illustrated for [2-M](X)$_2$ and [3-M](X)$_2$ (M = Fe, Ru, X = $PF_6^-$; M = Co, X = $BF_4^-$).*

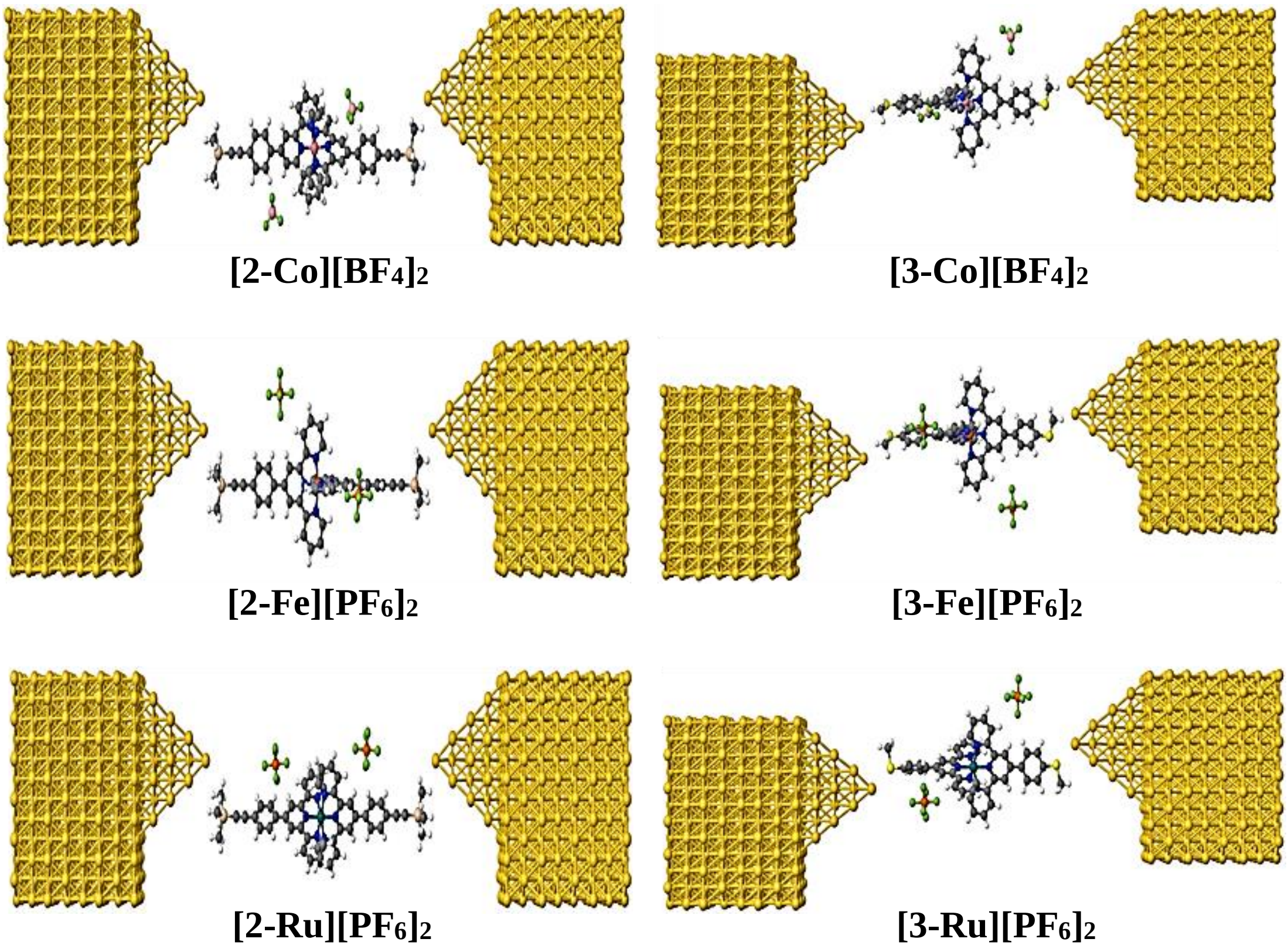

*Figure 6.7: The relaxed geometries, denoted Type III, illustrated for [2-M](X)$_2$ and [3-M](X)$_2$ (M = Fe, Ru, X = $PF_6^-$; M = Co, X = $BF_4^-$).*

To obtain realistic values of molecular conductance for the Type III junctions with a $Me_3SiC{\equiv}C$- anchor group, the binding energies of these structures were computed for a range of different molecular orientations within the junction (defined by the angle Θ, ∠$C_{ipso}$-Si-Au) as shown in figure 6.8.

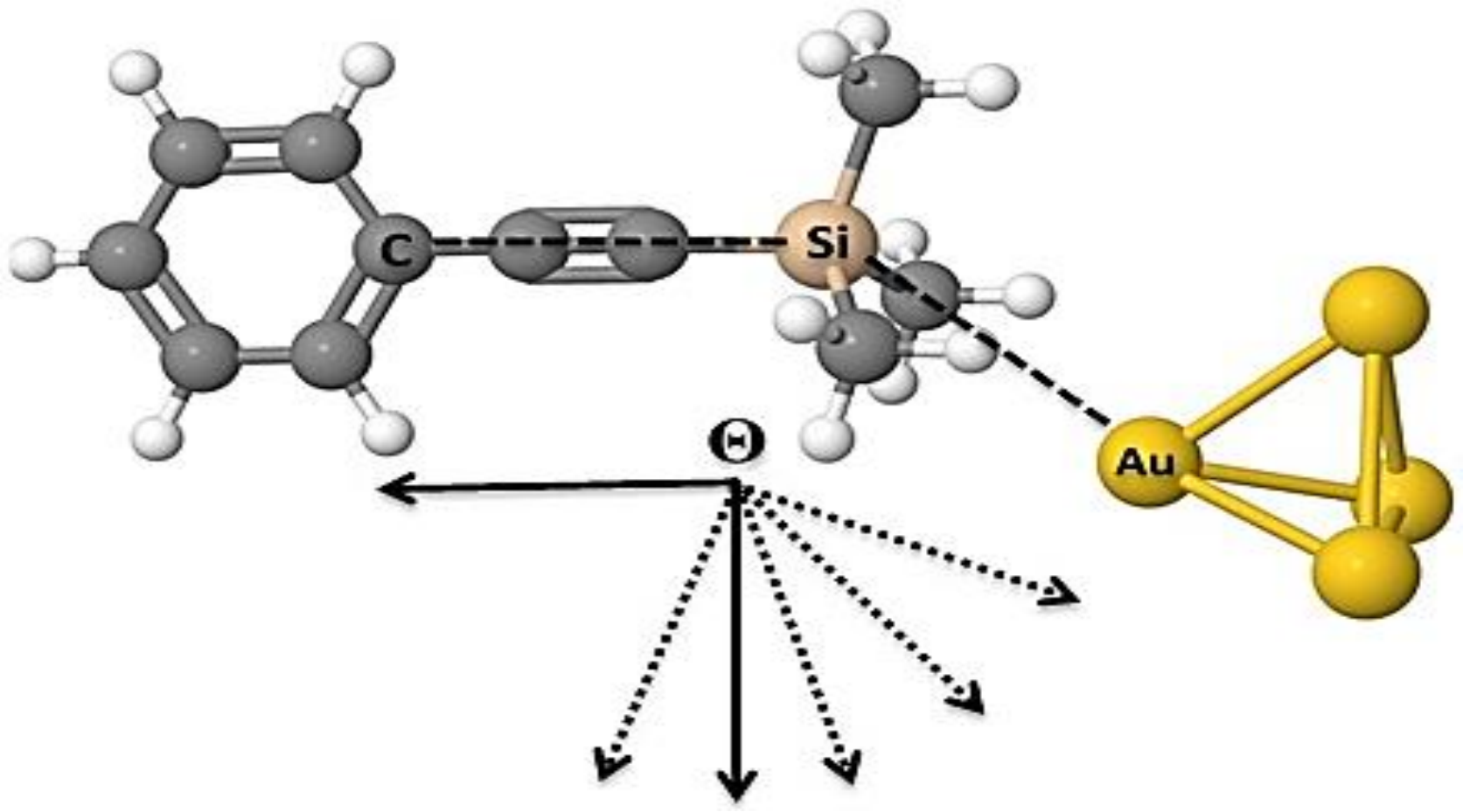

*Figure 6.8: A pictorial representation of the angle Θ used to describe the various geometries within Type III junctions for complexes [2-M]$^{2+}$.*

Within this range of different conformations of Type III junctions, the maximum binding energy varied by about 0.1 eV from 105º to 125º depending on the metal ion involved (figure 6.7). Allowing for room-temperature thermal fluctuations of ~25 meV, suggests that the optimal conformation of the angle Θ within the junction may vary from as little as 100º for [2-Co] to as much as 130º for [2-Fe] and [2-Ru].

However, since the results do not depend strongly on the angle, the plot has been chosen for the results for the case of Θ = 110º as a representation of the results for a range of nearby angles. Results for binding energies (figure 6.9) and transport properties (figures 6.13, 6.14, and 6.15) (vide infra) are shown for Type III junctions with Θ = 110º.

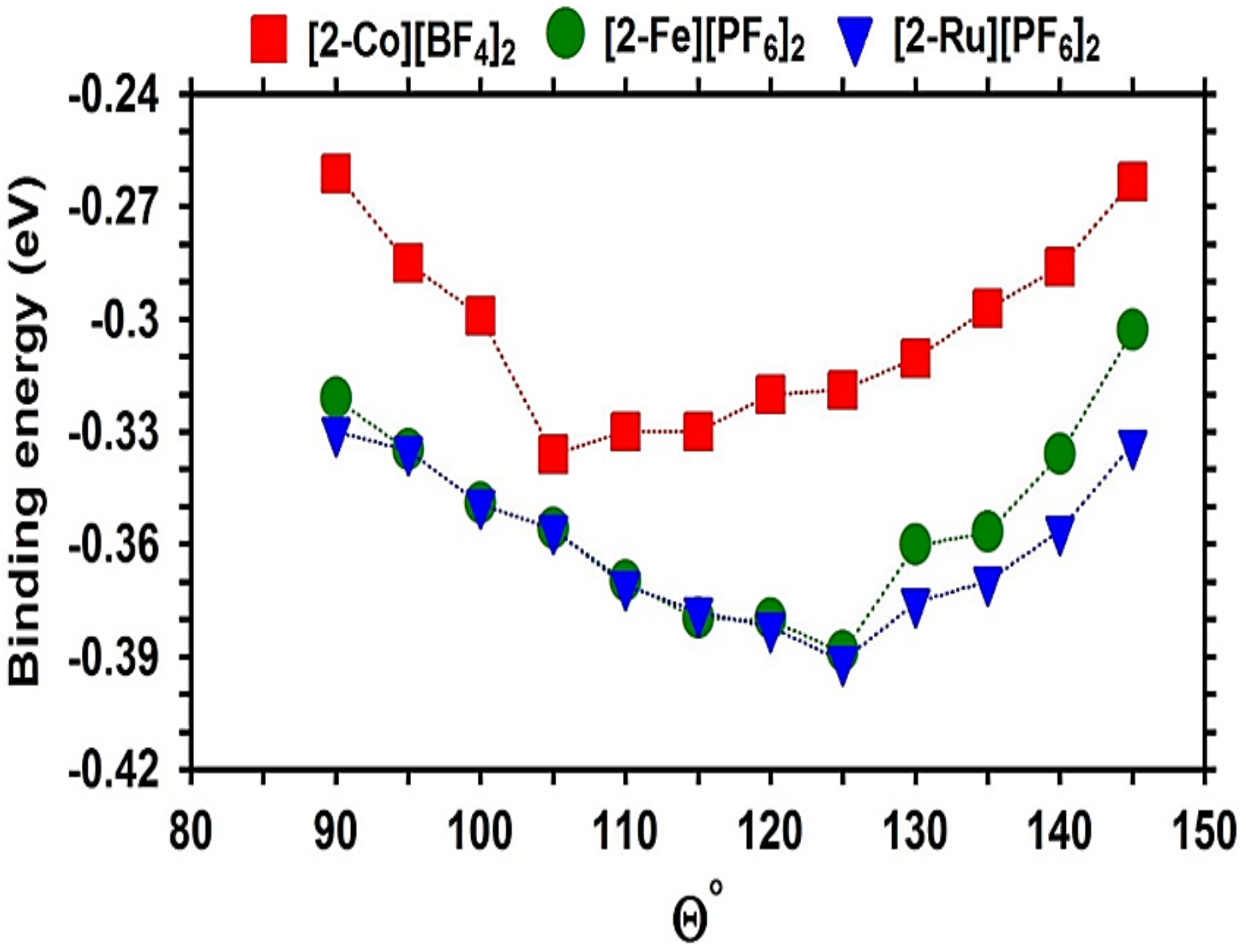

*Figure 6.9: Plots illustrating the binding energies as a function of the angle between Au-Si-C atoms, Θ, (figure 6.10) for the Type III configurations with the $Me_3SiC{\equiv}C$ anchored compounds [2-M](X)$_2$.*

Figure 6.9 shows that the minimum energy (i.e equilibrium) structures feature angles which vary between 105 and 125° for complexes [2-M](X)$_2$. A survey across these various structures shows that the calculated $G/G_0$ (figures 6.13, 6.14, and 6.15) does not depend strongly on the angle and therefore an angle of 110° has been chosen as a representative angle (lying between 105 – 125°). Plots of the conductance $G/G_0$ for angles of θ = 90, 95, 100, 105 and 110° that demonstrate this.

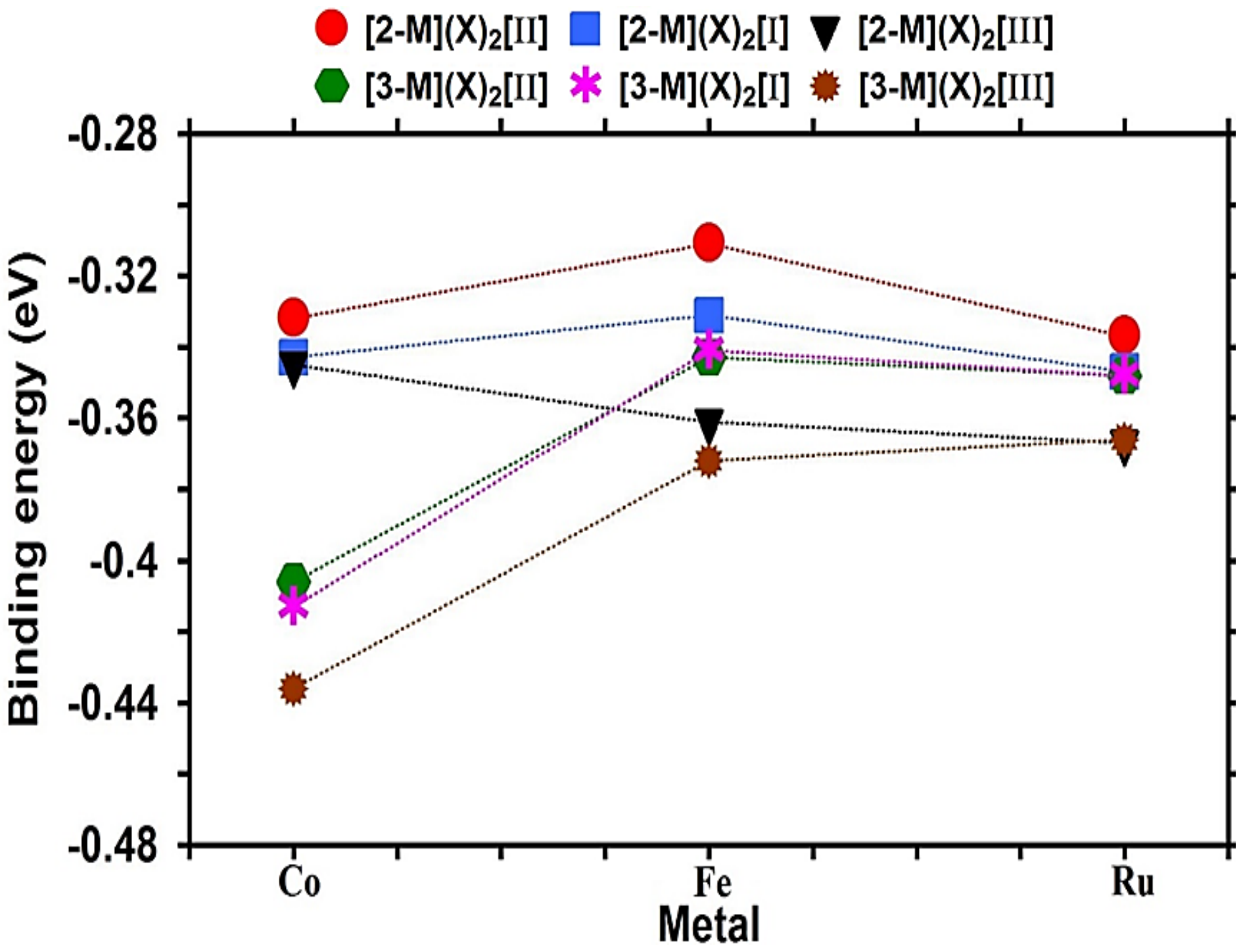

*Figure 6.10: Plots of the binding energy of [2-M](X)$_2$ and [3-M](X)$_2$ for three types of junction configurations, Type I, Type II and Type III.*

The calculated binding energies for the various complexes [2-M](X)$_2$ and [3-M](X)$_2$ in various junction models I, II, III (Θ = 110º) are plotted in figure 6.10, with two trends immediately apparent. Firstly, the MeS-based structures ([3-M]$^{2+}$) bind more strongly than the Me$_3$SiC≡C-anchored structures ([2-M]$^{2+}$). Secondly, the binding energies of the Type III electrode geometry are higher than the other junctions. For these Type III junctions, the order of the binding energies with Me$_3$SiC≡C-anchor groups ([2-M](X)$_2$) is Ru > Fe > Co, whereas with MeS-anchor groups (i.e. compounds [3-M](X)$_2$), the order of binding energies is Co > Fe > Ru, which are broadly consistent with the conductance trends (figure 6.12).

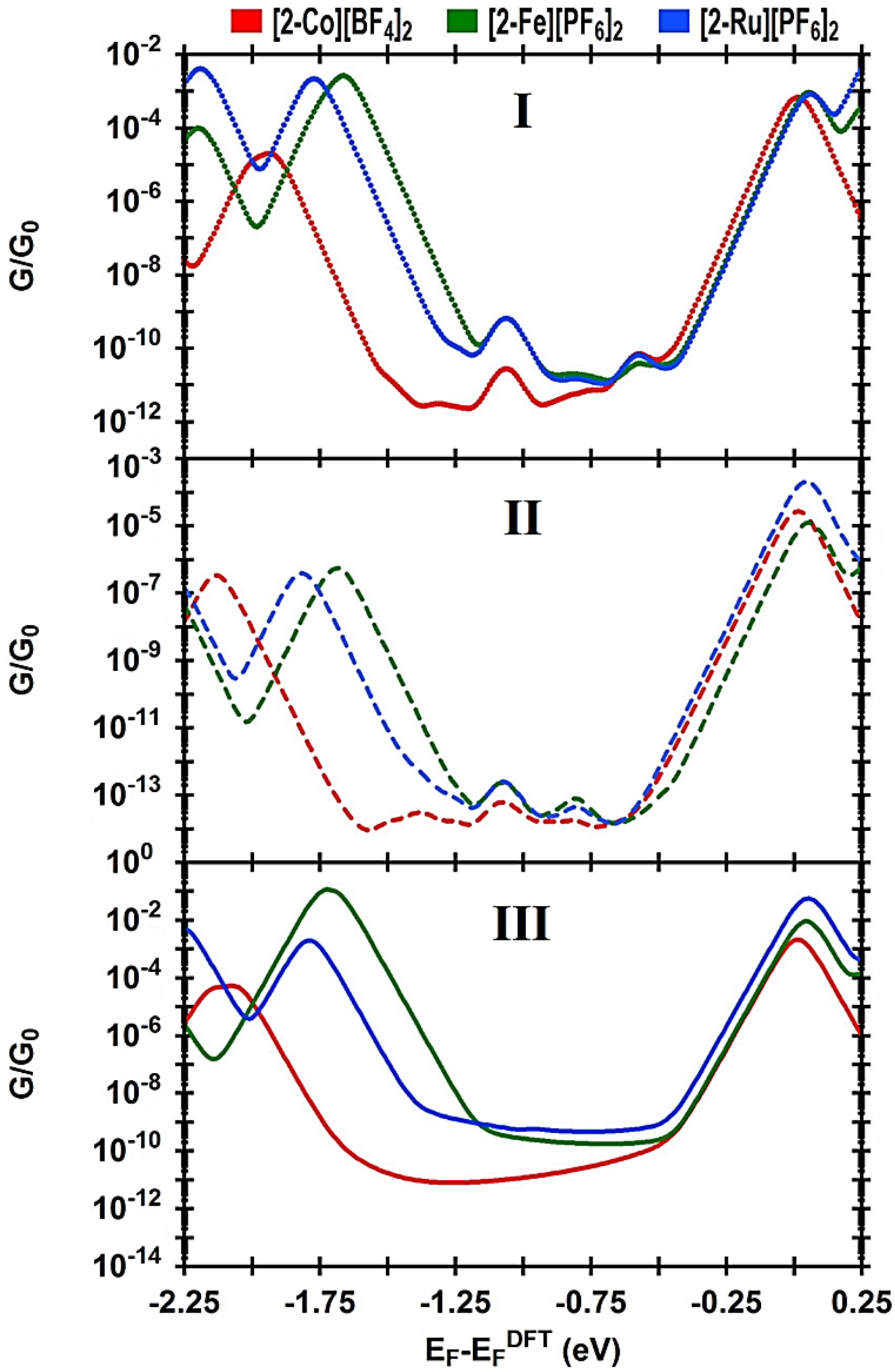

*Figure 6.11: Plots of the theoretical conductances as a function of the Fermi energy for [2-M](X)$_2$ with three different molecular junctions (Type I, II, III).*

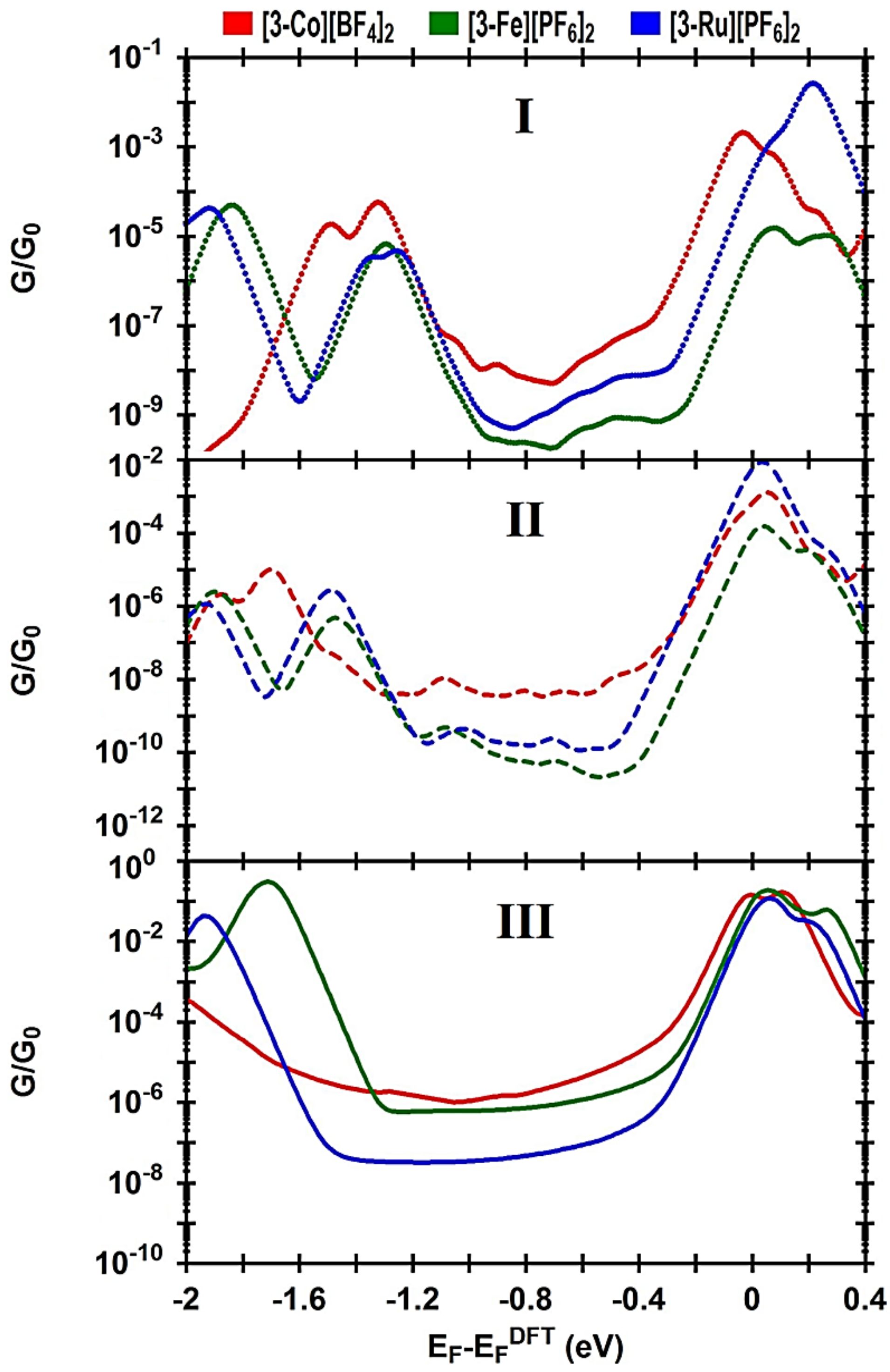

*Figure 6.12: Plots of the theoretical conductance as a function of the Fermi energy for all [3-M](X)$_2$ structures with three different molecular junctions (Type I, II, III).*

The most stable trimethylsilylethynyl-based configurations in the single-molecule junctions formed with complexes $[2\text{-M}]^{2+}$ are not, as might have been expected, based on previous proposals drawn from studies of self-assembled monolayers of trimethylsilylethynyl functionalized long-chain hydrocarbons on flat Au(111) surfaces [71, 72, 74, 86]. Rather than a five-coordinate silicon species chemisorbed to a flat terrace, the Type III junctions are most stable, and the silicon centre maintains an approximately tetrahedral geometry (table 6.3). It seems that for the single-molecule experiments, in the absence of additional dispersion forces present in the self-assembled mono-layer films, which might give additional energetic preference to alternative contact geometries, [7, 74] the most stable trimethylsilylethynyl | gold contacts are best described in terms of a molecule physisorbed at a defect site.

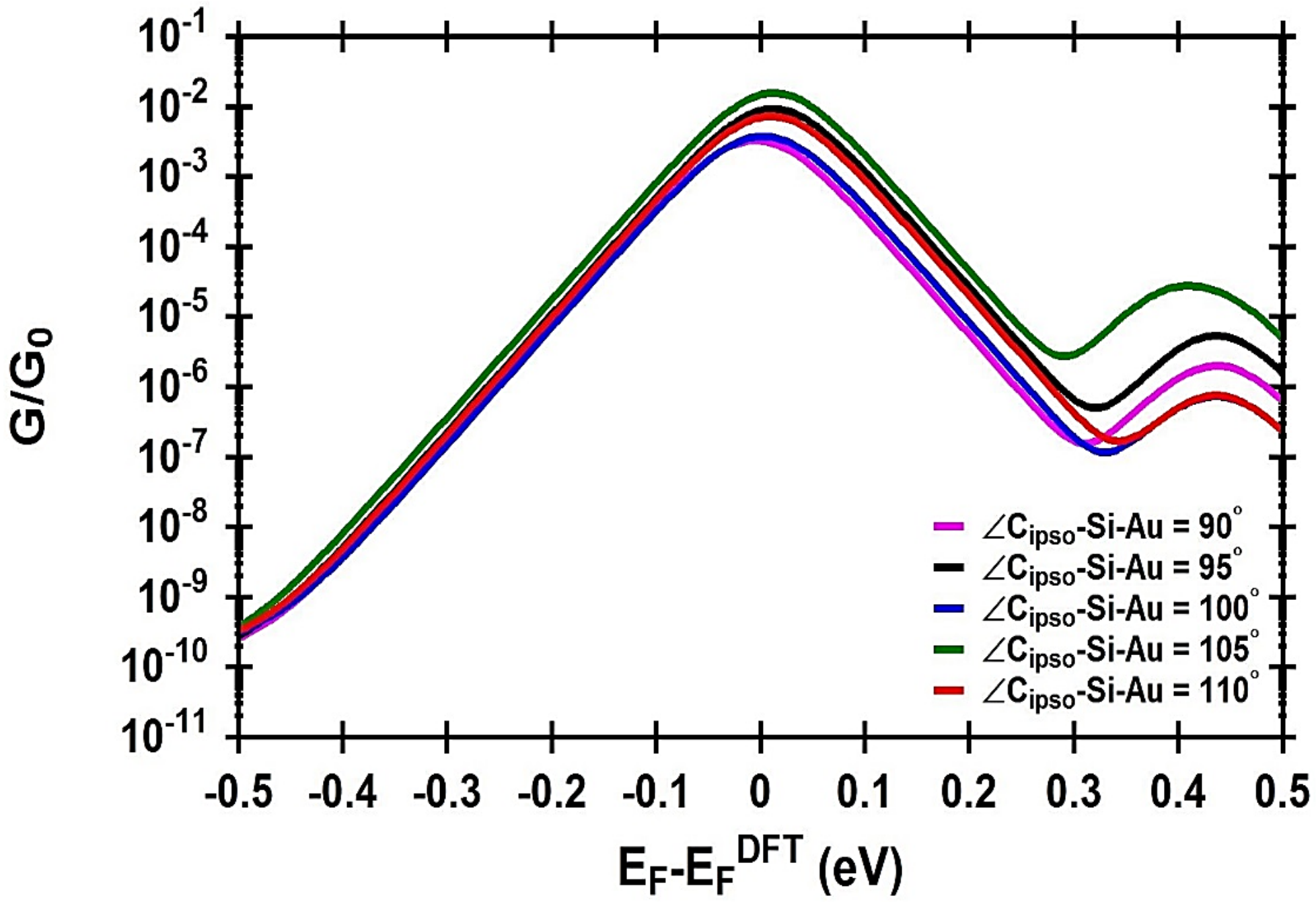

*Figure 6.13: The calculated conductance as a function of Fermi energy for [2-Co]$(BF_4)_2$ structure with $\theta$ = 90, 95, 100, 105, 110°.*

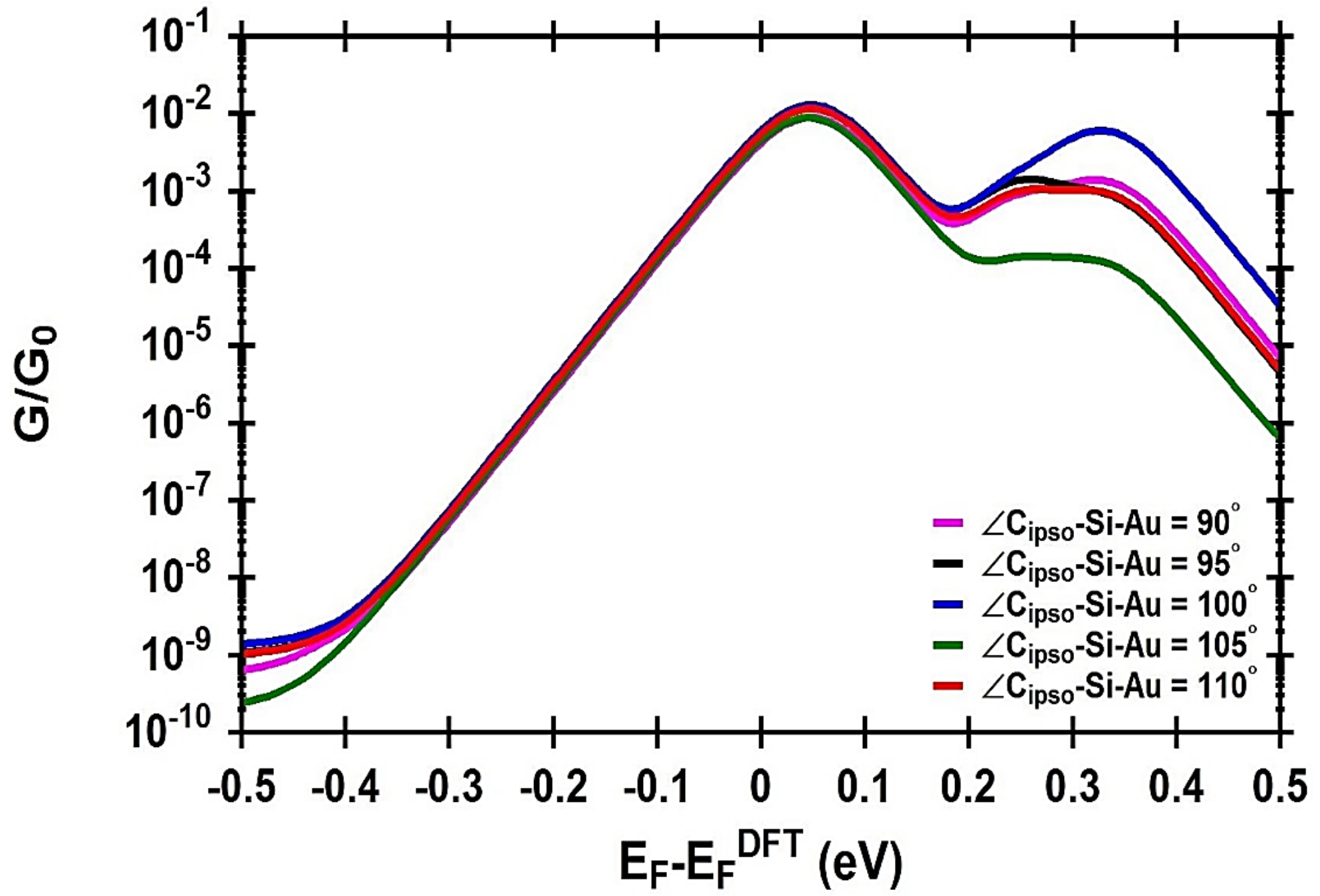

*Figures 6.14: The calculated conductance as a function of Fermi energy for [2-Fe]($PF_6$)$_2$ structure with θ = 90, 95, 100, 105, 110°.*

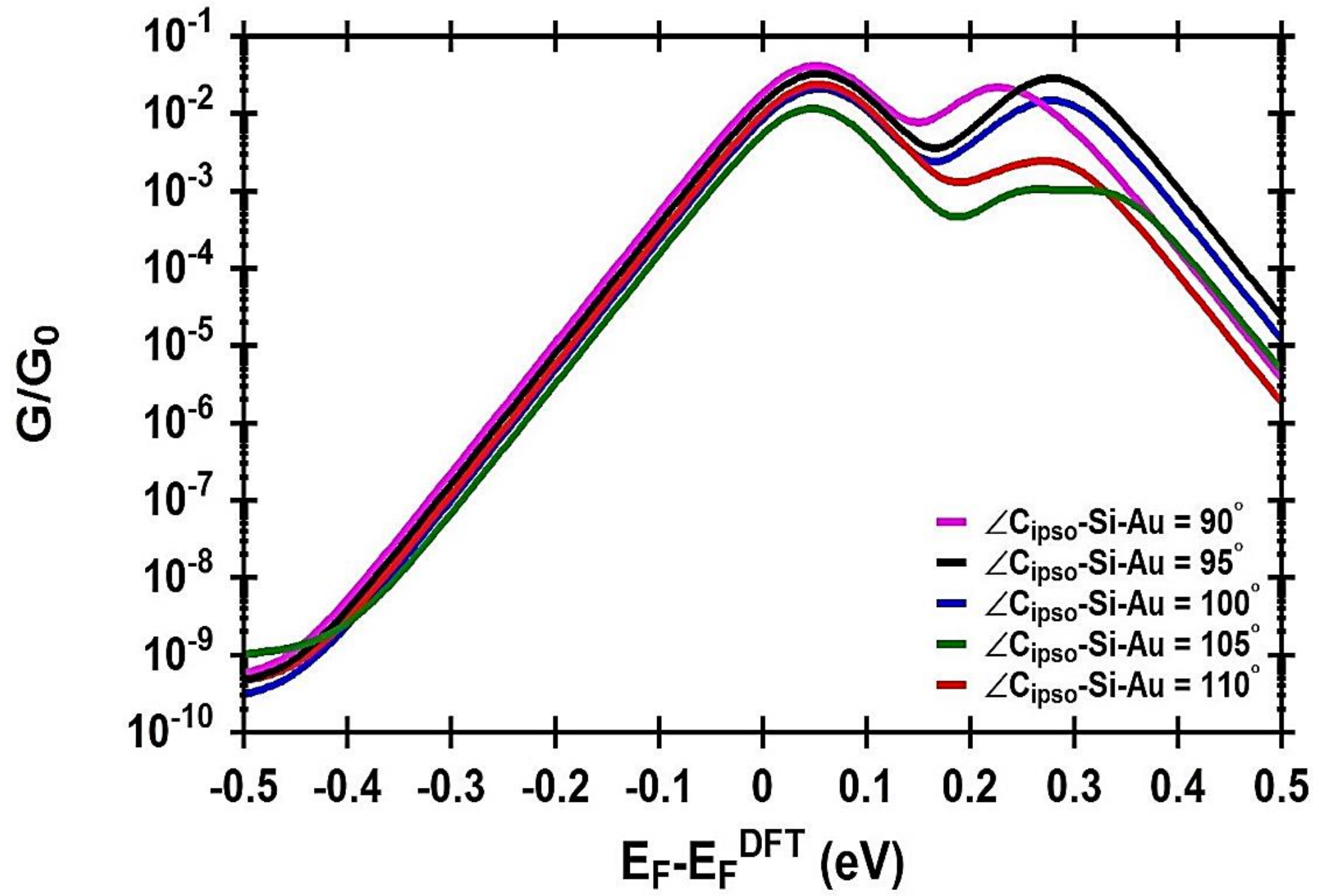

*Figure 6.15: The calculated conductance as a function of Fermi energy for [2-Ru]($PF_6$)$_2$ structure with θ = 90, 95, 100, 105, 110°.*

*Table 6.3: Summary of bond lengths (Å) and angles at Si (º) of the dication [2-M]$^{2+}$ in Type III junctions.*

| **Bond Type** | **Type III junction** | | |
|---|---|---|---|
| | [**2-Fe**]$^{2+}$ | [**2-Ru**]$^{2+}$ | [**2-Co**]$^{2+}$ |
| C≡C | 1.234 | 1.234 | 1.234 |
| Si-C≡ | 1.824 | 1.825 | 1.819 |
| Si-$C_{methyl}$ | 1.907 | 1.887 | 1.889 |
| ∠$C_{methyl}$-Si-$C_{alkyne}$ | 109.7 | 109.1 | 109.3 |

Regarding the relaxed geometries of the molecular junctions formed by the SMe contacted molecules [3-M]$(X)_2$, it has been noted that whilst the thiolate ($RS^-$) to gold interaction has been studied extensively, [87, 54 ] the thioether ($R_2S$) to gold interaction has been less thoroughly explored. In the Type III contacted thioether systems, the compounds [3-M]$^{2+}$ sit close to the apex of each pyramid-shaped model gold electrode, with a Au-S distance of 0.3 Å, and an Au-S-$C_{ipso}$ angle of 103.74º. These geometries compare with compounds such as [$Ph_3PAuSMe_2$][$CF_3SO_3$] (Au-S, 2.323(2) Å; Au-S-$C_{methyl}$ 106.7(2), 104.7(2)º), [88] and as such the sulfur-gold interaction is well approximated in terms of a coordination-type interaction (chemisorption) between the sulfur donor atom of the thio-ether and the gold atoms near the apex of the pyramid.

The calculated conductances as a function of the Fermi energy for complexes [2-M]$(X)_2$ and [3-M]$(X)_2$ within the three different molecular junctions Type I, II and III with $\Theta$ = 110º, are shown in figures 5.11 and 5.12. For both molecular contacts, it is clear the conductances of the structures with the Type III configurations are the highest, which correlates with their more favourable binding energies (figure 6.10). This is consistent with the relatively simple conductance histograms observed for trimethylsilylethynyl and SMe contacted molecules described here and elsewhere, [28, 61 – 63] and might be attributed to molecules bound at surface defects contributing predominantly to the conductance histograms.

Figures 6.16 and 6.17 show that the conductances of $[3\text{-M}](X)_2$ series are higher than that of $[2\text{-M}](X)_2$ for the structures denoted Type III as shown in figure 6.7. To some extent, there is a relationship between the distance $\chi$ and the conductance, since the conductance is raised with increasing of the distance $\chi$. In addition, the conductance order of all structures in terms of the junction type is $G_{TypeIII} > G_{TypeI} > G_{TypeII}$.

For each of the most energetically favorable Type III structures of $[2\text{-M}](X)_2$ and $[3\text{-M}](X)_2$, the room temperature electrical conductance G was calculated as described in the computational methods section, and plotted against the Fermi level ($E_F$-$E_F^{DFT}$) (figures 5.11 and 5.12). Since the DFT-predicted value of the Fermi energy is not usually reliable it has been treated as a free parameter, and it has been adjusted to improve the agreement with experiment.

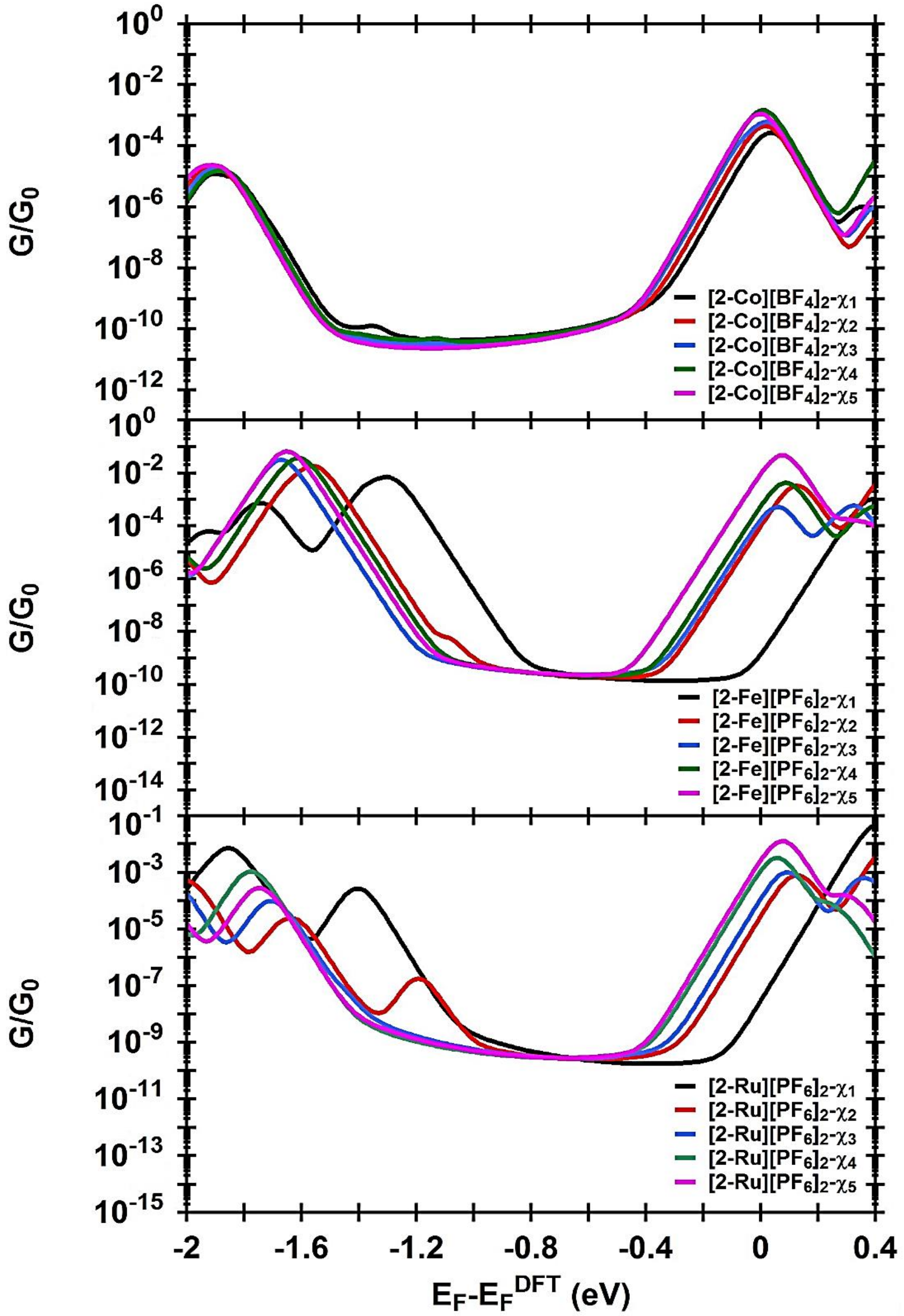

*Figure 6.16: The calculated conductances as a function of the Fermi energy for all [2-M](X)$_2$ junctions (Type III) in five different $\chi$.*

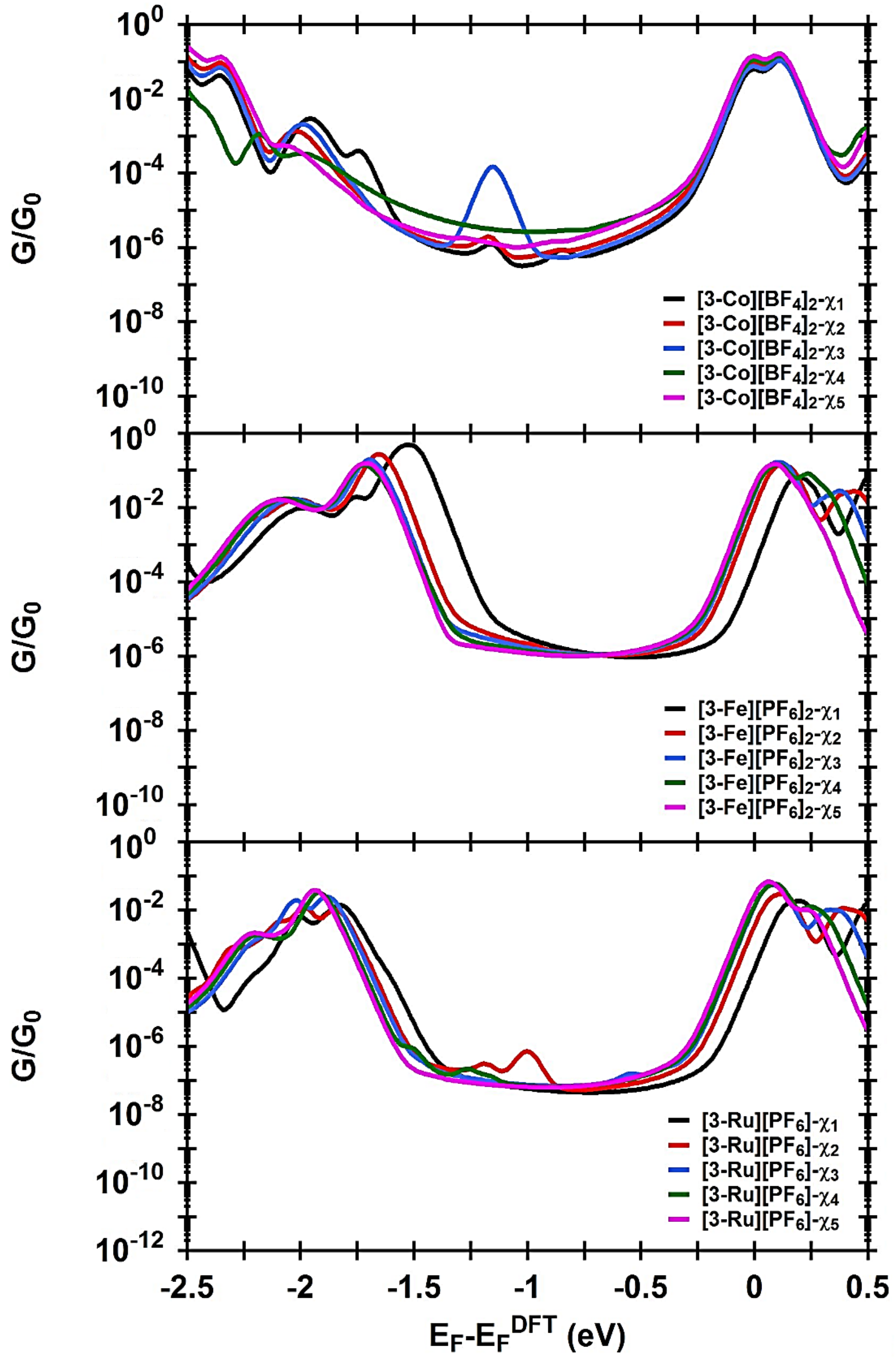

*Figure 6.17: The calculated conductance as a function of the Fermi energy for all [3-M](X)$_2$ junctions (Type III) in five different $\chi$.*

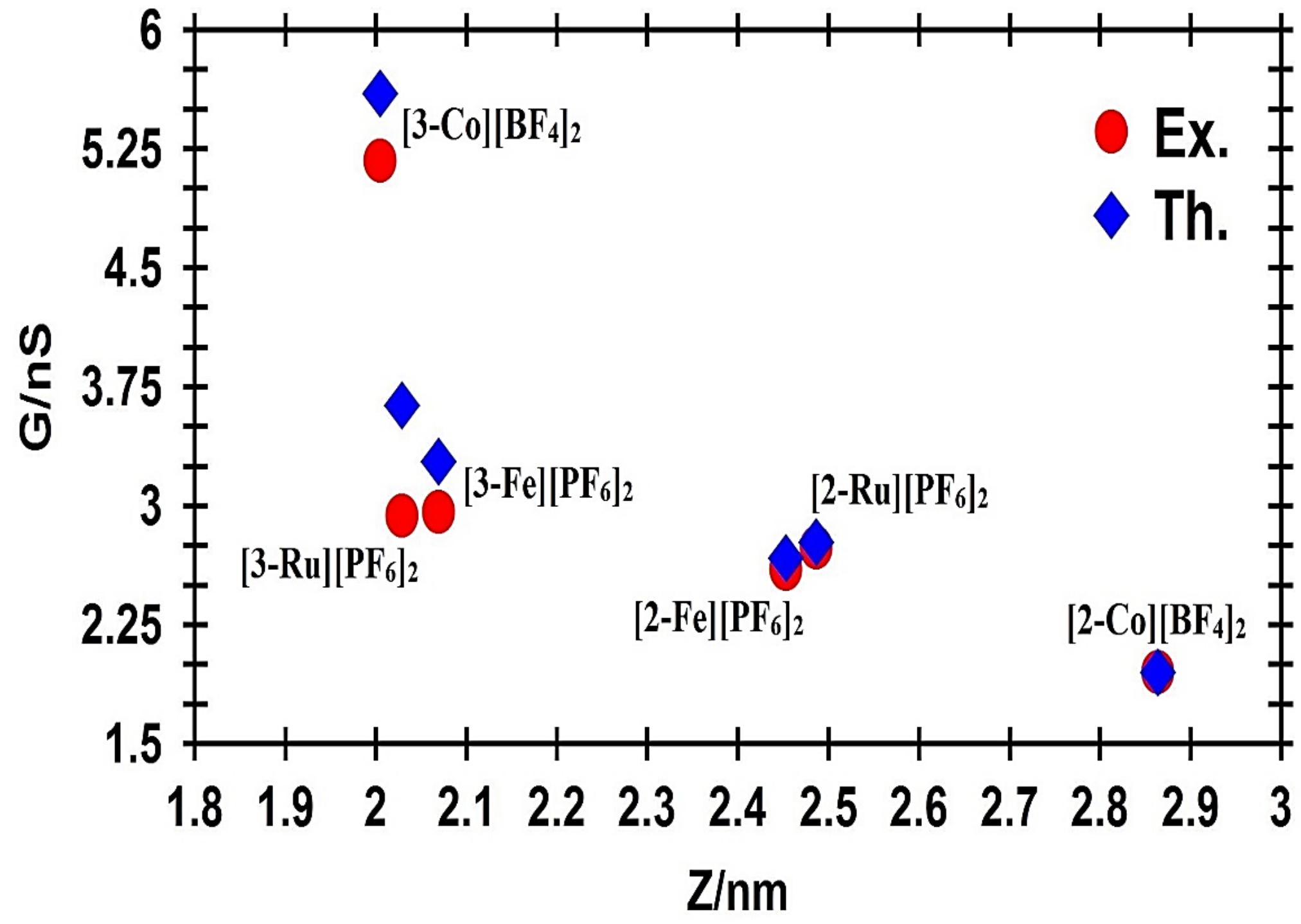

*Figure 6.18: A comparison between experimental and theoretical conductances G, plotted against the relaxed electrode separations Z as shown in table 5.1.*

By adjusting this single parameter, better agreement with six experimental conductance values is achieved, which gives us confidence that this is a reasonable procedure [80]. Given that the adjustment of -0.14 eV is rather small in this case, it could be concluded that the DFT-predicted value is actually rather close to the experimental value, and both experimental and computational results are close to the LUMO. For $E_F - E_F^{DFT} = -0.14$ eV, the conductances follow the small experimental trends with remarkably high degree of correlation (table 6.1 and figure 6.18). Thus, despite the changes in metal and surface contacting group, the observed conductances of these metal complexes, which span a relatively small range of values (from (1.4 ± 0.6) – (4.1 ± 1.0) nS, table 6.1), follow the trends in the edge of the LUMO resonances (figures 6.11, 6.12). In contrast, there is a more pronounced variation in the position of the HOMO resonances, but such variations are not consistent with the conductance data. LUMO-based conduction mechanisms have also recently been noted for a somewhat related Cu(phenanthroline) complex [41]. Interestingly, a family of SMe contacted oligo(thiophene-*S*,*S*-dioxides) have been shown to shift from LUMO to HOMO mediated conductance mechanisms as a function of increasing molecular length [62].

# 6.4. Summary

In this Chapter the electronic properties of bis(terpyridyl) complexes have been investigated experimentally and theoretically. The single-molecule conductance of bis(terpyridyl) complexes [2-M]$(X)_2$ and [3-M]$(X)_2$ display trends that are more closely associated with the binding energy and ligand structure than the nature of the metal ion.

The calculated conductances as a function of the Fermi energy for complexes series within the three different molecular junctions Type I, II and III with $\Theta = 110^{o}$ illustrated that for both molecular contacts, it is clear the conductance of the structures with the Type III configurations are the highest, which correlates with their more favourable binding energies. The limited role of the metal ion can be traced in part to the LUMO-based conduction mechanisms that arise from the use of trimethylsilylethynyl and, in the geometries adopted here, thiomethyl binding groups.

The optimised structures of the molecular junctions and considerations of the calculated conductance profiles indicate that the most conductive trimethylsilylethynyl contact to the gold electrodes is best described in terms of physisorption at defect sites, explaining the simple conductance profiles observed for compounds contacted through this group. The thiomethyl moiety contacts the gold electrodes in a more chemisorbed fashion, again at defect sites on the gold electrodes in the most conductive junctions. The conductances of [2-M]$(X)_2$ are found to be lower than those of [3-M]$(X)_2$, which is consistent with single phenylene and triple-bond spacers acting as tunnelling barriers.

Electrochemical computations clarified a crucial difference between electrochemical gating and the electrostatic gating.

# Chapter 7

# Experimental and computational studies of the single molecule conductance of Ru(II) and Pt(II) trans-bis(acetylide) complexes

## 7.1. Introduction

Measurements of the electrical characteristics of a wide variety of saturated, conjugated and redox active organic compounds have served to drive the development of concepts and techniques in molecular electronics [1 – 3]. However, metal complexes offer several potential advantages over organic compounds as components in molecular electronic devices, including redox activity at moderate potentials, ready tuning of frontier molecular orbital energy levels to better match the Fermi levels of metallic electrodes and magnetic properties [4, 5]. Consequently, attention has been turned to the construction and study of metal complexes [6 – 14], clusters [15 – 17], extended metal atom chains [18 – 20], and organometallic acetylide species [21 – 33] within molecular junctions.

In the case of purely organic oligo(aryleneethynylene)-based compounds with pyridyl contacting groups, the molecular conductance, as determined by single molecule STM break junction (STM-BJ) experiments, decreases with length, initially in line with the exponential decay expected for a tunneling mechanism before shifting to a shallower length dependence more indicative of an incoherent hopping mechanism of charge transport for compounds of ca. 3 nm in length [34].

Conductance values range from $10^{-4.5}G_0$ (2.45 nS) for the 1.6 nm long ‘3-ring’ oligoarylenes $NH_4C_5C{\equiv}CC_6H_2R_2C{\equiv}CC_5H_4N$ ($R{=}OC_6H_{13}$) decreasing by approximately three orders of magnitude for the 3.0 nm long ‘5-ring’ system $NH_4C_5C{\equiv}C(C_6H_2R_2C{\equiv}C)_3C_5H_4N$ ($10^{-6.7}G_0$, 0.015 nS), and thereafter falling only slightly to $10^{-6.9}G_0$ (0.01 nS) in an analogous 5.8 nm long ‘9-ring’ system.

In cases where direct comparison is possible, it has generally been found that the incorporation of a ruthenium metal centre such as $Ru(dppm)_2$ [33] or $Ru(dppe)_2$ [28] within a π-conjugated wire-like structure leads to a 2 – 5 fold increase in conductance with the conductance value measured likely also being dependent on the nature of the molecule-electrode contacting group (e.g. *trans*-$Ru(C{\equiv}CC_6H_4SAc)_2(dppm)_2$ STM break junction 19±7 nS [33]; *trans*-$Ru(C{\equiv}CC_6H_4C{\equiv}CSiMe_3)_2(dppe)_2$ *I*(*s*) method ($5.1{\pm}0.99{\times}10^{-5}G_0$ / 3.9±0.8 nS) [28]; *trans*-$Ru(C{\equiv}C\text{-}4\text{-}C_5H_4N)_2(dppe)_2$ STM-BJ $(2.5{\pm}0.4){\times}10^{-4}G_0$ / 19±3 nS. In contrast, earlier studies have shown that the Pt(II) complex *trans*-$Pt(C{\equiv}CC_6H_4SAc)_2(PPh_3)_2$ behaves rather more as an insulating species when bound within a mechanically controlled break junction (MCBJ), with resistances (5 – 50 GΩ; 0.2 – 0.02 nS) some three orders of magnitude larger than comparable organic compounds $AcSC_6H_4C{\equiv}CArC{\equiv}CC_6H_4SAc$ (Ar = 9,10-$C_{14}H_8$, 1,4-$C_6H_2$-2-$NH_2$-5-$NO_2$) being reported [21]. A later study with a range of *trans*-$Pt(C{\equiv}CC_6H_4SAc)_2(PR_3)_2$ complexes (R = Cy, Ph, OEt) revealed little effect of the supporting phosphine or phosphite ligand on the through-molecule conductance, although curiously the conductance for these Pt complexes measured in a cross-wire junction was reported to be some 2 – 3 fold greater than that of the simple oligo(phenyleneethynylene) $AcSC_6H_4C{\equiv}CC_6H_4C{\equiv}CC_6H_4SAc$ [22].

This chapter turns the attention to a family of linearly-conjugated, wire-like organometallic complexes featuring *trans*-$Ru(C{\equiv}CR)_2(dppe)_2$ and $Pt(C{\equiv}CR)_2(PPh_3)_2$ moieties embedded within the oligo(aryleneethynylene) backbone of ca. 3 nm molecular length, and describes the results of single molecule conductance studies based on the $I(s)$ method. These metal complexes are substantially more conductive than their purely organic analogs of comparable molecular length, with detailed computational investigation indicating that the enhanced conductance arises from conductance through the tails of the LUMO resonances. The conductance values obtained from the Pt and Ru systems are remarkably similar, suggesting that the readily-synthesized platinum complexes may have an important role to play in the further development of metal complexes for applications in single molecule electronics.

This chapter presents all theoretical details and experimental conductance measurements as a part of a published paper. For more details regarding the experimental methods and synthesis details see *Oday A. Al-Owaedi, David C. Milan, Marie-Christine Oerthel, Sören Bock, Dmitry S. Yufit, Judith A. K. Howard, Simon J. Higgins, Richard J. Nichols, Colin J. Lambert, Martin R. Bryce, Paul J. Low. Experimental and computational studies of the single molecule conductance of Ru(II) and Pt(II) trans-bis(acetylide) complexes. Organometallics, 2016, 35(17): p. 2955 – 2954.*

# 7.2. Experimental and Theoretical Methods

## 7.2.1 Experimental Methods

Conductance values of the compounds and the break-off distances were obtained using the scanning tunnelling microscopy, (current-distance ($I(s)$) technique). All details of experimental methods are presented in ref. [35].

## 7.2.2. Theoretical Methods

To avoid the duplication, the same computational methods that were used to relax the molecules (in a gas phase) in chapter 5, were utilized in this work as well.

To provide further insight into the experimentally observed trends obtained using the $I(s)$ technique, and to better evaluate the properties and behavior of these molecular junctions, calculations using a combination of DFT and a non-equilibrium Green's function formalism were also carried out. For the transport calculations, eight layers of (111)-oriented bulk gold with each layer consisting of 6×6 atoms and a layer spacing of 0.235 nm were used to create the molecular junctions as shown in figure 7.5, and described in detail elsewhere [36]. These layers were then further repeated to yield infinitely-long current-carrying gold electrodes. Each molecule was attached to two (111) directed gold electrodes; one of these electrodes is pyramidal, representing the STM tip, while the other is a planar slab representing the electrode formed by the idealized Au(111) substrate in the $I(s)$-based molecular junction. The molecules and first layers of gold atoms within each electrode were then allowed to relax, to yield the optimal junction geometries shown in figure 7.5. From these model junctions the transmission coefficient, $T(E)$, was calculated by first obtaining the corresponding Hamiltonian and overlap matrices using SIESTA [37, 38] and then using the GOLLUM code [36]. To determine $E_F$, we compared the predicted values of all molecules with the experimental values and chose a single common value of $E_F$ which gave the closest overall agreement. This yielded a value of $E_F - E_F^{DFT} = -0.07\ eV$, which is used in all theoretical results.

# 7.3. Results and Discussion

Single-molecule measurements using both organic and organometallic compounds have clearly shown that the electronic properties of the prototypical metal | molecule | metal junctions are strongly influenced by not only the chemical structure of the molecular backbone, but are also critically dependent on the combination of the surface and contacting groups [39 – 45]. The pyridyl-terminated compounds 1-Ru and 1-Pt together with the analogous methyl thioether-terminated compounds 2-Ru and 2-Pt were chosen to explore both the relative effects of the Ru(dppe) vs $Pt(PPh_3)_2$ fragments on molecular conductance, and the influence of the electrode-molecule contact in a comparable set of compounds. The pyridyl and methyl thioether moieties are already established as surface contacting groups in single-molecule studies of oligoynes and oligo(phenyleneethynylenes) [9, 34, 39, 46 – 49].

From analysis of the conductance traces (figure 7.1), break-off distances of 3.1 nm (1-Ru) and 3.0 nm (1-Pt) can be determined (table 1). These values are in a good agreement with the calculated molecular lengths (figure 7.5), which are consistent with the contact of these molecules being almost normal to the electrode surface via the pyridine lone pair. In contrast, shorter break off distances are determined for the methyl thioether complexes 2-Ru (2.4 nm) and 2-Pt (2.5 nm), which is compatible with rather more tilted arrangements of the molecule in the junction as might be expected from the geometry of the sulfur lone pairs in the thioether [50]; this interpretation has been supported by studies of the DFT-optimized junctions described in more detail below.

*Table 7.1. The frontier orbital energies (eV), experimental (Exp. $G/G_0$) and calculated conductances (Th. $G/G_0$) at $E_F$ - $E_F^{DFT}$ = – 0.07 eV, experimental 95$^{th}$ percentile break-off distance $Z^*$ (nm), molecular length from the DFT-optimized junctions $L = d_{r \ldots r}$ (nm), where r = N or S atoms, bond length between the top gold atoms of gold electrodes and the anchor atoms in the relaxed junctions, X (nm).*

| Molecule | $E_{HOMO}$ (eV) | $E_{LUMO}$ (eV) | Exp. $G/G_0$ | Th. $G/G_0$ | $Z^*$ (nm) | L (nm) | X (nm) | Contacting Group |
|---|---|---|---|---|---|---|---|---|
| 1-Ru | -4.42 | -1.46 | $4.5\times10^{-6}$ | $5.4\times10^{-6}$ | 3.1 | 2.9 | 0.23 | $4\text{-}C_5H_4N$ |
| 1-Pt | -4.69 | -1.48 | $9.8\times10^{-6}$ | $8.7\times10^{-6}$ | 3.0 | 2.86 | 0.23 | $4\text{-}C_5H_4N$ |
| 2-Ru | -4.18 | -1.07 | $1.8\times10^{-5}$ | $1.8\times10^{-5}$ | 2.4 | 2.65 | 0.245 | $4\text{-}C_6H_4SMe$ |
| 2-Pt | -4.40 | -1.12 | $1.8\times10^{-5}$ | $1.78\times10^{-5}$ | 2.5 | 2.68 | 0.245 | $4\text{-}C_6H_4SMe$ |

The conductance histograms constructed from 500 molecular junction formation traces with characteristic plateaus are shown in figures 7.1 and 7.2. The peak conductance values from these histograms together with key data are summarized in table 7.1. These conductance histograms reveal pronounced conductance peaks at 0.4±0.18 nS (1-Ru), 0.8±0.5 nS (1-Pt), 1.4 ±0.4 nS (2-Ru) and 1.8± 0.6 nS (2-Pt), and within each pair of compounds featuring the same contacting group these values are indistinguishable. The two- to four-fold increase in conductance values of 2-Ru and 2-Pt compared with 1-Ru and 1-Pt further indicates the important role of the contacting group in the electrical response of the junction.

However, in contrast to the thiolate-contacted molecules derived from *trans*-$Ru(C{\equiv}CC_6H_4SAc)_2(dppm)_2$ (STMBJ) [33] and *trans*-$Pt(C{\equiv}CC_6H_4SAc)_2(PPh_3)_2$ (MCBJ) [21], the differences in conductance as a function of the metallic moiety are negligible, and the platinum complexes are as conductive (or resistive) as the ruthenium analogs. The values for 1-Ru and 1-Pt whilst low are at least an order of magnitude higher that the 'five-ring' organic compound $NH_4C_5C{\equiv}C(C_6H_2R_2C{\equiv}C)_3C_5H_4N$ (R = $OC_6H_{13}$; $10^{-6.7}G_0$, 0.015 nS) of comparable molecular length (3 nm) (MCBJ data) [34].

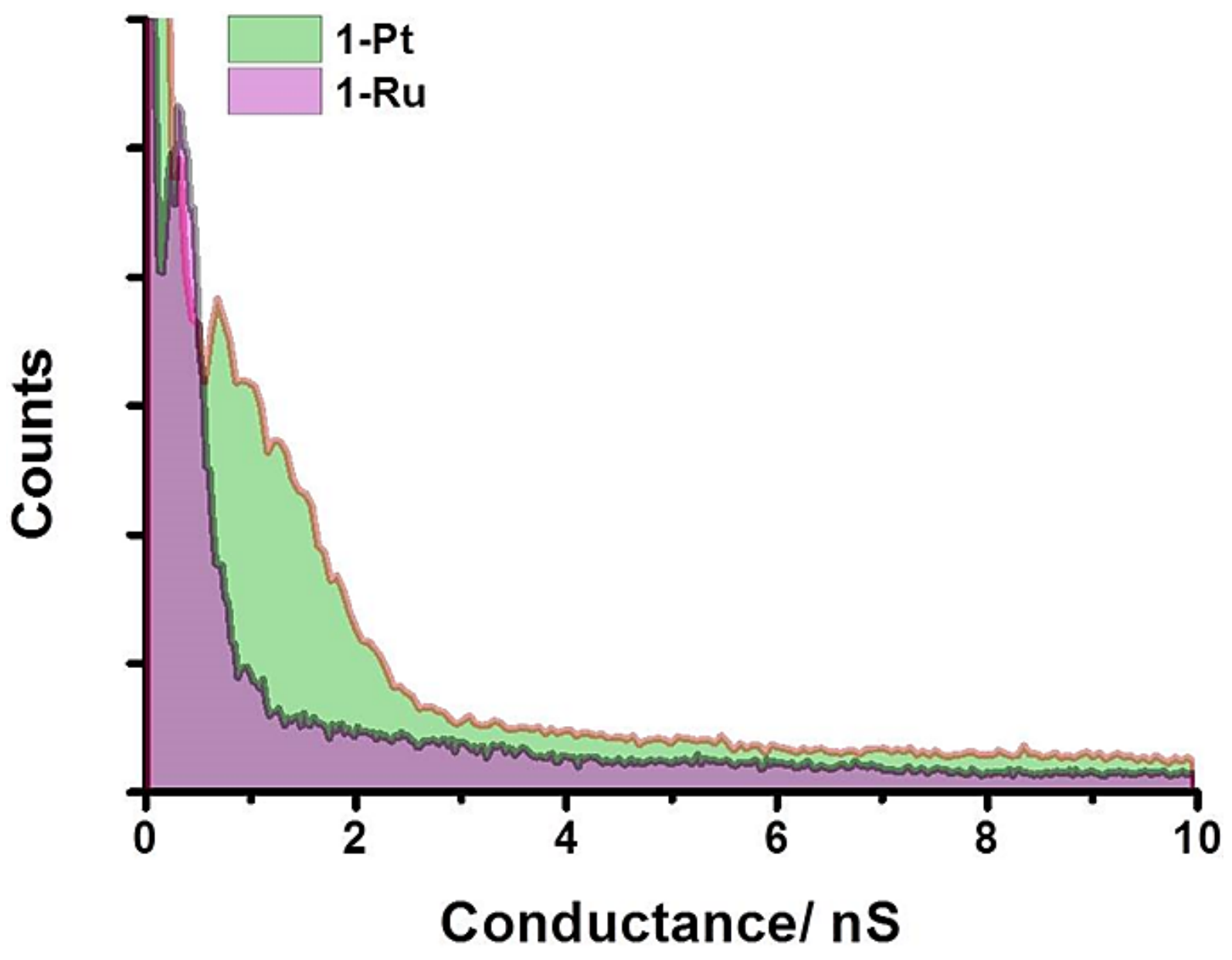

*Figure 7.1: I(s) conductance histograms of 1-Ru and 1-Pt constructed from 500 traces.*

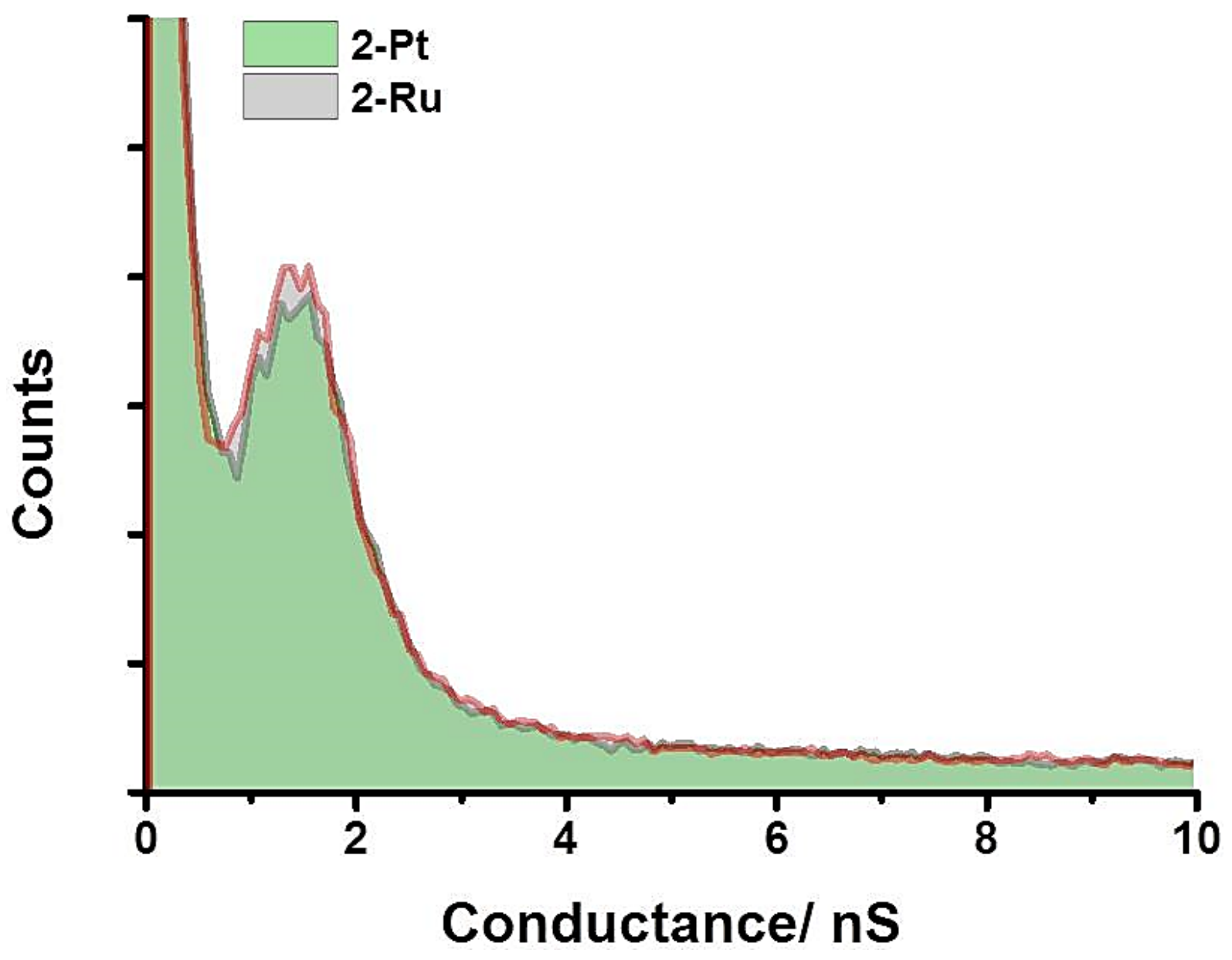

*Figure 7.2: I(s) conductance histograms of 2-Ru and 2-Pt constructed from 500 traces.*

In seeking to better understand these trends in conductance behavior, the electronic properties of the molecules and the electrical behavior of the junctions have been investigated by using DFT-based methods.

Plots of the highest occupied and lowest unoccupied molecular orbital (HOMO and LUMO, respectively) are given in figure 7.3, and analysis of the energy and distribution of the frontier molecular orbitals is summarized in tables 7.1 and 7.2.

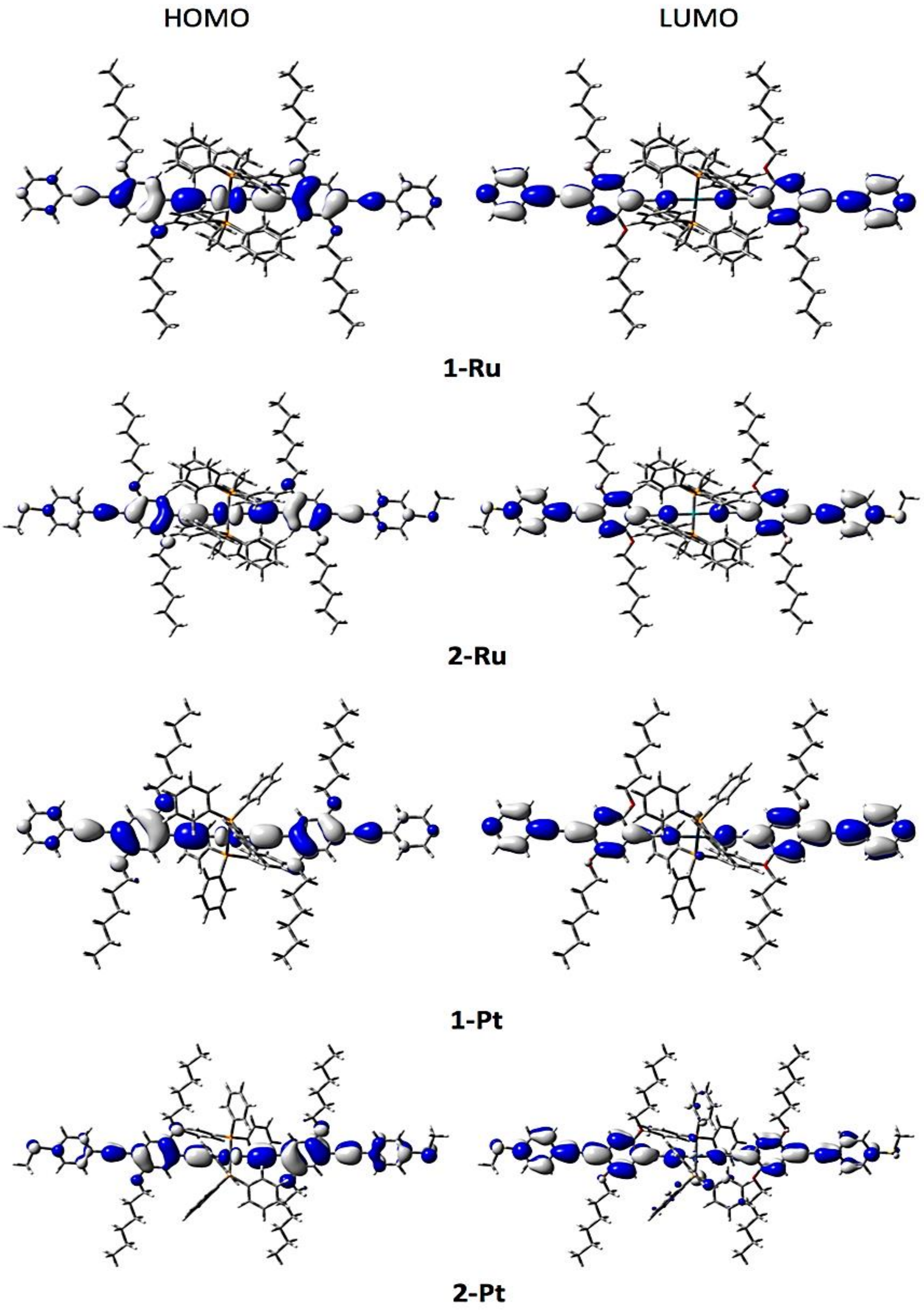

*Figure 7.3. The relaxed molecules in a gas phase and iso-surfaces of the HOMOs and LUMOs for 1-Ru, 1-Pt, 2-Ru and 2-Pt*

*Table 7.2: Composition (%) of the HOMOs and LUMOs of 1-Ru, 1-Pt, 2-Ru and 2-Pt.*

| | | **1-Ru** | |
|---|---|---|---|
| | Ru | dppe | $C{\equiv}CC_6H_4(OC_6H_{13})_2C{\equiv}CC_5H_4N$ |
| **LUMO** | 0 | 2 | 98 |
| **HOMO** | 25 | 3 | 72 |
| | | **1-Pt** | |
| | Pt | $PPh_3$ | $C{\equiv}CC_6H_4(OC_6H_{13})_2C{\equiv}CC_5H_4N$ |
| **LUMO** | 2 | 3 | 95 |
| **HOMO** | 6 | 2 | 92 |
| | | **2-Ru** | |
| | Ru | dppe | $C{\equiv}CC_6H_4(OC_6H_{13})_2C{\equiv}CC_6H_4SMe$ |
| **LUMO** | 0 | 2 | 97 |
| **HOMO** | 22 | 3 | 76 |
| | | **2-Pt** | |
| | Pt | $PPh_3$ | $C{\equiv}CC_6H_4(OC_6H_{13})_2C{\equiv}CC_6H_4SMe$ |
| **LUMO** | 4 | 10 | 86 |
| **HOMO** | 5 | 1 | 94 |

The HOMOs of the ruthenium complexes display the familiar pattern of dπ-pπ interactions along the metal-ethynyl axis [51], and extend along the molecular backbone. The nodal pattern of the HOMOs in the Pt complexes is similar, with a smaller metal contribution (figure 7.3). The LUMOs are also delocalized over the molecular backbones and can largely be described as the π* system of the diethynylarylene ligands with little (Pt) or no (Ru) metal character. These varying metal contributions are reflected in the relative orbital energies, with the significant Ru contribution to the HOMO in 1-Ru and 2-Ru resulting in these orbitals lying some ca. 0.25 eV higher in energy than in the Pt analogues 1-Pt and 2-Pt. The largely organic π* based LUMOs lead to less significant differences in LUMO energies, which differ by only 0.02 – 0.05 eV (table 7.1).

However, these frontier orbital distributions per se do not provide evidence relating to the mechanisms of conductance, which is instead dominated by the alignment of the key molecular orbitals with the Fermi level of the electrodes. As noted by Georgiev and McGrady in computational studies of the conductance properties of extended metal atom chain complexes, the dominant conductance channel need not necessarily be associated with a molecular orbital evenly distributed along the molecular backbone; for example, a dominant conduction channel in $Cr_3(dpa)_4(NCS)_2$ (dpa = dipyridylamide) is derived from a non-bonding combination of metal $d_{z^2}$ orbitals directed along the Cr-Cr-Cr axis and localized on the terminal chromium atoms [52].

To provide further insight into the experimentally observed trends obtained using the *I*(*s*) technique, and to better evaluate the properties and behavior of these molecular junctions, calculations using a combination of DFT and a non-equilibrium Green's function formalism were also carried out. It is well-known that the Fermi energy predicted by DFT is often not reliable, and as such the room temperature electrical conductance G was computed for a range of Fermi energies $E_F$; the calculated G is plotted as a function of $E_F - E_F^{DFT}$ in figure 7.4. A single common value of $E_F$ which gave the closest overall agreement between theory and experiment has been chosen.

Interestingly, a small value of $E_F - E_F^{DFT} = -0.07\ eV$ (figure 7.4), which has been chosen and used in all of the theoretical results described below. Thygesen *et al.* [53], have discussed similar situations for $C_{60}$ contacted molecular wires, and shown that critical molecular orbitals can become pinned close to the Fermi level due to the partial charge transfer and leading to good quantitative agreement between calculated and experimentally determined conductance. As shown below, the LUMO states of 1-M and 2-M (M = Ru, Pt) tail near the Fermi level in a manner similar to the Thygesen system, and partial charge transfer may also be responsible for the good agreement observed here.

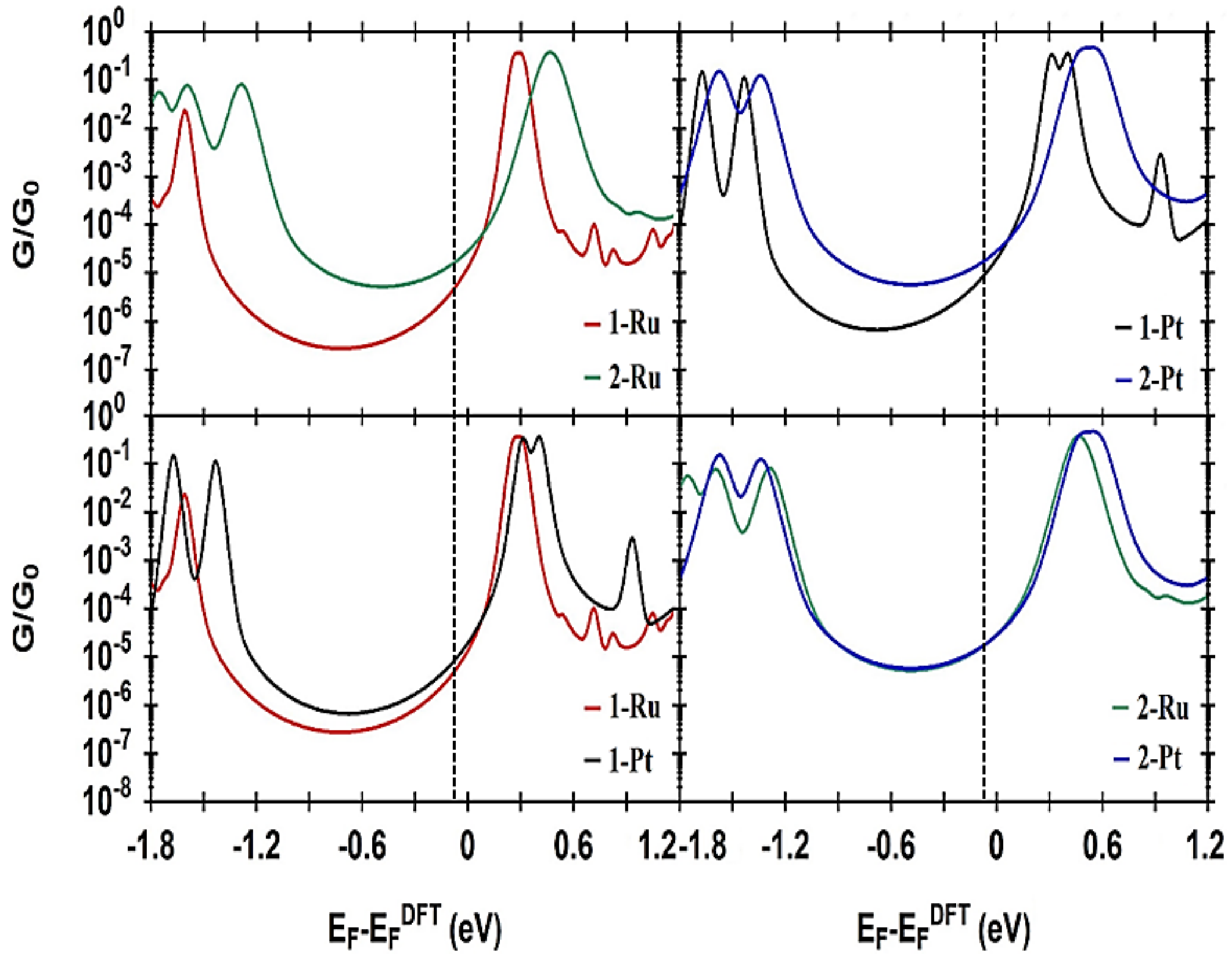

*Figure 7.4: Plots showing selected comparisons of calculated conductance as a function of the Fermi energy for molecular junctions 1-Ru, 1-Pt, 2-Ru and 2-Pt. Black dashed lines show the chosen Fermi energy ($E_F$ = -0.07 eV).*

The results of the room temperature conductance calculations are summarized in table 7.1 and comparisons between pairs of molecules according to contacting group and metal complex fragment are illustrated in figure 7.4. It is immediately apparent that the conductance of the methyl thioether-contacted molecules 2-M is approximately three to four times higher than the analogous pyridine contacted species 1-M, in good agreement with the experimental trends (figure 7.4, top row, table 7.1). The greater conductance of the methyl thioether-contacted compounds 2-M likely arises from the greater Au-S bond strength and the broadening of the LUMO resonances arising from these interactions versus the pyridine-contacted analogues 1-M.

More surprising is the limited influence of the metal-phosphine fragment on the molecular conductance (figure 7.4, bottom row), which can be explained by the relative energy of the Fermi level and the molecular LUMOs together with a conductance mechanism based on a tunneling process through the tails of the respective LUMO states. Although tunneling through pyridine-terminated compounds is usually attributed to LUMO-based transport [45, 54, 55], the methyl thioether contact has been shown to permit both HOMO- and LUMO-based conductance mechanisms, depending on the nature of the molecular backbone [56]. Here it appears that the similar conductance values obtained from both series of compounds reflects the similar nature, energy and composition of the LUMOs, which provide a conductance channel between the electrodes. This contrasts with the recently-reported single-molecule conductance studies of *trans*-Ru(C≡$CC_5H_4N$)$(LL)_2$ (LL = dppe, dmpe, $\{P(OMe)_3\}_2$) with the shorter alkynyl pyridine ligands in which the ligand π* levels are likely to be much higher in energy than the extended alkynyl-based ligands in compounds 1-M and 2-M, and a HOMO-mediated conductance channel is proposed [27].

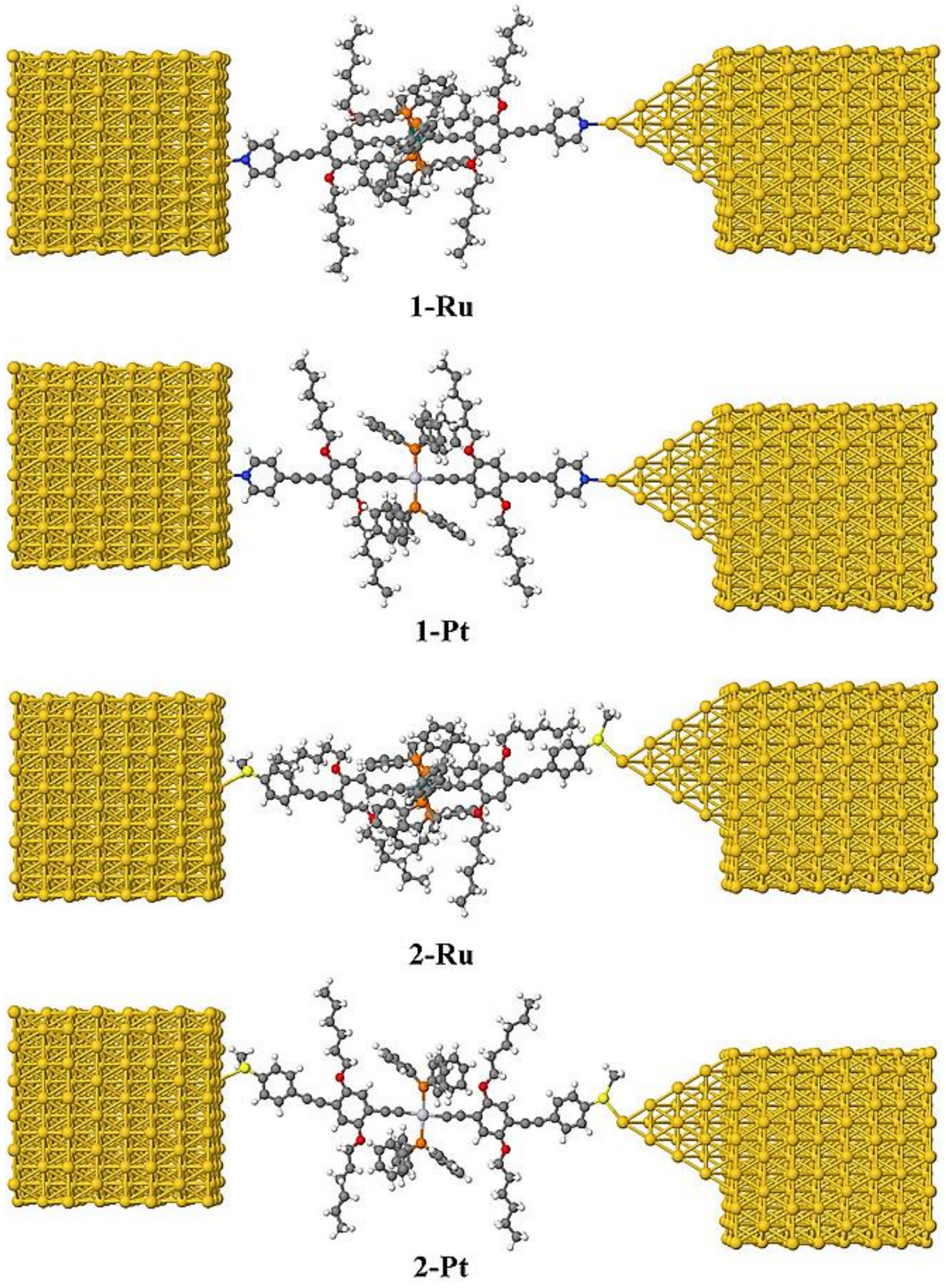

*Figure 7.5: The relaxed geometries of molecular junctions of 1-Ru, 1-Pt, 2-Ru and 2-Pt.*

The optimized junction geometries conform well to a description of the pyridine contacted compounds 1-Ru and 1-Pt forming point contacts between the pyridine nitrogen atom and the under-coordinated gold atoms of the gold electrodes.

As expected, figure 7.5 shows that the methyl thioether contacted compounds 2-Ru and 2- Pt are not oriented normal to the idealized, flat electrode surface within the molecular junction. Rather, they are tilted within molecular junctions to accommodate the directionality of the lone pairs of electrons on the sulfur atoms that bind to the gold electrodes [50, 49]. The calculated molecular lengths and experimental break-off distances are consistent with these interpretations (table 7.1).

## 7.4. Summary

In summary, the single molecule conductance of two pairs of *trans*-bis(alkynyl) organometallic complexes based on $Ru(dppe)_2$ and $Pt(PPh_3)_2$ fragments and methyl thioether and pyridyl surface contacting groups have been studied theoretically and experimentally. Perhaps surprisingly, the nature of the metal moiety is a less significant point of chemical control over the electrical properties of the junction, with $Pt(PPh_3)_2$ based complexes being essentially as conductive (or as resistive) as the analogous $Ru(dppe)_2$ derivatives. The conductance of these compounds is more dependent on the position of the LUMO resonance with respect to the Fermi level of the junction, and largely influenced by the electrode-molecule contact. The energy and distribution of the molecular LUMOs are qualitatively similar in all of the compounds studied here and can be well described as the ethynylarylene ligand π* orbitals. Given the rather straight-forward synthetic chemistry associated with the preparation of long chain ethynylarylene ligands, this work opens new avenues for the design of metal-complex based molecular wires, including those based on readily- available *trans*-bis(alkynyl) Pt(II) complexes.

# Chapter 8

## Conclusions and Future Works

### 8.1. Conclusions

The electronic properties of various molecules (organic and organometallic molecules) have been studied in this thesis using density functional theory and the Green's function formalism which are described in chapters 2 and 3 respectively.

Chapter 4 documented studies of the charge transport of pyridyl terminated OPE derivatives using the MCBJ and STM-BJ techniques, DFT-based theory and analytic Green's functions, and investigated the interplay between QI effects associated with central and terminal rings in molecules of the type X-Y-X. The results demonstrated that the contribution to the conductance from the central ring is independent of the para or meta nature of the anchor groups and the combined conductances satisfy the quantum circuit rule $G_{ppp}/G_{pmp} = G_{mpm}/G_{mmm}$. For the simpler case of a two-ring molecule, the circuit rule $G_{pp}G_{mm} = G_{pm}^2$ is satisfied (see figure 4.14). It should be noted that the circuit rule does not imply that the conductance $G_{XYX}$ is a product of three measureable conductances associated with rings X, Y and X. Indeed the latter property does not hold for a single molecule. On the other hand, provided sample to sample fluctuations lead to a broad distribution of phases within an ensemble of measurements, a product rule for ensemble averages of conductances can arise. The qualitative relationship between the conductances agrees well with the simple QI picture of molecular conduction. It has been reported that destructive QI exists in benzene with the meta connectivity and is responsible for the observed reduction of conductance, whereas for para and ortho connectivities, constructive QI should be observed. The transmission coefficient calculations through junctions where the metal-ring connection is artificially blocked (figure 4.11) show that the artificially coupled pyridyl ring exhibits similar behaviour to the benzene ring, with destructive QI in the case of the meta coupling significantly reducing the conductance compared with para and ortho connectivities. The dashed curves in the bottom panel in figure 4.11 clearly demonstrate that when the conduction is through only the nitrogen atoms, the conductance of the meta isomer is much lower than in the para and ortho isomers. More realistically, however, in the presence of metal-ring overlap, the effect of varying the positions of the nitrogens in the anchors becomes much weaker, and as demonstrated by figure 4.10, the major changes in the molecular conductance are caused by the variations in the connectivity of the central ring. The dominant influence of the central ring is accounted for by the fact that the central ring is not in direct contact with electrodes and therefore no parallel conductance paths are present, which could bypass the ethynylene connections to the anchors.

In a sub nanometre scale molecular circuit, as in standard complementary metal-oxide-semiconductor (CMOS) circuitry, electrical insulation is of crucial importance. Destructive interference in a two-terminal device may not be desirable, because of the lower conductance. However, for a three-terminal device minimizing the conductance of the third terminal is highly desirable, because the third (gate) electrode should be placed as close to the molecule as possible, but at the same time, there should be no leakage current between the molecule and gate. One way of achieving this may be to use an anchor group with built-in destructive interference. Therefore, destructive QI may be a vital ingredient in the design of future three-terminal molecular devices and more complicated networks of interference-controlled molecular units.

Chapter 5 explored and studied the electronic properties of oligoyne-based molecular wires in three different mediums, theoretically and experimentally. It has been demonstrated that the changing of the solvent can lead to changes in both the conductance and the attenuation factor of oligoyne molecular bridges. DFT computations shown that both the molecular junction conductance and the decay constant depend in a very sensitive manner on the position of the contact Fermi energies within the HOMO – LUMO gap. In addition, it has been shown that the interactions between the solvent molecules and the oligoyne-bridges affected the structural features of these molecules, since all bridges have been bended, and that leads to different molecular lengths. By way of an example, the molecular length of 3 in MES-solvent is 0.993 nm, while in TCB and PC are 0.995 and 1.007 nm respectively.

Furthermore, it has been demonstrated that the structures with TCB and PC solvation exhibit slightly stronger Au – TMS contact binding than for mesitylene solvation (-0.41 eV for the structures with MES and -0.44 eV for the structures with TCB and PC).

Finally, these results shown that the solvent environment is an important variable to consider in interpreting conductance measurements and that the environment can give rise to dramatic changes in electronic properties of this kind of molecules.

Chapter 6 presented theoretical and experimental studies of electronic properties of the bis(terpyridyl) complexes. The single-molecule conductance of bis(terpyridyl) complexes [2-M](X)$_2$ and [3-M](X)$_2$ display trends that are more closely associated with the binding energy and ligand structure than the nature of the metal ion.

The calculated conductance as a function of the Fermi energy for the complex series within three different molecular junctions Type I, II and III with $\Theta = 110^{o}$ illustrated that for both molecular contacts, it is clear the conductance of the structures with the Type III configurations are the highest. This correlates with their more favourable binding energies. The limited role of the metal ion can be traced in part to the LUMO-based conduction mechanisms that arise from the use of trimethylsilylethynyl and, in the geometries adopted here, thiomethyl binding groups. The optimised structures of the molecular junctions and considerations of the calculated conductance profiles indicate that the most conductive trimethylsilylethynyl contact to the gold electrodes is best described in terms of physisorption at defect sites, explaining the simple conductance profiles observed for compounds contacted through this group. The thiomethyl moiety contacts the gold electrodes in a more chemisorbed fashion, again at defect sites on the gold electrodes in the most conductive junctions. The conductances of [2-M](X)$_2$ are found to be lower than those of [3-M](X)$_2$, which is consistent with single phenylene and triple-bond spacers acting as tunnelling barriers.

Chapter 7, probed the single molecule conductance of two pairs of *trans*-bis(alkynyl) organometallic complexes based on Ru(dppe)$_2$ and Pt(PPh$_3$)$_2$ fragments and methyl thioether and pyridyl surface contacting groups, theoretically and experimentally. Perhaps surprisingly, the nature of the metal moiety is a less significant point of chemical control over the electrical properties of the junction, with Pt(PPh$_3$)$_2$ based complexes being essentially as conductive (or as resistive) as the analogous $Ru(dppe)_2$ derivatives. The conductance of these compounds is more dependent on the position of the LUMO resonance with respect to the Fermi level of the junction, and largely influenced by the electrode-molecule contact. The energy and distribution of the molecular LUMOs are qualitatively similar in all of the compounds studied here and can be well described as the ethynylarylene ligand $\pi^*$ orbitals. Given the rather straight-forward synthetic chemistry associated with the preparation of long chain ethynylarylene ligands, this work opens new avenues for the design of metal-complex based molecular wires, including those based on readily- available *trans*-bis(alkynyl) Pt(II) complexes.

# 8.2. Future Works

In this thesis I have concentrated on the electrical conductance of molecular wires attached to gold electrodes. For the future, one can envisage extending this work in a number of directions. First it would be interesting to examine how results change when the gold electrodes are replaced by alternative metals such as platinum, palladium [1, 2] or iron [3] or combinations of electrode materials [4] . This is particularly important for the technological development of molecular electronics, because gold is not compatible with CMOS technology. Secondly, it would be of interest to extend my study to other transport properties, including thermopower [5 – 8], for which further studies of nanoscale and molecular-scale phonon transport will also be needed [9, 10]. Finally, during the past few years, graphene has emerged as a new electrode material for molecular electronics [11, 12 ] and it is therefore of interest to develop new anchor groups for binding to graphene [13], which preserve coherent electron transport across the molecule-graphene interface .